\documentclass[fleqn,usenatbib]{mnras}

\usepackage{newtxtext,newtxmath}

\usepackage[T1]{fontenc}

\DeclareRobustCommand{\VAN}[3]{#2}
\let\VANthebibliography\thebibliography
\def\thebibliography{\DeclareRobustCommand{\VAN}[3]{##3}\VANthebibliography}

\usepackage{graphicx}	
\usepackage{amsmath}	

\usepackage{xcolor}
\usepackage[version=4]{mhchem}

\usepackage{subcaption}         
\usepackage{lscape}             
\usepackage{placeins}           

\title[The molecule-rich disc of HD\,35929]{MINDS: The molecule-rich disc of the Herbig star HD\,35929 revealed with JWST/MIRI}

\author[T. Kaeufer et al.]{Till Kaeufer$^{1}$\thanks{E-mail: t.kaeufer@exeter.ac.uk},
Rens Waters$^{2,3}$,
Danny Gasman$^{4}$,
Milou Temmink$^{5}$,
Hyerin Jang$^{2}$,
Ewine F. van Dishoeck$^{5,6}$,\newauthor
Manuel G\"udel$^{7,8}$,
Thomas Henning$^4$,
Alessio Caratti o Garatti$^{9}$,
Inga Kamp$^{10}$,
Aditya M. Arabhavi$^{10}$,\newauthor
Pac\^{o}me Esteve$^{11}$,
Sierra L. Grant$^{12}$,
Jayatee Kanwar$^{13}$,
Nicolas T. Kurtovic$^{6}$,
Giulia Perotti$^{14,4}$,\newauthor
Kamber Schwarz$^{4}$,
Lucas M. Stapper$^{4}$,
and Beno\^{i}t Tabone$^{11}$
\\
$^{1}$Department of Physics and Astronomy, University of Exeter, Exeter EX4 4QL, UK\\
$^{2}$Department of Astrophysics/IMAPP, Radboud University, PO Box 9010, 6500 GL Nijmegen, The Netherlands\\
$^{3}$SRON Netherlands Institute for Space Research, Niels Bohrweg 4, NL-2333 CA Leiden, The Netherlands\\
$^4$Max-Planck-Institut f\"{u}r Astronomie (MPIA), K\"{o}nigstuhl 17, 69117 Heidelberg, Germany\\
$^{5}$Leiden Observatory, Leiden University, PO Box 9513, 2300 RA Leiden, The Netherlands\\
$^{6}$Max-Planck Institut f\"{u}r Extraterrestrische Physik (MPE), Giessenbachstr. 1, 85748, Garching, Germany\\
$^7$Dept. of Astrophysics, University of Vienna, T\"urkenschanzstr. 17, A-1180 Vienna, Austria\\
$^8$ETH Z\"urich, Institute for Particle Physics and Astrophysics, Wolfgang-Pauli-Str. 27, 8093 Z\"urich, Switzerland\\
$^{9}$INAF – Osservatorio Astronomico di Capodimonte, Salita Moiariello 16, 80131 Napoli, Italy\\
$^{10}$Kapteyn Astronomical Institute, Rijksuniversiteit Groningen, Postbus 800, 9700AV Groningen, The Netherlands\\
$^{11}$Universit\'e Paris-Saclay, CNRS, Institut d’Astrophysique Spatiale, 91405, Orsay, France\\
$^{12}$Earth and Planets Laboratory, Carnegie Institution for Science, 5241 Broad Branch Road, NW, Washington, DC 20015, USA\\
$^{13}$Department of Astronomy, University of Michigan, 1085 S. University, Ann Arbor, MI 48109, USA\\
$^{14}$Niels Bohr Institute, University of Copenhagen, NBB BA2, Jagtvej 155A, 2200 Copenhagen, Denmark\\
}

\date{Accepted 2025 November 17. Received 2025 October 31; in original form 2025 September 5}

\pubyear{\the\year{}}

\begin{document}
\label{firstpage}
\pagerange{\pageref{firstpage}--\pageref{lastpage}}
\maketitle

\begin{abstract}
Our knowledge of the chemical composition of the gas in the inner disc of intermediate-mass young stars is limited, due to the lack of suitable instrumentation. The launch of JWST has provided a significant improvement in our ability to probe gas in these inner discs.
We analyse the gas composition and emitting conditions of the disc around HD\,35929, a young intermediate-mass Herbig star, using MIRI/MRS data. Our goal is to constrain the chemistry and kinematics of the gas phase molecules detected in the inner disc.
We use iSLAT to examine the observed molecular lines and DuCKLiNG to detect, fit, and analyse the molecular emission.
We find gas phase H$_2$O, CO, CO$_2$, and OH in the disc, as well as HI recombination lines. Surprisingly, we also detect gas phase SiO in the fundamental $v$=1-0 vibrational band. We derive column densities and temperature ranges of the detected species, arising from the inner $\sim0.2\,\rm au$, hinting towards a compact and very warm disc. The molecular column densities are much higher than found in lower mass T~Tauri discs.
In general, the molecular composition is consistent with an O-rich gas from which silicate-rich solids condense and the strong gas phase molecular line emission suggests a low dust opacity. The unexpected detection of gas phase SiO at the source velocity points to an incomplete condensation of rock forming elements in the disc, suggesting chemical disequilibrium and/or an underestimate of the gas kinetic temperature.
\end{abstract}

\begin{keywords}
protoplanetary discs -- infrared: general -- techniques: spectroscopic -- methods: data analysis
\end{keywords}



\section{Introduction}
\label{sec:intro}
Herbig Ae/Be stars (hereafter referred to as Herbig stars)  \citep{herbig_spectra_1960}  are intermediate-mass  (1.5 to $\sim8$ M$_{\odot}$) pre-main-sequence (PMS) stars surrounded by a planet forming disc \citep[for a recent review, see][]{brittain_herbig_2023}.  Herbig stars represent a relatively old (a few to ~10 Myrs) population of PMS stars, as their spectral types are between B and F \citep{vioque_gaia_2018}. Nevertheless, compared to lower mass T~Tauri stars, their disc masses are higher \citep{stapper_complete_2025}.  The younger counterparts to Herbig stars are the intermediate mass T Tauri stars \citep{2004AJ....128.1294C,valegard_what_2021}.

Because of their brightness, Herbig star discs have been studied extensively. While atomic emission lines are frequently detected in UV, optical and infrared spectra (e.g., HI Balmer lines, [OI]), little is known about the molecular gas composition of the inner disc.  Previous studies using the Infrared Space Observatory (ISO) and the Spitzer Space Telescope showed that the mid-infrared (mid-IR) spectra of Herbig stars are characterized by strong dust emission from amorphous and crystalline silicates, and by Polycyclic Aromatic Hydrocarbons \citep[e.g.,][]{malfait_spectrum_1998, meeus_iso_2001, juhasz_dust_2010, acke_spitzers_2010}. However, Herbig star discs often lack the molecular line emission that characterizes the mid-IR spectra of lower mass T~Tauri stars \citep[e.g.][]{pontoppidan_spitzer_2010}. This has been attributed to the bright mid-IR to far-IR continuum emission from the disc, which hampers the detection of weak emission lines at low spectral resolution \citep{antonellini_understanding_2015, antonellini_water_2016}. Also, the strong UV field of warm Herbig stars may photodissociate molecules in the disc \citep{fedele_water_2011}. 

In the 1 to 5 $\mu$m wavelength range, Herbig discs regularly show CO ro-vibrational line emission when observed at high spectral resolution \citep[e.g.][]{brittain_warm_2007}. The CO is excited both thermally as well as through IR pumping and UV fluorescence, and originates both from the innermost disc regions, close to the dust sublimation radius \citep[thermal excitation and IR pumping,][]{blake_high-resolution_2004}, and from the outer disc surface (UV fluorescence). The inner disc CO emission is correlated with disc geometry as noted by \citep{van_der_plas_structure_2008}: discs classified as Group~I \citep{meeus_iso_2001}, which show large gaps and dust-depleted inner discs, have large inner radii of CO ro-vibrational line emission. In contrast, Group~II discs have no large gaps and can show much smaller CO emitting inner radii, that coincide with the dust inner radius; these discs may show smaller, inner disc gaps at spatial scales of a few au \citep{menu_midi}.  \cite{hein_bertelsen_proposed_2016} show through modeling that the disc geometry affects the CO line widths, providing a tool to constrain  disc structure. \cite{banzatti_observing_2018,banzatti_scanning_2022} noted that the CO inner radius increases when the near-IR excess decreases. This is consistent with a picture in which the inner radius of CO increases as dust in the inner disc, or the inner disc surface layer, is depleted. A small group of Herbig stars, including HD\,35929, shows CO first overtone emission, indicative of a very hot and dense inner disc molecular reservoir \citep{ilee_investigating_2014}. In addition to CO, both H$_2$O and OH have been detected in Herbig star discs at near-IR wavelengths \citep{fedele_water_2011,adams_water_2019,Banzatti2023}. OH emission was found to occur more frequently in Group~I discs \citep{brittain_study_2016}. In only one Herbig star, HD~101412, CO$_2$ was detected using \emph{Spitzer} \citep{pontoppidan_spitzer_2010}. 

The launch of the James Webb Space Telescope (JWST) has opened the possibility to search for weak mid-IR molecular gas line emission arising from the inner regions of planet forming discs. Initial results show a wide diversity in the inner disc chemistry, with molecules as H$_2$, CO, H$_2$O, CO$_2$, OH, C$_2$H$_2$, and HCN commonly detected in solar mass T Tauri discs, while a rich variety of hydrocarbons has been found in discs surrounding lower mass ($<$ 0.2-0.3 M$_{\odot}$) objects \citep[e.g.][]{Perotti2023,grant_minds_2023,arabhavi_abundant_2024,pontoppidan_high-contrast_2024,colmenares_jwstmiri_2024, arabhavi_minds_2025,Temmink2025,Arulanantham2025}.  Most well-studied Herbig stars are too bright for JWST/MIRI. For that reason, only four Herbig stars were included in the MIRI mid-INfrared Disk Survey \citep[MINDS,][]{kamp_chemical_2023,henning_minds_2024} program. Here we present the MIRI spectrum of one of these, HD\,35929.  This paper is organised as follows: in Section~\ref{sec:obs_perspect} we discuss the observation and it's context. This includes the properties of HD\,35929 and its evolutionary stage (Sect.~\ref{sec:sed}), the data reduction of the JWST/MIRI data (Sect.~\ref{sec:obs}), the inventory of atoms, molecules, and dust identifiable by eye (Sect.~\ref{sec:inventory}), and the kinematic information from spectrally resolved lines (Sect.~\ref{sec:kinematics}). Section~\ref{sec:model} describes a modelling approach to analyse the data (introduced in Sect.~\ref{sec:fitting}) with the results of the fitting described in Section~\ref{sec:results}.  In Section \ref{sec:discussion} we discuss the implications of our findings. Finally, Section \ref{sec:conclusions} contains the conclusions of this study.

\section{An observational perspective}
\label{sec:obs_perspect}
\subsection{Properties of HD\,35929}
\label{sec:sed}

\begin{figure}
     \centering
     \includegraphics[width=1.0\linewidth]{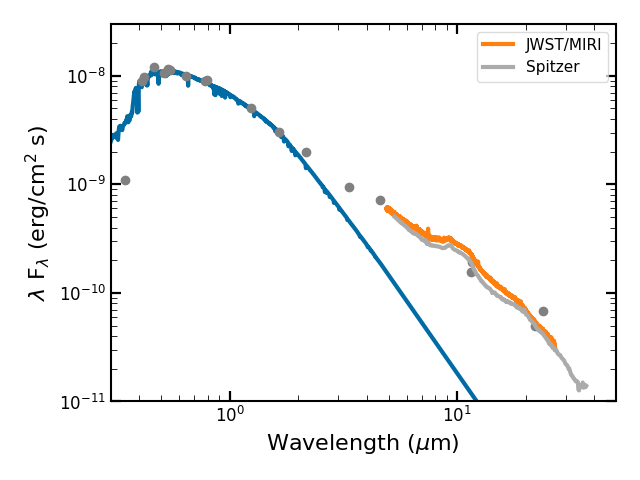}
     \caption{Spectral Energy Distribution of HD\,35929. The blue curve is a Kurucz model with T$_{\rm eff}$ = 6500\,K and log g = 3.5, reddened with A$_V$ = 0.20 mag. The orange and grey lines represent the JWST/MIRI and Spitzer/IRS spectra, respectively.  Also shown are photometric points (Str\"omgren, GAIA, 2MASS, IRAC, WISE, and IRAS).}
     \label{fig:SED}
 \end{figure}

HD\,35929 is an F2 IV-V star located in the direction of the Orion OB1 association. The distance of 380\,pc \citep{Gaia2020} places it firmly in the Orion star-forming region.  In the HR diagram, the star is located above the main sequence \citep{fairlamb_spectroscopic_2015,vioque_gaia_2018,valegard_what_2021}. These studies used PMS evolutionary tracks to constrain the stellar mass and age to 2.9\,M$_{\odot}$ and 1.1-1.7\,Myrs, respectively. From optical spectroscopy, an effective temperature of 7000\,K has been inferred, and a metallicity of [Fe/H] =-0.2 \citep{miroshnichenko_fundamental_2004}. However, we note that the GAIA derived temperature is only 6400\,K, which would imply a significantly lower [Fe/H] of $\sim$ -0.4. 
We have used the ARIADNE software tool \citep{vines_span_2022} to fit the stellar Spectral Energy Distribution (SED), shown in Fig.~\ref{fig:SED}. Fitting a grid of Kurucz model atmospheres \citep{Kurucz1979} to these data results in a temperature of 6500\,K, a stellar radius of 6.7\,R$_{\odot}$ and a luminosity of 74\,L$_{\odot}$, in good agreement with previous studies, and would suggest a metallicity of [Fe/H] of about -0.4. \cite{miroshnichenko_fundamental_2004} suggested that HD\,35929 is a post-main-sequence star rather than a young PMS star. Unfortunately, post- and pre- main sequence evolutionary tracks intersect in this part of the HR diagram, which therefore cannot help to distinguish post- and pre-main-sequence stars. We note that HD\,35929 is not closely associated with a cluster in Orion \citep{2023AJ....166..183V}, and that it is classified as a $\delta$ Scuti pulsating star \citep{marconi_pulsation_2000} of the $\gamma$ Dor class \citep{barcelo_forteza_unveiling_2020}. 

The SED (Fig.~\ref{fig:SED}) shows that the IR excess of HD\,35929 is modest compared to other Herbig stars, and the very steep negative slope of the IR continuum classifies it as a Group II disc. \cite{Stapper2025A&A...693A.286S} using ALMA find a dust mass in HD\,35929 of 0.23\,M$_{\oplus}$ and an upper limit to the disc radius of 17 au. The gas accretion rate onto HD\,35929 has been measured using different diagnostics \citep{fairlamb_spectroscopic_2015, grant_tracing_2022,ilee_investigating_2014,wichittanakom_accretion_2020}, leading to  mass accretion rates between $10^{-6.8}$ and $10^{-6.31}$ M$_{\odot}$yr$^{-1}$. Adopting the canonical gas-to-dust ratio of 100, this leads to an unrealistic  short disc lifetime of a 3-4 years. We will return to this point in Sect.~\ref{sec:dis}. \cite{grady_ensuremathbeta_1996} report Type III P Cygni emission line profiles in the MgII 2802.7 Å, 2795 Å lines as well as strong emission from CIV and OI. \cite{valenti_iue_2000} report far-UV line emission from SI, OI, and SiII.

Near-IR ro-vibrational line emission from CO in the fundamental v=1-0 and v=2-0 overtone bands has been detected \citep{ilee_investigating_2014,banzatti_scanning_2022}, pointing to a compact rotating ring of dense gas with high densities and high column densities, probably from a dust-free region with a radius of 0.21\,au, well inside the dust sublimation radius \citep[estimated at 0.4\,au,][]{banzatti_scanning_2022}. The \emph{Spitzer} IRS spectrum of HD\,35929 shows weak silicate emission, dominated by large ($2-5\,\rm \mu m$) amorphous grains with SiO$_2$, SiO$_3$ and SiO$_4$ composition \citep{juhasz_dust_2010}, while no evidence for $0.1\, \rm \mu m$ sized amorphous silicates was found. The absence of small grains suggests a low dust opacity, possibly exposing more gas above the optically thick dust layers. \cite{lazareff_structure_2017} marginally resolve the inner disc in the H band using Pionier at the Very Large Telescope Interferometer (VLTI) and find a size of $\sim$ 0.21\,au, and a poorly constrained disc inclination angle of $31.8 \pm 16\,\rm  degrees$.

\subsection{Observations and data reduction}
\label{sec:obs}

HD\,35929 was observed with JWST/MIRI \citep{wright_mid-infrared_2023,argyriou_jwst_2023} as part of the MINDS GTO program. The observation used a point source positive dither pattern without target acquisition, for a duration of approximately 22 minutes per band. The data were reduced using version 1.16.1 of the standard JWST pipeline \citep{pipeline}. The fringes were corrected using both the \texttt{CALWEBB\_SPEC2} fringe flat and residual fringe correction steps, along with a final residual fringe correction on the extracted spectrum. The spectrum was extracted per band from the cube images, using a double the Full Width at Half Maximum (FWHM) circular aperture and an annulus to estimate the background. The pipeline outlier detection step was applied, and outliers were replaced using the \texttt{pixel\_replace} option in \texttt{CALWEBB\_SPEC3}. Because the spectrum is full of emission lines, an estimate of the flux uncertainties was difficult to make. The S/N of the data, derived from the MIRI pipeline, ranges between $\sim100$ and ~$\sim800$ (for the wavelength range up to $25\,\rm \mu m$).  

\subsection{Atomic, molecular and dust inventory}
\label{sec:inventory}

\begin{figure*}
    \centering
    \includegraphics[width=1.0\linewidth]{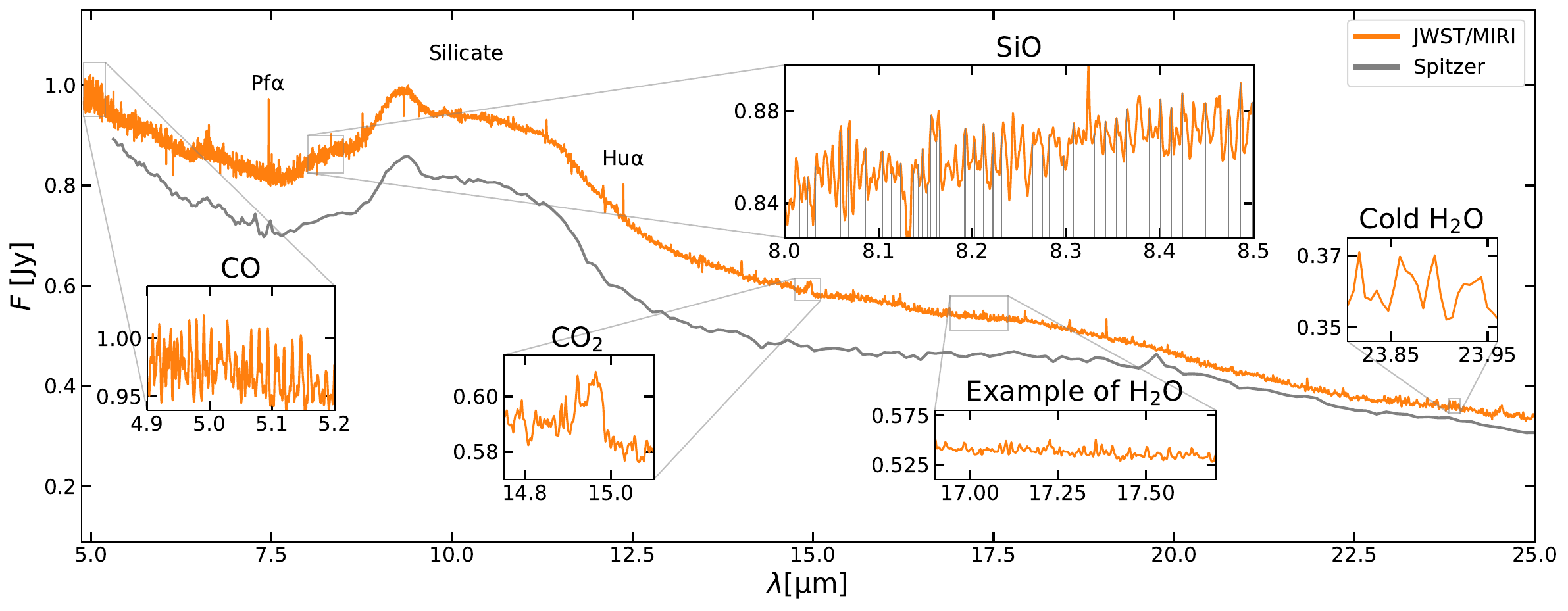}    \caption{JWST/MIRI MRS spectrum (orange) of HD\,35929. The Spitzer/IRS spectrum (grey) is shown for comparison. The insets highlight the emission from \ce{CO}, \ce{CO2}, \ce{H2O}, and most notably the SiO fundamental vibrational mode. The vertical lines in the SiO panel indicate the positions of the rotational lines with upper level energies lower than 3600\,K.}
    \label{fig:MIRI}
\end{figure*}

The MIRI spectrum shown in Fig.~\ref{fig:MIRI} is dominated by thermal emission from warm silicate grains in the disc. The $10\, \rm \mu m$ silicate band shows a narrow peak near $9.2\, \rm \mu m$, often associated with silicates with an SiO$_2$ composition \citep{sargent_silica_2008}. Superimposed is prominent emission from ro-vibrational lines of H$_2$O, and at longer wavelengths many weaker H$_2$O rotational lines.  Other molecules that can easily be identified are CO v=1-0, v=2-1 and v=3-2 P branch emission (Fig.~\ref{fig:rot-co} in Appendix~\ref{sec:CO}), SiO v=1-0, the CO$_2$ bending mode branch at $14.98\, \rm  \mu m$ and pure rotational lines of OH. We do not find evidence for \ce{H2} line emission. To our knowledge, this is the first detection of gas phase SiO vibrational line emission in a disc orbiting an intermediate-mass young star. Therefore, Sect.~\ref{sec:detect-mol}-\ref{sec:mol_conditions} highlight the robustness of this detection. In addition, a number of HI recombination lines are detected, as well as the [FeII] $\lambda$ 17.395 $\mu$m fine structure line. The JWST/MIRI and \emph{Spitzer}/IRS spectra (Fig. \ref{fig:MIRI}) show 5-25 percent differences in continuum flux, that are likely larger than calibration uncertainties. The ro-vibrational water band near $6.5\, \rm \mu m$ is clearly visible in both datasets. We note that the persistent disc emission between the \emph{Spitzer} and JWST/MIRI epochs, separated by 2 decades, is inconsistent with the disc lifetime mentioned above. In the next sections, we will describe how we quantify the gas emission from the disc starting with the line kinematics in Sect.~\ref{sec:kinematics}.

\subsection{Emission line kinematics}
\label{sec:kinematics}

The large line widths ( $\sim 100\, \rm  km~s^{-1}$) of the double-peaked CO ro-vibrational lines in HD\,35929 reported by \cite{banzatti_scanning_2022} suggest that CO is in a rotating gaseous disc observed at a non-zero inclination, consistent with the inclination inferred by \cite{lazareff_structure_2017}. Therefore, the molecules and atomic lines detected in the MIRI data at much lower spectral resolution may be spectrally resolved \citep{banzatti_water_2025}. For instance, the $4.9-5.2\, \rm \mu m$ spectral region indeed shows broad CO lines, while the presence of CO v=2-1 and v=3-2 hot band emission lines results in significant line blending (Fig.~\ref{fig:rot-co}). Similar line kinematic broadening and line blending is seen for the other detected molecules. To quantify the kinematics of the gas, we used iSLAT \citep{Jellison2024} to fit Gaussian line profiles to individual emission lines of molecules and HI gas, avoiding line blends as much as possible. The choice of water lines was based on \cite{banzatti_water_2025}, probing both pure rotational as well as ro-vibrational lines, and different excitation levels. We accounted for the instrumental spectral resolution given by \cite{pontoppidan_high-contrast_2024}. 

\begin{figure}
    \centering
    \includegraphics[width=1.0\linewidth]{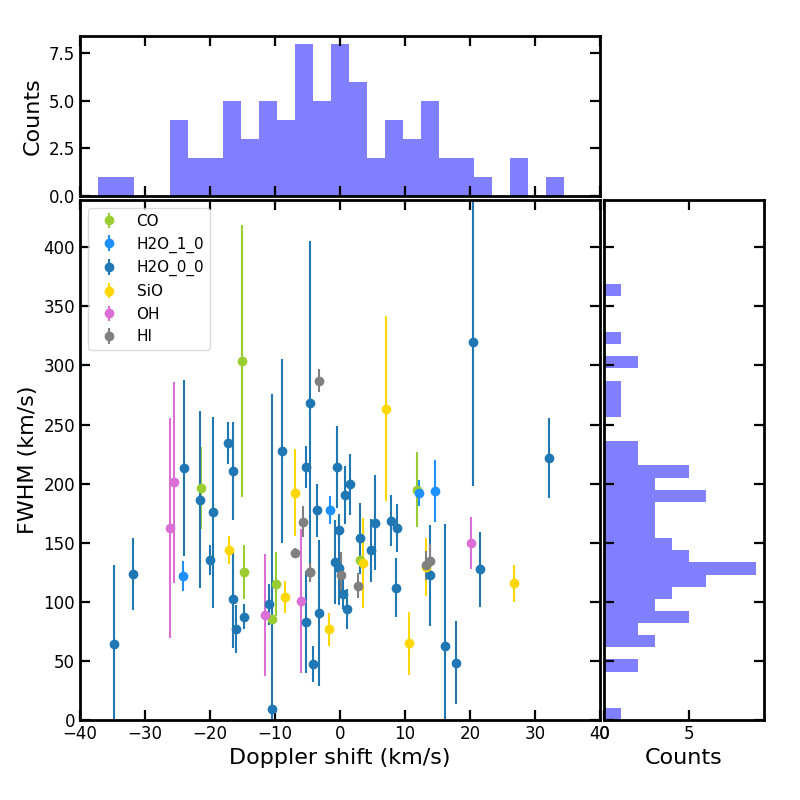}
    \caption{Full Width at Half Maximum line widths of selected emission lines as a function of their Doppler shift. A stellar radial velocity of 22 km~s$^{-1}$ was adopted.}
    \label{fig:FWHM}
\end{figure}

The results are listed in Table~\ref{tab:lines} (Appendix~\ref{sec:meas-emis-line}). We show the instrument corrected FWHM as a function of the Doppler shift in Fig.~\ref{fig:FWHM}, where we applied a stellar heliocentric radial velocity of $22\, \rm km~s^{-1}$ (H. van Winckel, private comm.).  It is clear that the emission lines are spectrally resolved, consistent with line widths obtained from ground-based high spectral resolution observations by \cite{banzatti_scanning_2022}. We find FWHM values between $\sim 50$ and $\sim 300\, \rm  km~s^{-1}$, but many lines are clustered near $\sim 130-140\, \rm  km~s^{-1}$, with a tail towards higher values. We cannot exclude the possibility that some of the broader lines are still affected by blends, and the lower range of FWHM is more likely to represent the actual kinematic broadening. This is supported by the width of the strongest HI lines, whose width is not affected much by any molecular line contribution (Table \ref{tab:lines}), and by the spectrally resolved v=1-0 CO lines reported by \cite{banzatti_scanning_2022} that show a FWHM of 125 km~s$^{-1}$. The FWHM of the emission lines is similar for different molecular species as well as for the HI lines, and does not depend on the upper level energy or the wavelength of the transition. This suggests that the line emitting region is similar for all species. The median Doppler shift of the lines is $-4\, \pm 36~ \rm km~s^{-1}$; taking only lines with a FWHM less than $200\,\rm  km~s^{-1}$, this becomes $-2\ \pm 13~, \rm km~s^{-1}$. This small non-zero value shows that the bulk of the gas is not flowing out or accreting, and is consistent with rotational broadening. Finally, we mention the [FeII] fine structure line observed at 17.9372 $\mu$m, with a FWHM of $ 161\ \pm~ 12\ \rm km~s^{-1}$ at a velocity of $-60\ \pm 5~\ \rm km~s^{-1}$. 

The large line widths imply that the (molecular) gas must be close to the star. \cite{lazareff_structure_2017} derive an uncertain disc inclination angle of $31.8 \pm 16\, \rm  degrees$. Assuming a typical projected rotational velocity of 65 km/s, and adopting the stellar parameters given in Section \ref{sec:sed}, this corresponds to a distance scale of 0.17\,au or 5.3 R$_*$. This is roughly consistent with the spatial scale of Br$\gamma$ emission detected in Herbig stars \citep{gravity_collaboration_gravity_2019}. We conclude that the gas is located close to, or inside the dust sublimation distance of 0.4 au. Next, we further analyse the gas emission using 1D modelling.

\section{A modelling perspective}
\label{sec:model}

\subsection{Fitting process}
\label{sec:fitting}
We fitted the JWST/MIRI-MRS spectrum of HD\,35929 using the Dust Continuum Kit with Line Emission from Gas \citep[DuCKLinG\footnote{\url{https://github.com/tillkaeufer/DuCKLinG}}, ][]{kaeufer_bayesian_2024}. While DuCKLinG has the capability of simultaneous fitting both for the gas and dust composition \citep[e.g.][]{kaeufer_bayesian_2024,kaeufer_disentangling_2024}, in this study, we focus on the gas emission and defer an analysis of the dust mineralogy to a subsequent study. Therefore, the gas emission of the JWST/MIRI spectrum is isolated using the method introduced by \cite{Temmink_continuum_2024}, resulting in the continuum-subtracted spectrum shown in Fig.~\ref{fig:mol-fit}. In addition, we selected two wavelength windows that show the strongest molecular emission ($4.9-9.0\,\rm \mu m$ and $13.5-25\,\rm \mu m$). These regions are highlighted in Fig.~\ref{fig:mol-fit}. We excluded the region around the silicate feature ($9.0-13.5\,\rm \mu m$) from the fit due to poor fitting quality in this region which might originate from unaccounted molecular emission.

The continuum-subtracted spectrum is analysed using DuCKLinG's gas component (fitting all examined molecules simultaneously), which is described in detail by \cite{kaeufer_bayesian_2024}. In brief, this component is based on a large grid (in column density and temperature) of 0D local thermodynamic equilibrium (LTE) slab models \citep[as introduced by][]{tabone_rich_2023,arabhavi_abundant_2024} that can be interpolated and integrated to mimic the emission along a radial temperature and column density power law or can be used to obtain the flux for models with single values for both quantities. The total gas emission is simply the sum of all molecular fluxes (accounting for line overlap for each molecule but not between molecules). We note that the \ce{SiO} slab models use molecular line data from the ExoMol database \citep{Buldyreva2022,Guest2024,Tennyson2024}, while the others are based on HITRAN \citep{Gordon2022}. The slab models are similar to those used in other studies using JWST/MIRI-MRS observations of planet-forming discs of the MINDS, XUE and JDISCS programs.
In this study, we convolved all slab models with the spectral resolution of JWST/MIRI-MRS limited to values not larger than $R=2500$ (for $\lambda<18.0\,\rm \mu m$ $R=2500$ and $R=2000$ beyond that). This accounts for the fact that the spectrum of HD\,35929 shows significant rotational broadening (Sect.~\ref{sec:kinematics}). Additionally, the radial velocity of HD\,35929 of $22\, \rm km \ s^{-1}$ is high enough to leave a visible shift on the observed spectrum. Therefore, we shift the observed spectrum to its rest frame before comparing it to models, which are rebinned to the observed wavelength points using the SpectRes Python package \citep{Carnall2017}.

We fitted the observation using MultiNest \citep{Feroz2008,Feroz2009,Feroz2019} through the Python interface PyMultiNest \citep{Buchner2014}. This Bayesian inference tool uses multimodal nested sampling to determine the Bayesian evidence and posterior distribution for a given likelihood function. The details about the algorithm and the numerical implementation are described in the provided references above.

The fitting quality is determined using a Gaussian likelihood function $\mathcal{L}$ that evaluates the difference between model flux ($F_{\rm i,\rm model}$) and observed flux ($F_{\rm i,\rm obs}$) with respect to the given uncertainty $\sigma_{\rm i}$ per data point $i$ with a given weight ($w_i$) for all wavelengths points ($N_{\rm obs}$)

\begin{align}
    \mathcal{L} = \prod_{i=1}^{N_{\rm obs}}
    \left[\frac{1}{\sqrt{2\pi\, \sigma_{i}^2}}
    \exp{\left(-\frac{\left(F_{i,\rm model} 
    -F_{i,\rm obs}\right)^2}
    {2\,\sigma_{i}^2}\right)}\right]^{w_i} \ .
    \label{eq:likelihood}
\end{align}

As detailed by \cite{kaeufer_bayesian_2024}, the uncertainty $\sigma_{i}$ in this description accounts not just for the observational uncertainty, but also for the mismatch of the model to the observation and is treated as a free parameter ($\sigma$, which is wavelength independent) that is optimised during the fitting.

The likelihood function treats every observational point independently, which effectively results in wavelength regions with more points having a greater influence on the resulting likelihood value. To mitigate this effect, we determine a weight for every point. These weights ($w_i$) are proportional to the difference in wavelength between two consecutive data points in the spectrum with $\sum_i w_i= N_{\rm obs}$ resulting in an uniform contribution by wavelength to the likelihood function. For the case in which the Bayesian evidence is compared between fits (Sect.~\ref{sec:detect-mol} and Sect.~\ref{sec:mol_complex}), we opted to set all weights to $1$ to not influence the evidence calculations, while fits analysing the emitting conditions all follow the weight description above.

Next to the likelihood function, priors need to be defined for all fitted parameters (Table~\ref{tab:priors}). To remain agnostic about the value of the parameters, we choose uniform priors based on the limits of the slab grid created by \cite{arabhavi_abundant_2024} when possible. While the maximum ($T_{\rm max}^{\rm mol}$) and minimum ($T_{\rm min}^{\rm mol}$) temperature parameters for every molecule ($\rm mol$) use uniform priors in linear space ($\mathcal{U}$), the large possible variation in column density leads us to choose uniforms priors in logarithmic space ($\mathcal{J}$) for the column density at the maximum ($N_{\rm tmax}^{\rm mol}$) and minimum temperature ($N_{\rm tmin}^{\rm mol}$). For \ce{CO2} and \ce{C2H2} the upper limit for both column density priors are set to $10^{21}\,\rm cm^{-2}$. This is based on first tests where very optically thick \ce{CO2} and \ce{C2H2} components were fitted. To exclude these extreme cases, which do not fit any clear molecular features, from our analysis we lowered the upper column density limit for \ce{CO2} and \ce{C2H2}. For \ce{SiO} the upper column density limit is set to $10^{20}\,\rm cm^{-2}$ for the same reason. Similarly, the upper temperature limits for \ce{CO} were extended to from $1500\,\rm K$ to $2500\,\rm K$ since first tests showed a \ce{CO} temperature convergence toward the upper temperature limit. The prior of the slope of the radial temperature power law ($q_{\rm emis}$) is chosen to ensure a radial decrease in temperature and to include extracted slopes from thermochemical models \citep[e.g.][]{Fedele2016,brittain_herbig_2023}. The emitting area of every molecule is not treated as a Bayesian parameter but due to its linear nature optimised with non-negative least square fitting for every parameter combination during the fitting \citep[see][]{kaeufer_bayesian_2024,kaeufer_disentangling_2024}. The uncertainty parameter $\sigma$ includes typical observational uncertainties and could potentially be adjusted if the parameter's posterior distribution converges towards the priors' edges.

\begin{table}

    \caption{Prior distributions of all free Bayesian parameters. For \ce{CO}, the temperatures priors are extended up to $2500\,\mathrm{K}$. For \ce{CO2} and \ce{C2H2}, the column density priors range only up to $10^{21}\,\mathrm{cm^{-2}}$ to avoid extremely optically thick cases. For \ce{SiO}, the upper limit is set to $10^{20}\,\mathrm{cm^{-2}}$ for the same reason. The terms $\mathcal{U}(x,y)$ and $\mathcal{J}(x,y)$ denote uniform and log-uniform priors in the range from $x$ to $y$, respectively.}
    \label{tab:priors}
    \centering
    \begin{tabular}{l|l}
\hline \hline & \\[-1.9ex] 
Parameter & Prior  \\ \hline  
  & \\[-1.9ex] 
$T_{\rm max}^{\rm mol}\,\rm [K]$ & $\mathcal{U}(25,1500)$ \\ 
  & \\[-1.9ex] 
$T_{\rm min}^{\rm mol}\,\rm [K]$ & $\mathcal{U}(25,1500)$ \\  
 &  \\[-1.9ex] 
$N_{\rm tmin}^{\rm mol}\,\rm [cm^{-2}]$ & $\mathcal{J}(10^{14},10^{24})$   \\ 
 &  \\[-1.9ex] 
$N_{\rm tmax}^{\rm mol}\,\rm [cm^{-2}]$ & $\mathcal{J}(10^{14},10^{24})$ \\
 &  \\[-1.9ex] 
$q_{\rm emis}$ & $\mathcal{U}(-2,-0.1)$ \\
 &  \\[-1.9ex] 
$\sigma\,\rm [mJy]$ & $\mathcal{J}(0.1,10)$ \\
 \hline
    \end{tabular}
        
\end{table}

\subsection{Modelling results}
\label{sec:results}

We aim to describe the molecular emission seen in the JWST/MIRI spectrum of HD\,35929. For that, we need to first determine which molecules are needed and if radial power laws in temperature and column density result in significant improvements of the fitting quality.

\subsubsection{Detected gas phase molecules\label{sec:detect-mol}}

To evaluate the evidence for different molecules we run Bayesian retrievals (with an evidence tolerance of $0.5$) including and excluding individual molecules, following the procedure outlined by \cite{kaeufer_disentangling_2024}. The evidence for every molecule is calculated by the logarithm of the Bayes factors ($\ln{B}$) between a run with and without the molecules. The Bayes factor is a measure calculating the differences in global evidence for two fits. Values of $\ln{B}<1$, $1<\ln{B}<2.5$, $2.5<\ln{B}<5$, $5<\ln{B}<11$, and $11<\ln{B}$ are interpreted as no evidence, weak evidence, moderate evidence, strong evidence, and very strong evidence, respectively, for one model over another \citep{Trotta2008}. The preference of the original model is indicated by a positive sign of $\ln{B}$. A negative value indicates a preference for the variant model.

We start with a fit including five molecules for which features are visually identified in the spectrum: \ce{H2O}, \ce{CO}, \ce{CO2}, \ce{SiO}, and \ce{OH} (see Sect.~\ref{sec:inventory}). This setup is called the original model. We quantify the evidence for these molecules by excluding them (one at a time) from the fit. Additionally, we test the evidence for \ce{C2H2} and \ce{HCN} by adding them to the original model. All molecules are treated as 0D slabs (no temperature and column density range) for this test to reduce the fit complexity. In this case, the priors for the temperature and column density follow Table~\ref{tab:priors}. The evidence for more complex molecular conditions is tested in Sect.~\ref{sec:mol_complex}.

\begin{table}
    \centering
    \caption{Bayes factors quantifying the evidence for different molecules in the JWST/MIRI spectrum of HD\,35929. The first columm (Mol.) lists the included or excluded molecule compared to the original model that includes \ce{H2O}, \ce{CO}, \ce{CO2}, \ce{SiO}, and \ce{OH}. The logarithm of the Bayes factor between the fit with and without a molecule ($\ln{B}$) indicates if there is a preference (Pref.) for the new fit over the original one. The evidence-column shows the interpretation of $\ln{B}$ \citep{Trotta2008}.}
    \label{tab:bayes_mols}
\begin{tabular}{l|l|l|l}
\hline
\hline
Mol. &  $\ln{B}$ & Pref. & Evidence \\ \hline
 w/o \ce{H2O} & -4456.31 & No & Very strong \\ 
 w/o \ce{CO} & -1746.55 & No & Very strong \\ 
 w/o \ce{CO2} & -171.61 & No & Very strong \\ 
 w/o \ce{SiO} & -415.02 & No & Very strong \\ 
 w/o \ce{OH} & -30.17 & No & Very strong \\ 
 \ce{C2H2} & 1.96 & Yes & Weak \\ 
 \ce{HCN} & 0.90 & Yes & None \\ 
\hline
\end{tabular}

\end{table}

The results are listed in Table~\ref{tab:bayes_mols}. The logarithm of the Bayes factors is negative for all five runs excluding one of the molecules that were visually identified (\ce{H2O}, \ce{CO}, \ce{CO2}, \ce{SiO}, and \ce{OH}). The large values ($\ln{B}>>11$) indicate that fits excluding these molecules result in significantly worse residuals, which henceforth is interpreted as strong evidence for their detection.

We include all these molecules in the further fitting process. However, we note that \ce{OH} is most likely not in LTE contrary to the assumption of the model \citep[e.g.][]{carr_oh_2014,tabone_oh_2024}. Therefore, we refrain from interpreting the extracted conditions for \ce{OH}.

The inclusion of \ce{HCN} results in an inconclusive Bayes factor ($\ln{B}=0.90$), which shows that the inclusion of \ce{HCN} is not needed to explain the data. Henceforth, we exclude \ce{HCN} from further fitting. 

\begin{figure}
    \centering
    \includegraphics[width=1.0\linewidth]{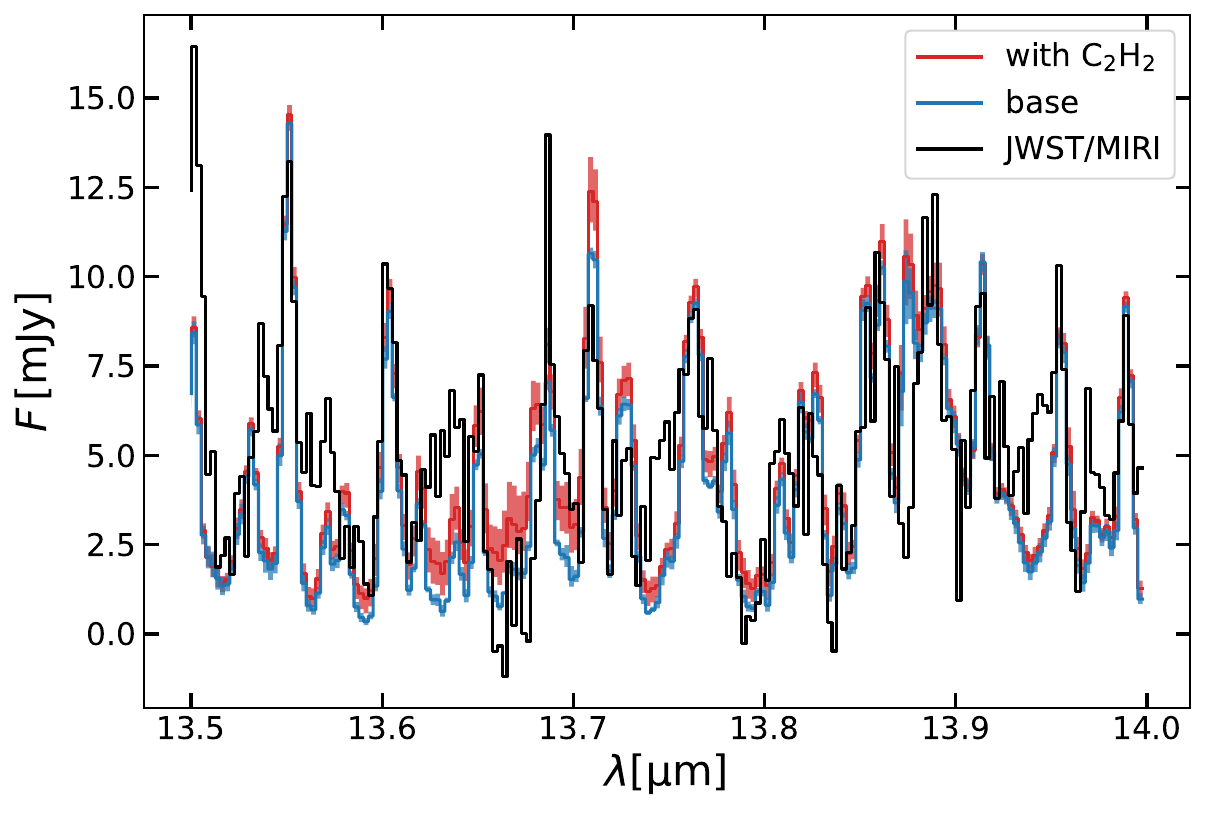}
    \caption{Flux from the base model (blue) and a model including \ce{C2H2} (red) compared to the observation (black) in the region where \ce{C2H2} emission is expected. The lines indicate the median posterior flux, with the shaded region indicating the $1\sigma$ level.}
    \label{fig:c2h2-nondetection}
\end{figure}

Similarly, the addition of \ce{C2H2} does not significantly change the evidence ($\ln{B}=1.96$). Since the value of the Bayes factor is slightly larger than the one for \ce{HCN}, we visually inspected the two fits (with and without \ce{C2H2}). As seen in Fig.~\ref{fig:c2h2-nondetection}, the slight improvement in fit quality for the fit including \ce{C2H2} is not driven by a fitted molecular feature but just by a slightly better match to the noise in the data. Due to the additional component, the water model shifts to slightly different best parameters which explains the flux changes at prominent water lines. Together with the lack of a clearly identifiable $Q$-branch, we conclude that the observation does not hold any proof for a \ce{C2H2} detection in HD\,35929, which is therefore excluded from further fitting.

\subsubsection{Molecular complexity\label{sec:mol_complex}}

After determining which molecules are visible in the JWST/MIRI spectrum of HD\,35929, we analyse the complexity of their emitting conditions (temperature and column density ranges over fixed conditions). Thermochemical models show that some molecules emit lines along an extended surface layer that cannot easily be described by a single temperature and column density \citep[e.g.][]{woitke_2024,Kanwar2025}. Therefore, several studies used series of slab models to mimic this behaviour \citep[e.g.][]{Romero_2024,Temmink2025}. We aim to quantify the need for this increase in complexity by following \cite{kaeufer_disentangling_2024}. Similarly to the detection runs, Bayes factors are used to quantify the evidence for different setups. We start with every molecule as a 0D slab and compare that to fits where individual molecules are treated as temperature power laws or temperature and column density power laws. The Bayes factor between those fits quantifies if the added parameters are justified by a significant increase in fitting quality.

\begin{table}
    \centering
    \caption{Bayes factors quantifying the evidence for different model choices in the JWST/MIRI spectrum of HD\,35929. The first columm (Mol.) lists the molecule for which the complexity is changed compared to the original model. The logarithm of the Bayes factor between the base model and the model with increased complexity ($\ln{B}$) indicates if there is a preference (Pref.) for the new fit over the original one. The evidence-column shows the interpretation of $\ln{B}$ \citep{Trotta2008}. The best model shown in the last column includes \ce{CO}, \ce{CO2}, \ce{H2O}, and \ce{OH} with temperature and column density power laws with an additional 0D \ce{SiO} component.}

    \label{tab:bayes_complex}
\begin{tabular}{l|p{1cm}|p{1cm}|l|l|l}
\hline
\hline
Mol. & $T$ range & $N$ range & $\ln{B}$ & Pref. & Evidence \\ \hline
 \ce{CO} & Yes & No & 3.18 & Yes & Moderate \\ 
 \ce{CO} & Yes & Yes & 13.84 & Yes & Very strong \\ 
 \ce{CO2} & Yes & No & 4.73 & Yes & Moderate \\ 
 \ce{CO2} & Yes & Yes & 5.61 & Yes & Strong \\ 
 \ce{H2O} & Yes & No & 180.12 & Yes & Very strong \\ 
 \ce{H2O} & Yes & Yes & 192.93 & Yes & Very strong \\ 
 \ce{OH} & Yes & No & 0.30 & Yes & None \\ 
 \ce{OH} & Yes & Yes & 2.46 & Yes & Weak \\ 
 \ce{SiO} & Yes & No & -0.60 & No & None \\ 
 \ce{SiO} & Yes & Yes & -0.17 & No & None \\ 
 Best &  &  & 214.47 & Yes & Very strong \\ 
\hline
\end{tabular}

\end{table}

The Bayes factors and their interpretations are listed in Table~\ref{tab:bayes_complex}. For \ce{SiO}, neither the inclusion of a temperature power law nor the additional inclusion of a column density power law significantly increases the quality of the fitting. Therefore, we conclude that the emission of \ce{SiO} in HD\,35929 originates in a region with near-constant conditions. For all other molecules, there is evidence for the inclusion of a column density power law. For \ce{OH} this evidence is only weak, with no evidence towards a temperature range without a simultaneous addition of a column density range. The evidence for \ce{CO} increases moderately for a temperature gradient but the addition of a column density power law improves the fit significantly. \ce{CO2} shows similar evidence for the temperature power law with and without a column density range. The strongest evidences can be found for \ce{H2O} ($\ln{B}=180.12$ and $\ln{B}=192.93$). The drastic increase in evidence for the inclusion of a temperature power law conclusively shows that water is emitting from a range of conditions in discs and is poorly approximated by 0D slab models \citep[consistent with previous studies, e.g.][]{Romero_2024,temmink_minds_2024}. 

Based on these results, we select the best fit which includes column density and temperature power laws for \ce{CO}, \ce{CO2}, \ce{H2O}, and \ce{OH}, while \ce{SiO} is described by constant column density and temperature. As shown in Table~\ref{tab:bayes_complex}, this fit's Bayes factor ($\ln{B}=214.47$ with $w_i=1$ for all $i$) is as expected better than all individual complexity enhancements.

\subsubsection{Emission contributions and conditions\label{sec:mol_conditions}}

\begin{figure}
    \centering
    \includegraphics[width=1.0\linewidth]{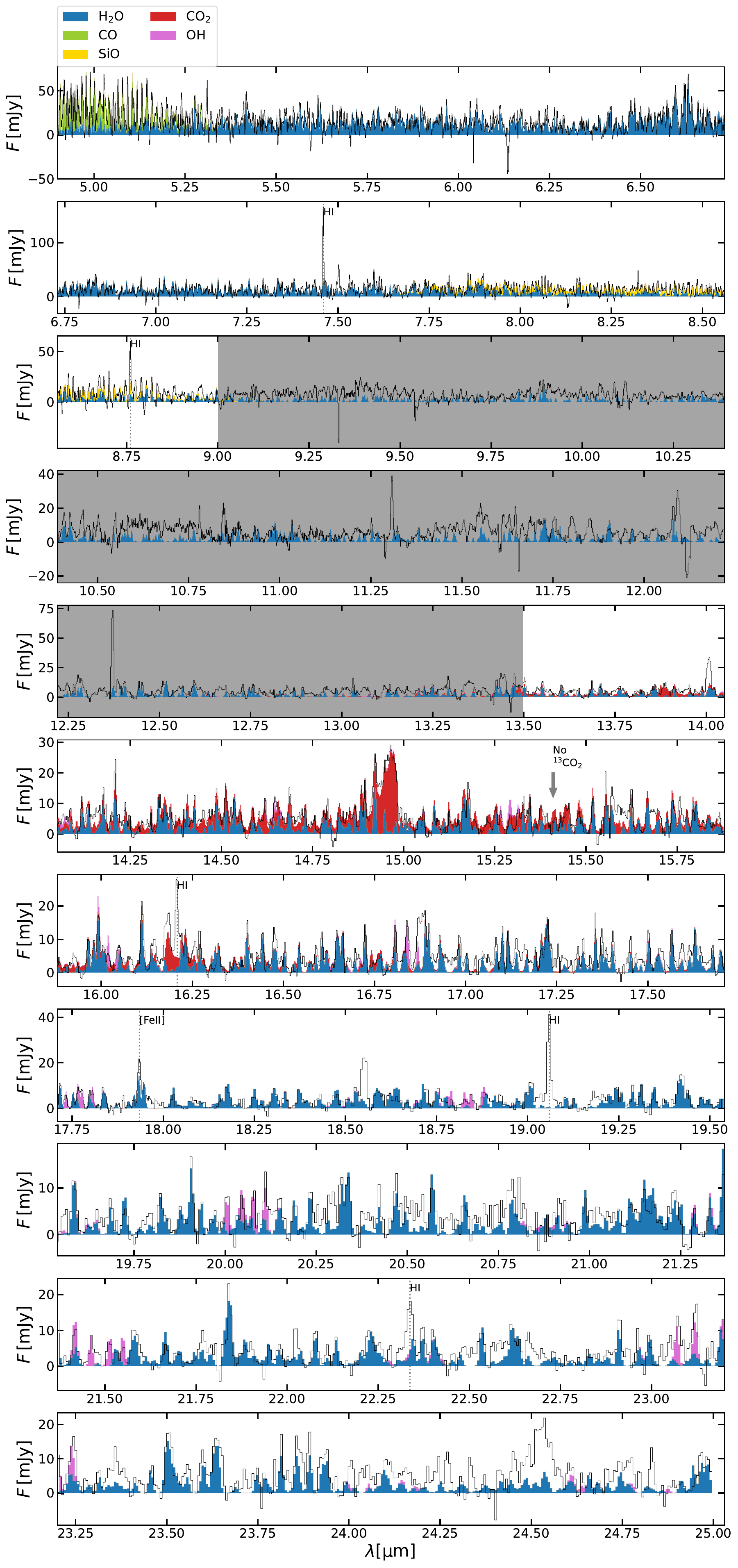}
    \caption{Continuum subtracted MIRI spectrum, using the 4.9 - 25 $\mu$m wavelength range, but excluding the 10 $\mu$m silicate band region between 9.0 and 13.5 $\mu$m (excluded region marked in grey). Overplotted are the cumulative fluxes from H$_2$O, CO, SiO, CO$_2$, and \ce{OH} from the median probability model. A few unfitted atomic lines and the non-detection of \ce{^{13}CO2} are labelled. The apparent feature around $18.55\,\rm \mu m$ can be traced back to a bad pixel artifact.}
    \label{fig:mol-fit}
\end{figure}

\begin{figure}
    \centering
    \includegraphics[width=1.0\linewidth]{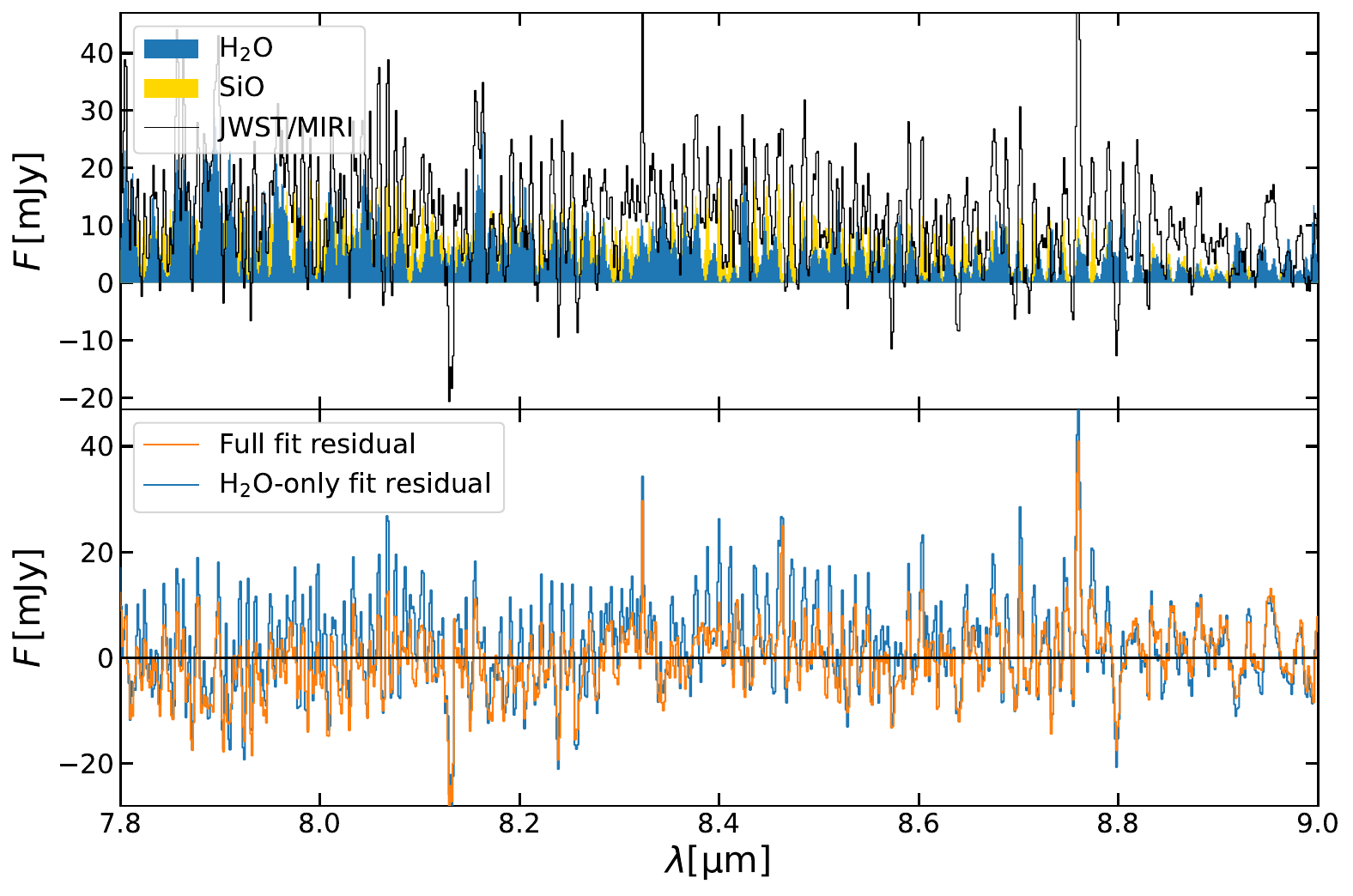}
    \caption{Residual in the wavelength range dominated by SiO of the full fit compared to a water-only fit.}    \label{fig:mol-sio-zoom}

\end{figure}

We run the retrieval with the molecules and the molecular complexities determined in Sect.~\ref{sec:detect-mol} and Sect.~\ref{sec:mol_complex}, respectively. This setup is identical to the `Best' model listed in Table~\ref{tab:bayes_complex}, with the inclusion of wavelength-dependent weights (Eq.~\ref{eq:likelihood}). The modelled flux for this model compared to the observation is shown in Fig.~\ref{fig:mol-fit}. While large parts of the spectrum are very well fitted, at long wavelengths there are several features that the model cannot explain (e.g. around $23.4\,\rm \mu m$ and $24.5\,\rm \mu m$ ). These might be unaccounted molecular emission or mismatches in the continuum estimation. However, the spectrum is dominated by the emission of water, but clear features of \ce{CO} and \ce{CO2} are also visible at $5\,\rm\mu m$ and $15\,\rm\mu m$, respectively. Additionally, multiple \ce{OH} lines appear between $15\,\rm\mu m$ and $25\,\rm\mu m$. 
While all of these molecules are commonly detected in JWST/MIRI spectra, the detection of \ce{SiO} is more surprising. The upper panel of Fig.~\ref{fig:mol-sio-zoom} displays a zoom-in to the region dominated by \ce{SiO}. While the fit is not perfect, it becomes clear that several emission features in the observations are well matched by \ce{SiO}. To add to the significance of that, we run a fit with \ce{H2O}-only to the wavelength region dominated by \ce{SiO} ($7.5 -9.0\,\rm \mu m$) to conclusively show that the emission seen in the observation cannot be explained without \ce{SiO}. The residual of this fit is compared to the residual of the full fit in the lower panel of Fig.~\ref{fig:mol-sio-zoom}. The mean of the absolute values of the residuals are smaller for the full fit ($\sim5.11\,\rm mJy$) compared to the \ce{H2O}-only fit ($\sim6.43\,\rm mJy$). Additionally, the latter shows many clear residual lines which are well fitted by \ce{SiO} in the full fit. Therefore, these lines are most likely due to \ce{SiO}.

\begin{figure}
    \centering
    \includegraphics[width=1.0\linewidth]{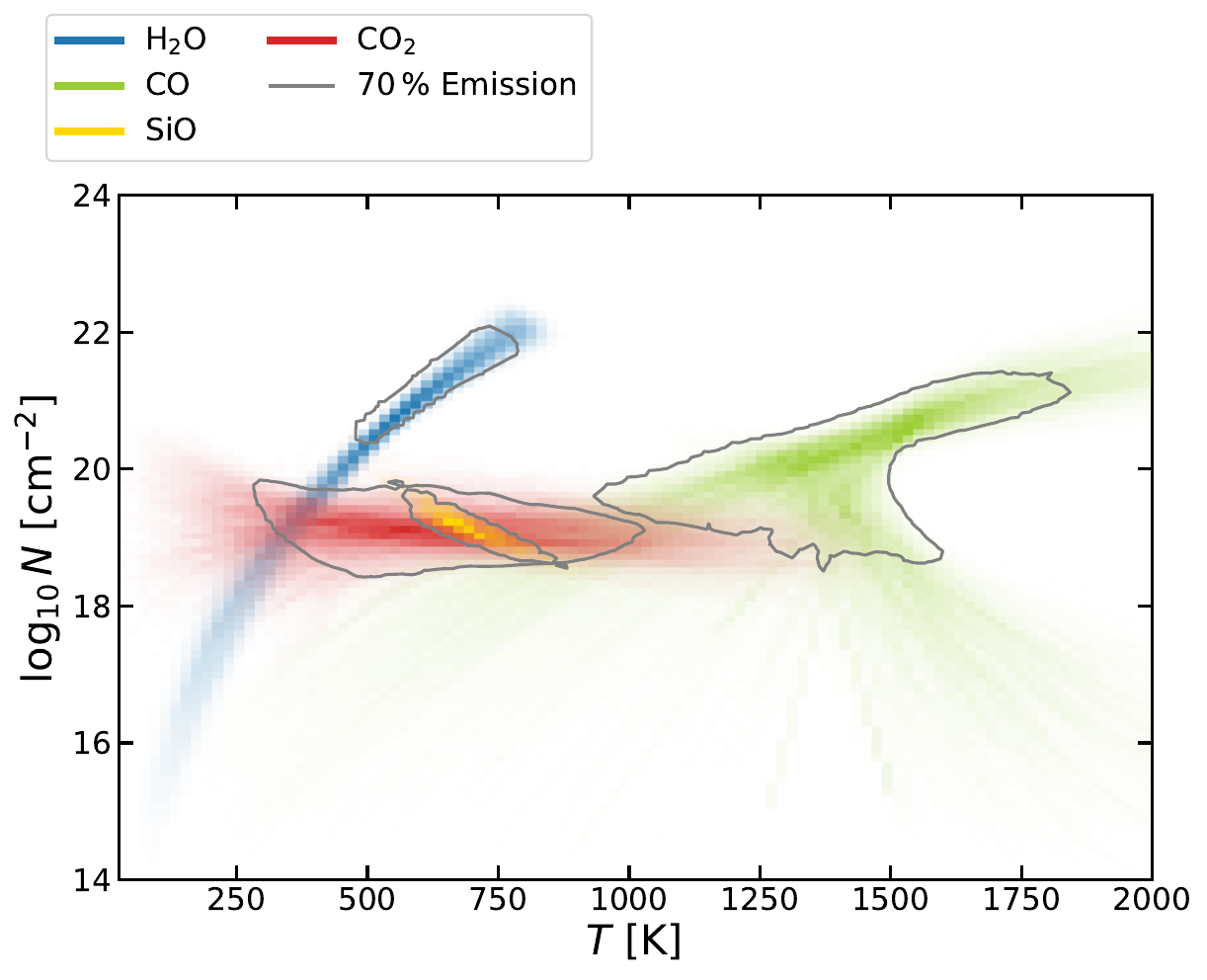}
    \caption{Probability distribution of the molecular column densities and temperatures derived from the continuum subtracted MIRI spectrum. The colours show the emissions for the molecules listed in the legend, while the contours denote the region with the highest contribution to the respective molecular fluxes.}
    \label{fig:mol-con}
\end{figure}

Based on this full fit (Fig.~\ref{fig:mol-fit}), we analyse the molecular conditions of the models of the posterior. Fig.~\ref{fig:mol-con} shows the posterior distributions of column density and temperature for all fitted molecules (except \ce{OH} which is excluded due to its non-LTE origin). The coloured regions indicate the full temperature and column density ranges for all posterior models.
Additionally, the region which dominates the integrated flux is plotted as contours. For this, the radial cumulative flux per molecule is determined for every model. The radii which enclose $15\,\%$ and $85\,\%$ of the cumulative flux are chosen as the inner and outer radius of the effective emitting region, following \cite{kaeufer_bayesian_2024} based on typical definitions used by thermochemical models \citep[e.g.][]{woitke_2024}. After deriving this region for every individual model, the distribution of these regions over the full posterior is taken and indicated in Fig.~\ref{fig:mol-con}.

All molecules show signs of high column densities ($>10^{18}\,\rm cm^{-2}$), with \ce{CO2} and \ce{SiO} emitting at column densities of $\sim10^{19}\,\rm cm^{-2}$ and \ce{CO} and \ce{H2O} emitting from even higher column densities ($>10^{20}\,\rm cm^{-2}$). Given these high column densities, we searched for isotopologue features in the spectrum (Appendix~\ref{sec:isos}), but no clear features were found. \ce{CO} shows emission from the highest temperatures ($\sim1500\,\rm K$ and beyond) hinting towards an origin close to the star, where temperatures are highest. While \ce{CO2} emits from a wide temperature range (from $\sim250\,\rm K$ to $\sim1000\,\rm K$), the emission of \ce{SiO} is confined to a narrow temperature range between $500\,\rm K$ and $900\,\rm K$.

Next to the column density and temperature, we retrieve the emitting area of every molecule. For \ce{CO}, \ce{SiO} and \ce{CO2}, the emitting area where most of the emission (defined as the range between $15\,\%$ and $85\,\%$ of the radial cumulative flux) originates, are equivalent to discs with radii of $0.12\,\rm au$, $0.22\,\rm au$, and $0.14\,\rm au$, respectively. For \ce{CO} and \ce{CO2}, which are fitted with radial profiles these regions extend outwards no more than $~0.15\,\rm au$. Even though these radii are purely based on the derived temperature slope and emitting area, all radial extents are consistent with the close in physical extent of $0.17\,\rm au$ derived by the line kinematics (Sect.~\ref{sec:kinematics}). We caution however that the spatial scales derived from FWHM measurements reflect typical distance scales while the slab models provide radiating surfaces, and so their similarity may be coincidental. 

To increase our confidence in the high retrieved column densities, we examined \ce{CO2} and \ce{H2O} in more detail. Fitting a narrow wavelength range around the \ce{CO2} Q-branch ($14.8-15.0\,\rm\mu m$) with \ce{CO2} and \ce{H2O} results in \ce{CO2} column densities of $\sim10^{18}\,\rm cm^{-2}$. While this is lower than the value retrieved fitting the full wavelength range ($\sim10^{19}\,\rm cm^{-2}$), it shows that high column densities for \ce{CO2} are retrieved independently of the exact fitting method. Similarly, we examine water by fitting different wavelength ranges. 

The ro-vibrational water lines, which have to be fitted to constrain the \ce{SiO} properties, contain non-LTE effects, which limits the conclusions drawn from LTE models \citep{Banzatti2023}. Therefore, the rotational water lines are regularly used to analyse the water spectrum observed by JWST/MIRI \citep[e.g.][]{Romero_2024,Temmink2025}. 
To gather a more accurate picture of the emission conditions of water, we repeat the fits presented in Sect.~\ref{sec:fitting} for the wavelength range of $13.5-25\,\rm \mu m$, which excludes the ro-vibrational water. Since this wavelength also excludes \ce{CO} and \ce{SiO} emission we exclude these molecules from the fit. 
All other species are included using temperature and column density power laws. The resulting fit is displayed in Fig.~\ref{fig:rot-water-fit} in Appendix~\ref{sec:rot-water}.

We perform an additional fit that is only focused on the unblended rotational water lines determined by \cite{banzatti_water_2025}. For this fit, narrow wavelength windows around each line are selected \citep[based on Table~5 of][]{banzatti_water_2025} and water is included as the only molecule. The resulting fit is shown in Fig.~\ref{fig:unblended-water-fit} in Appendix~\ref{sec:rot-water}.

\begin{figure}
    \centering
    \includegraphics[width=1.0\linewidth]{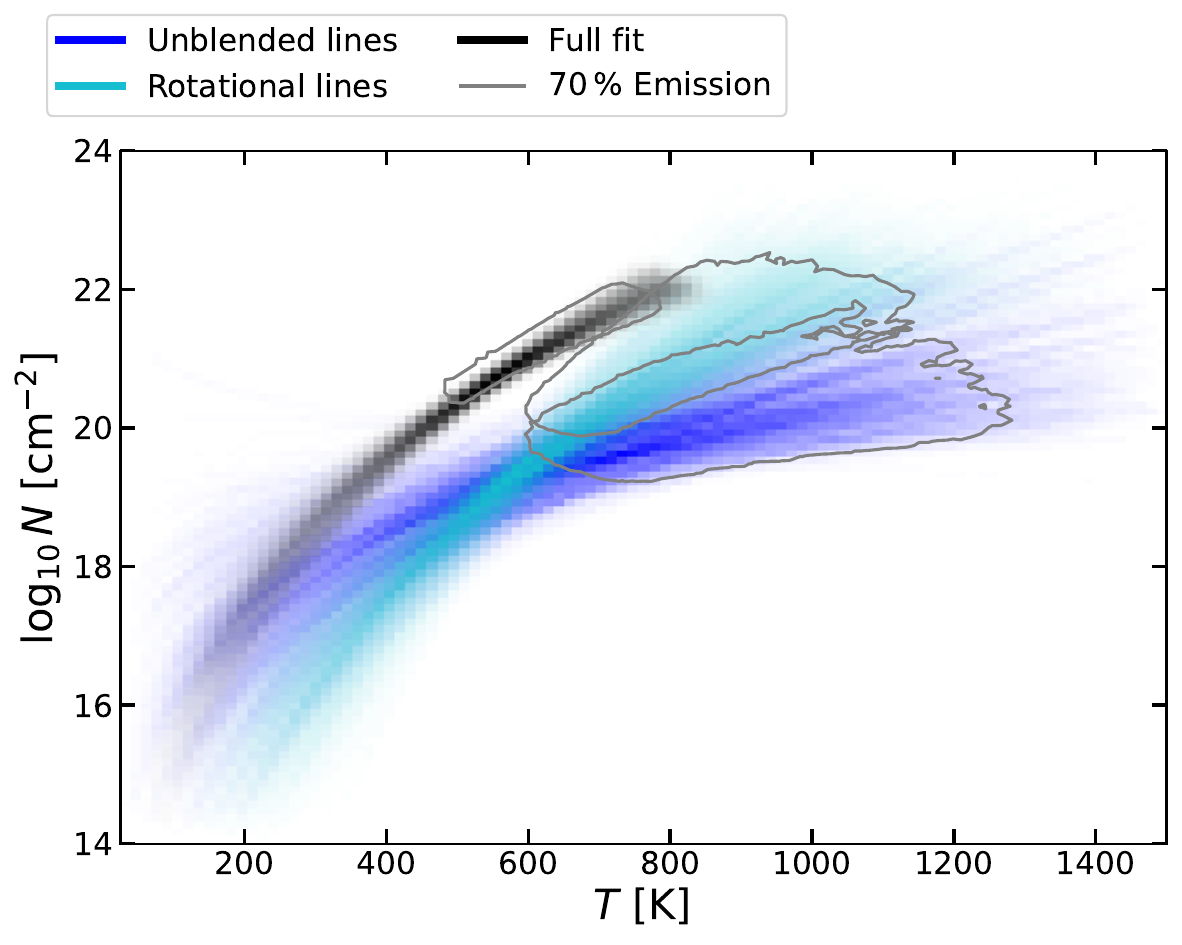}
    \caption{Probability distribution of the molecular column densities and temperatures derived from the fit to the full spectrum (same as water in Fig.~\ref{fig:mol-con}, black), the rotational water spectrum (cyan), and the unblended rotational lines (bright blue).}
    \label{fig:water-con}
\end{figure} 

The extracted emission conditions from the full water fit, the fit of the rotational lines, and the unblended rotational lines are shown in Fig.~\ref{fig:water-con}. It can be seen that there is a large overlap between the extracted conditions, albeit with significant spread. In all cases, water emits from very large column densities ($>10^{20}\,\mathrm{cm^{-2}}$). This is significantly higher than extracted column densities from T~Tauri stars, that typically range between 10$^{18}$ and 10$^{20}$ $\mathrm{cm^{-2}}$  \citep[e.g,][]{Temmink2025}. This means that many water lines are optically thick and huge columns of water are probed. Therefore, even the unblended water lines with the lowest Einstein-A coefficient (e.g. at $12.565\,\rm \mu m$, $20.662\,\rm \mu m$, and $23.318\,\rm \mu m$, as best seen in Fig.~\ref{fig:unblended-water-fit}) are detectable. While there is general overlap between the retrievals, the full fit probes somewhat different temperatures and column densities, which is likely an effect of the non-LTE condition of ro-vibrational water. The smaller column density and temperature spread along the power laws might be a consequence of fitting many water lines of different origin. 
There are differences between the rotational water fit and the fit to the unblended water lines. The unblended lines retrieve a shallower relation between temperature and column density, resulting in higher column densities (by $\sim 1\,\rm  dex$) at temperatures $<400\,\rm K$ and lower column densities (by $\sim 1\,\rm  dex$) at temperatures $>600\,\rm K$. This might be an effect of the composition upper level energies of the unblended lines. Due to the significant noise in line-free regions for this spectrum, we think that fitting the unblended water lines results in the most accurate description of \ce{H2O} emission conditions. However, despite of these differences, there seems to be a consensus of high columns of water that are dominated by warm temperatures ($>500\,\rm K$) independent on the retrieval method. 
Analysing the water radial extent based on the fit to the unblended water lines, we find an inner and outer radius of $0.09\,\rm au$ and $0.18\,\rm au$, respectively. Similar to \ce{CO}, \ce{SiO}, and \ce{CO2} these values are broadly consistent with the radial position estimated based on line kinematics (Sect.~\ref{sec:kinematics}).

The high column densities of (especially hot) water can be understood when comparing HD\,35929 to other water-rich sources. BP\,Tau and XX\,Cha are found to have similar emitting areas (radial profiles up to radii of about $\sim 2 \, \rm au$) and column densities ($\sim 10^{18}\,\rm cm^{-2}$) of their cold water component compared to HD\,35929 \citep{Temmink2025}. In Appendix~\ref{sec:compare-ttauri}, we highlight that the cold water lines in all three spectra look remarkably similar when scaled to the same distance. However, all hot water lines are systematically stronger for HD\,35929 compared to the T\,Tauri discs. Therefore, the column densities at higher temperatures must be significantly larger in HD\,35929 than the column densities at high temperature found for BP\,Tau and XX\,Cha which are slightly larger than $10^{18}\,\rm cm^{-2}$. This is consistent with retrieved column densities above $10^{20}\,\rm cm^{-2}$.

\begin{figure}
    \centering
    \includegraphics[width=1.0\linewidth]{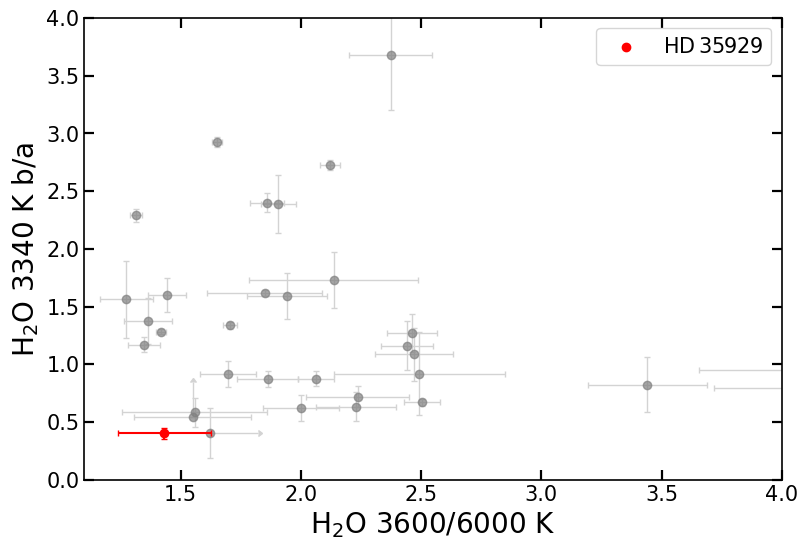} \\
    \includegraphics[width=1.0\linewidth]{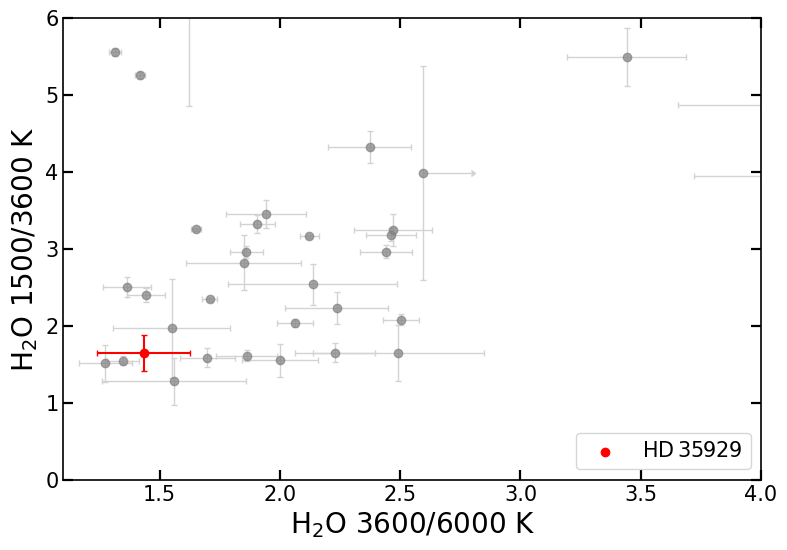}
    \caption{Water line diagnostics with the temperatures denoting the upper level energies \citep{banzatti_water_2025}. The x-axes show the ratio of water lines tracing warm water emission ($\sim 400 \, \rm K$) compared to hot water ($\sim 850 \, \rm K$). The vertical axis of the upper panel shows the ratio of two lines with similar upper level energies but different Einstein-A coefficients tracing the column densities of $\sim 400\, \rm K$ water. The y-axis of the lower panel shows the ratio between water lines tracing cold emission ($\sim 200\, \rm K$) and warm emission. The grey dots are observations taken by JWST/MIRI \citep{banzatti_water_2025,Temmink2025} with the red dot indicating HD\,35929.}
    \label{fig:water-line-diagnostics}
\end{figure}

Additionally, the conclusions regarding the column density and temperature can be tested without relying on models using the water line ratios introduced by \cite{banzatti_water_2025}. The integrated line strengths were derived directly from the spectrum using iSLAT \citep{Jellison2024}. Fig.~\ref{fig:water-line-diagnostics} shows the resulting line diagnostic diagrams. The column density tracer (\ce{H2O} $3340\,\rm K$ a/b) has a value of $0.40$ which indicates high column densities and is very low compared to other objects observed by JWST/MIRI. Additionally, the low ratios of the cold/warm tracer (\ce{H2O} $1500/3600\,\rm K$: $1.65$) and warm/hot tracer (\ce{H2O} $3600/6000\,\rm K$: $1.43$) indicate an overall lack of cold water. This confirms model-independently the dominance of hot water with high column densities.

\section{Discussion}
\label{sec:discussion}

\subsection{Presence of gas phase SiO}

The visual evidence (Fig.~\ref{fig:mol-sio-zoom}) together with the Bayesian evidence presented in Sect.~\ref{sec:mol_conditions} leaves little doubt about the detection of gas phase SiO. The high excitation temperature and column density, combined with the width of the lines, strongly suggest a location very close to the star, near or within the silicate dust sublimation radius of HD\,35929. To our knowledge, this is the first detection of ro-vibrational gas phase emission from SiO in a disc surrounding an intermediate-mass young star. Gas phase SiO has been suggested in the Spitzer spectrum of the young HD\,172555 debris disc \citep{lisse_abundant_2009}; however, \cite{johnson_self-consistent_2012} questioned this, and a recent study by \cite{Samland2025} shows no evidence for gas phase SiO in the JWST/MIRI spectrum of that source.

Gas phase SiO is detected in molecular clouds and could play a role in the formation of silicates in cold environments \citep{reber_sio_2008, krasnokutski_formation_2014, yang_directed_2018,Rouille_cold_silicate}. Gas phase SiO can also be produced in outflow or jet induced shocks that destroy silicate dust grains \citep{guilloteau_first_1992,bally_protostellar_2016,gusdorf_sio_2008}. Gas phase SiO first overtone ro-vibrational line emission was found in dusty discs surrounding evolved luminous, massive B[e] supergiant stars \citep{Be_SiO_Kraus_2015}. These objects also show prominent CO ro-vibrational emission. Recently, gas phase SiO was detected in the fundamental ro-vibrational band towards two young protostars \citep{gelder_jwst_2024}. In these cases, the SiO is likely associated with shocks related to a disc wind or a jet, and it is not within the silicate dust sublimation radius.  \cite{mcclure_refractory_2025} presented the detection of warm gas phase SiO in the innermost disc of a Class I protostar, where the high mass accretion rate causes the inner disc solids to evaporate. We note that no evidence for a jet in HD\,35929 has been reported in the literature, but the type III P~Cygni profiles detected in the UV point to an outflow \citep{grady_ensuremathbeta_1996}.  The [FeII] emission reported in Section \ref{sec:inventory} is blueshifted and may be formed in a lower density environment, possibly associated with the outflow detected in the UV. We checked that the SiO emission is not spatially extended.

We conclude that the SiO is probably associated with the high density, warm/hot inner disc, near or inside the dust sublimation radius. Under these conditions, SiO can be formed via a gas phase reaction between Si and OH, further leading to the formation of SiO$_2$, which may condense into solid SiO$_2$ or other forms of silicates. The high column densities derived in Sect.~\ref{sec:mol_conditions} suggest that shielding by dust, and/or by molecules such as H$_2$O,  and self-shielding may explain the survival of SiO close to this UV-bright star \citep{heays_photodissociation_2017}. 

Chemical equilibrium models \cite[e.g., GGCHEM,][]{woitke_equilibrium_2018} show that in an oxygen-rich gas H$_2$O, CO, CO$_2$, OH, and SiO form at high temperature and pressure; these are indeed the molecules we detect. However, when the formation of solids is taken into account, such chemical equilibrium models predict silicon to be depleted at temperatures below ~1700 K (at 1 bar pressure), reaching a depletion of many orders of magnitude at temperatures below 1000\,K, which is well above the retrieved temperature derived in Sect.~\ref{sec:mol_conditions} \citep{woitke_equilibrium_2018}. This suggests that the SiO is not fully condensing, indicating a non-equilibrium chemistry \citep[which is not uncommon in protoplanetary discs as shown by][]{Kanwar2025}, or the derived SiO excitation temperature  does not represent the true SiO gas kinetic temperature. In this context, it is interesting to note that \cite{juhasz_dust_2010} derived a high abundance of solid amorphous SiO$_2$ in HD\,35929, as evidenced by the prominent 9.2-9.3\,$\mu$m peak in the silicate emission, and the blue shoulder at about 8\,$\mu$m. 

\subsection{Nature of the disc surrounding HD\,35929}
\label{sec:dis}

The MIRI spectrum of HD\,35929 is remarkable because it is rich in molecular line emission. As mentioned in Section~\ref{sec:intro}, previous observations of Herbig star discs using the Spitzer Space Telescope showed a low detection rate of molecular species compared to lower mass T~Tauri stars \citep{pontoppidan_spitzer_2010}. This was attributed to a bright dust continuum in Herbig stars and/or a modest spectral resolution and sensitivity \citep{antonellini_understanding_2015} in combination with \ce{H2O} photodissociation by the strong stellar UV field \citep{fedele_water_2011}.  Indeed, the SED of HD\,35929 shows a steep spectral index at IR wavelengths, i.e. weak IR dust continuum emission. The dust is depleted in the smallest, (sub-) $\mu$m sized grains and lacks a substantial cold dust reservoir. In addition, the S/N of the HD\,35929 MIRI data is very high (between 100 and 800). Therefore, the JWST/MIRI data of HD\,35929 provide optimal conditions to detect the emission of molecules from its inner disc. 

Near-IR interferometric observations of Herbig stars have revealed the presence of emission arising from close to, or within the (silicate) dust sublimation radius, which is pressure dependent and often taken at a temperature of 1500\,K \citep{dullemond_passive_2001,woitke_equilibrium_2018}. The nature of this emission has been debated. \cite{eisner_water_2007} used the Keck Interferometer to spatially and spectrally resolve the inner disc of the Herbig star MWC\,480 and found evidence for hot H$_2$O emission on a scale of $0.16\,\rm au$. However, \cite{najita_high-resolution_2009} showed with high spectral resolution observations that MWC\,480 lacks strong near-IR water line emission. \cite{Arulanantham2025} presented JWST/MIRI MRS data of MWC\,480 and found the presence of a compact warm H$_2$O reservoir (emitting area $0.2\,\rm au$) with a column density of 10$^{18.8}$ cm$^{-2}$. In addition, they report the presence of C$_2$H$_2$. The inner disc emission may also be due to hot refractory dust.  \cite{benisty_strong_2010} found strong inner disc emission from HD\,163296 on spatial scales of 0.1\,to 0.45\,au, and attributed this to refractory dust species. The nature of the material in these innermost disc regions remains unclear, but our results for HD\,35929 show that very dense and compact molecular gas reservoirs can exist on the spatial scales probed by the Keck and VLTI interferometers. In the light of this, the nature of the gas and dust in the inner $0.1-0.4\,\rm au$ of Herbig discs may need to be revisited.

However, the compact nature of the disc and the presence of high column density molecular gas emission set HD\,35929 apart from other (Herbig) discs that lack such emission \citep{Arulanantham2025} (but see below) and tend to show colder dust reservoirs of material at larger distances from the star. HD\,35929 shares some properties with that of the rapidly rotating 4 M$_{\odot}$ Herbig star 51~Oph \citep{ancker_composition_2001,Thi_2005}, with detections of CO, CO$_2$ and H$_2$O as well as silicate dust. The CO first overtone emission of 51~Oph is very compact, at a spatial scale of 0.1\,au \citep{gravity_collaboration_gravity_2021}, i.e. comparable to that of HD\,35929. The 2.5\,M$_{\odot}$ Herbig star HD\,101412 shows near-IR H$_2$O and OH as well as CO first overtone emission \citep{adams_water_2019} and \ce{CO2} emission \citep{pontoppidan_spitzer_2010}, in a disc with a dust-depleted innermost region and prominent mid-IR dust emission from material at distances of the order of 2\,au \citep{fedele_structure_2008}. \cite{adams_water_2019} suggest the presence of a close-in planet separating the innermost dust depleted disc from the cooler material further out. The HD\,35929 disc may represent a further evolutionary phase in the dispersal of discs surrounding intermediate-mass stars, in which only the innermost dust depleted region remains. 

The estimated disc lifetime of HD\,35929 is only a few years (Sect.~\ref{sec:sed}), based on the observed gas accretion rate, a gas-to-dust mass ratio of 100, and the solid mass reservoir of only 0.23 M$_{\oplus}$ \citep{Stapper2025A&A...693A.286S}. The disc lifetime could be longer if the gas accretion rate is variable, or the gas-to-dust mass ratio is (much) higher than nominal. The unusually high column density of water and other molecules derived in this study indeed suggest that we can see deep into the disc, possibly as a result of dust depletion.

\cite{grant_dotmmdisk_2023} noticed that a subset of Herbig stars, mostly of Group~II \citep{meeus_iso_2001}, show a similar disc lifetime problem. They propose that these Group~II discs may have experienced efficient pebble drift, and we detect these discs when they are at the brink of dispersal. Alternatively, the mass accretion rate may be overestimated because of contributions from a disc wind \citep{grant_dotmmdisk_2023}, or possibly a breakdown of the calibration of mass accretion rates from emission line diagnostics for higher-mass, warmer stars: the spatial scale of the Br$\gamma$ emission in Herbig stars is consistent with an origin in a disk wind \citep{gravity_collaboration_gravity_2024}. 

An alternative explanation for the unusual disc properties of HD\,35929 is a post-main-sequence evolutionary status of the star \citep{miroshnichenko_fundamental_2004}. In that scenario, the disc is the result of the injection of gas from the stellar photosphere into the disc, where it may viscously spread. if the gas density is high enough, molecules can form \citep{Thi_2005,Thi_steam2005A&A...438..557T} and eventually even dust. This would alleviate the disc lifetime problem. A possible mechanism for the formation of such a disc could be the combination of rapid rotation and pulsations, as is observed in more massive Be stars \citep{rivinius_classical_2024}.    

We can compare the JWST/MIRI observations of HD\,35929, which is located in Orion, to those of irradiated discs in the XUE sample in NGC6357 \citep{Macla_XUE01,Ramirez-taunnus2025,Jenny_XUE10}. The XUE stars are $1-2\,\rm Myrs$ old and most have masses between 2 and 3 M$_{\odot}$, i.e. comparable to HD\,35929. The XUE discs show both Group~I as well as Group~II SEDs, and line emission from a range of molecular species, including CO, H$_2$O, OH, CO$_2$, and C$_2$H$_2$. The XUE sample shows that the inner discs surrounding irradiated young intermediate-mass stars share gas and dust properties with lower mass discs in low mass star-forming regions. In one object, XUE10, very high CO$_2$ molecular gas column densities were found \citep{Jenny_XUE10}. However, most of the XUE sample discs show relatively weak molecular line emission, suggesting inner disc gas column densities that are lower than found for HD\,35929. This is in line with the small sample of Herbig stars observed using JWST/MIRI \citep{Arulanantham2025}. Clearly, the high sensitivity of JWST is needed to detect warm inner-disc gas in discs surrounding Herbig stars. More JWST observations of discs surrounding intermediate-mass stars in different irradiation environments are needed to establish their inner disc molecular reservoir, placing HD\,35929 into context.

\subsection{Limitations of the model}

Since the chosen modelling approach uses spectral templates, and is not based on a full thermochemical disc model, the resulting column densities cannot easily be translated into relative volume densities of the molecules probed since the molecules are not necessarily co-spatial. In addition, LTE was assumed in deriving the temperatures, which is likely not the case given the strong UV field of the star and the effects of IR pumping of the lines \citep[e.g.][]{gelder_jwst_2024}. We have also neglected possible radiative transfer effects that could occur in discs with low dust opacities, in which the layers from which the gas and dust emission originate may be cospatial, resulting in more complex spectra than can be allowed for in our models. In addition, the estimation of the dust continuum may result in neglecting optically thick gas pseudo-continuum emission, which would lead to an underestimate of the gas column densities. Clearly, a next step in the analysis of the MIRI data needs to involve a full thermochemical disc model. 

\section{Conclusions}
\label{sec:conclusions}

We have analysed the JWST/MIRI MRS spectrum of the Herbig star HD\,35929, observed in the context of the MINDS \citep{kamp_chemical_2023,henning_minds_2024} program. The main findings of our study can be summarized as follows:

\begin{itemize}
    \item We find a rich molecular line emission spectrum, dominated by warm/hot water, on top of a dust continuum with weak silicate emission bands. We also detect CO, CO$_2$, OH, SiO, HI and [FeII]. The continuum flux of the MIRI and Spitzer spectra show modest differences, indicating that the disc is relatively stable over a $\sim20$ year time span.
    \item Many of the emission lines are spectrally resolved, placing the gas emission near the star at a typical distance of ~0.1-0.2 au, close to or inside the dust sublimation radius of 0.4 au.
    \item We have fitted the continuum subtracted spectrum with DuCKLinG, and find high column densities of water, and a clear lack of a cold water reservoir. Taken together with the very low mass and small size ($<$ 17 au, \citep{Stapper2025A&A...693A.286S} of the cold dust mass reservoir, the steeply declining SED, and the small angular scale observed in interferometric studies, we suggest that the disc is compact. 
    \item The detected gas phase SiO emission likely originates from the inner disc. The derived SiO excitation temperatures are significantly below the condensation temperature of SiO, pointing to non-equilibrium chemistry or an underestimate of the SiO kinetic temperature.
    \item  The nature of the HD\,35929 disc is unclear. Its estimated lifetime is only a few years, which is incompatible with the persistent IR emission over longer timescales, yet it is clearly gas-rich. A significant fraction of Herbig stars have short lifetimes \citep{grant_dotmmdisk_2023}, but not as short as HD~35929. We may be witnessing the final dispersal of a luminous ($\sim70\,\rm L_{\odot}$) Herbig star disc, with only a small, gas-rich and dust-poor innermost disc remaining. 
    
\end{itemize}

\section*{Acknowledgements}

It is a pleasure to thank H. van Winckel, C. Aerts, N. van Assen, Th. Postma, Jelke Bethlehem, and the XUE team, lead by M.-C. Ramirez-Tannus and A. Bik for valuable discussions concerning the nature of HD\,35929. We thank A. Banzatti for providing the iSHELL  CO observations of HD\,35929. 

This work is based on observations made with the NASA/ESA/CSA James Webb Space Telescope. The data were obtained from the Mikulski Archive for Space Telescopes at the Space Telescope Science Institute, which is operated by the Association of Universities for Research in Astronomy, Inc., under NASA contract NAS 5-03127 for JWST. These observations are associated with program \#1282. The following National and International Funding Agencies funded and supported the MIRI development: NASA; ESA; Belgian Science Policy Office (BELSPO); Centre Nationale d’Etudes Spatiales (CNES); Danish National Space Centre; Deutsches Zentrum fur Luft- und Raumfahrt (DLR); Enterprise Ireland; Ministerio De Econom\'ia y Competividad; Netherlands Research School for Astronomy (NOVA); Netherlands Organisation for Scientific Research (NWO); Science and Technology Facilities Council; Swiss Space Office; Swedish National Space Agency; and UK Space Agency.

T.K. acknowledges support from STFC Grant ST/Y002415/1.

M.T. acknowledges support from the ERC grant 101019751 MOLDISK.

E.v.D. acknowledges support from the ERC grant 101019751 MOLDISK and the Danish National Research Foundation through the Center of Excellence ``InterCat'' (DNRF150). I.K., A.M.A., and E.v.D. acknowledge support from grant TOP-1 614.001.751 from the Dutch Research Council (NWO).

T.H. and K.S. acknowledge support from the European Research Council under the Horizon 2020 Framework Program via the ERC Advanced Grant Origins 83 24 28. 

A.C.G. acknowledges support from PRIN-MUR 2022 20228JPA3A “The path to star and planet formation in the JWST era (PATH)” funded by NextGeneration EU and by INAF-GoG 2022 “NIR-dark Accretion Outbursts in Massive Young stellar objects (NAOMY)” and Large Grant INAF 2022 “YSOs Outflows, Disks and Accretion: towards a global framework for the evolution of planet forming systems (YODA)”.

I.K. acknowledges funding from H2020-MSCA-ITN-2019, grant no. 860470 (CHAMELEON).

G.P. gratefully acknowledges support from the Carlsberg Foundation, grant CF23-0481 and from the Max Planck Society.

L.M.S. has received funding from the European Research Council (ERC) under the European Union’s Horizon 2020 research and innovation programme (PROTOPLANETS, grant agreement No. 101002188).

BT acknowledges to support of the Programme National PCMI of CNRS/INSU with INC/INP cofunded by CEA and CNES.
\section*{Data Availability}

The analysed JWST/MIRI-MRS spectrum will be made public on the MINDS webpage (\url{https://minds.cab.inta-csic.es}). The DuCKLinG modelling setup can be found on github (\url{https://github.com/tillkaeufer/DuCKLinG}). Additional data
products are available from the authors upon reasonable request.



\bibliographystyle{mnras}
\bibliography{lib} 

@article{Be_SiO_Kraus_2015,
doi = {10.1088/2041-8205/800/2/L20},
url = {https://doi.org/10.1088/2041-8205/800/2/L20},
year = {2015},
month = {feb},
publisher = {The American Astronomical Society},
volume = {800},
number = {2},
pages = {L20},
author = {Kraus, M. and Oksala, M. E. and Cidale, L. S. and Arias, M. L. and Torres, A. F. and Fernandes, M. Borges},
title = {DISCOVERY OF SiO BAND EMISSION FROM GALACTIC B[e] SUPERGIANTS*},
journal = {\apjl}
}

@ARTICLE{Macla_XUE01,
       author = {{Ram{\'\i}rez-Tannus}, Mar{\'\i}a Claudia and {Bik}, Arjan and {Cuijpers}, Lars and {Waters}, Rens and {G{\"o}ppl}, Christiane and {Henning}, Thomas and {Kamp}, Inga and {Preibisch}, Thomas and {Getman}, Konstantin V. and {Chaparro}, Germ{\'a}n and {Cuartas-Restrepo}, Pablo and {de Koter}, Alex and {Feigelson}, Eric D. and {Grant}, Sierra L. and {Haworth}, Thomas J. and {Hern{\'a}ndez}, Sebasti{\'a}n and {Kuhn}, Michael A. and {Perotti}, Giulia and {Povich}, Matthew S. and {Reiter}, Megan and {Roccatagliata}, Veronica and {Sabbi}, Elena and {Tabone}, Beno{\^\i}t and {Winter}, Andrew J. and {McLeod}, Anna F. and {van Boekel}, Roy and {van Terwisga}, Sierk E.},
        title = "{XUE: Molecular Inventory in the Inner Region of an Extremely Irradiated Protoplanetary Disk}",
      journal = {\apjl},
         year = 2023,
        month = dec,
       volume = {958},
       number = {2},
          eid = {L30},
        pages = {L30},
          doi = {10.3847/2041-8213/ad03f8},
archivePrefix = {arXiv},
       eprint = {2310.11074},
 primaryClass = {astro-ph.SR},
       adsurl = {https://ui.adsabs.harvard.edu/abs/2023ApJ...958L..30R}
}

@ARTICLE{Jenny_XUE10,
       author = {{Frediani}, Jenny and {Bik}, Arjan and {Ram{\'\i}rez-Tannus}, Mar{\'\i}a Claudia and {Waters}, Rens and {Getman}, Konstantin V. and {Feigelson}, Eric D. and {Portilla-Revelo}, Bayron and {Tabone}, Beno{\^\i}t and {Haworth}, Thomas J. and {Winter}, Andrew and {Henning}, Thomas and {Perotti}, Giulia and {Brandeker}, Alexis and {Chaparro}, Germ{\'a}n and {Cuartas-Restrepo}, Pablo and {Hern{\'a}ndez A.}, Sebastian and {Kuhn}, Michael A. and {Preibisch}, Thomas and {Roccatagliata}, Veronica and {van Terwisga}, Sierk E. and {Zeidler}, Peter},
        title = "{XUE: The CO$_{2}$-rich terrestrial planet-forming region of an externally irradiated Herbig disk}",
      journal = {\aap},
         year = 2025,
        month = sep,
       volume = {701},
          eid = {A14},
        pages = {A14},
          doi = {10.1051/0004-6361/202555718},
archivePrefix = {arXiv},
       eprint = {2507.13921},
 primaryClass = {astro-ph.EP},
       adsurl = {https://ui.adsabs.harvard.edu/abs/2025A&A...701A..14F}
}

@ARTICLE{menu_midi,
       author = {{Menu}, J. and {van Boekel}, R. and {Henning}, Th. and {Leinert}, Ch. and {Waelkens}, C. and {Waters}, L.~B.~F.~M.},
        title = "{The structure of disks around intermediate-mass young stars from mid-infrared interferometry. Evidence for a population of group II disks with gaps}",
      journal = {\aap},
         year = 2015,
        month = sep,
       volume = {581},
          eid = {A107},
        pages = {A107},
          doi = {10.1051/0004-6361/201525654},
archivePrefix = {arXiv},
       eprint = {1506.03274},
 primaryClass = {astro-ph.SR},
       adsurl = {https://ui.adsabs.harvard.edu/abs/2015A&A...581A.107M}
}

@ARTICLE{Rouille_cold_silicate,
       author = {{Rouill{\'e}}, Ga{\"e}l and {J{\"a}ger}, Cornelia and {Henning}, Thomas},
        title = "{Separate Silicate and Carbonaceous Solids Formed from Mixed Atomic and Molecular Species Diffusing in Neon Ice}",
      journal = {\apj},
         year = 2020,
        month = apr,
       volume = {892},
       number = {2},
          eid = {96},
        pages = {96},
          doi = {10.3847/1538-4357/ab7a11},
archivePrefix = {arXiv},
       eprint = {2002.10728},
 primaryClass = {astro-ph.GA},
       adsurl = {https://ui.adsabs.harvard.edu/abs/2020ApJ...892...96R}
}

@ARTICLE{Thi_steam2005A&A...438..557T,
       author = {{Thi}, W. -F. and {Bik}, A.},
        title = "{Detection of steam in the circumstellar disk around a massive Young Stellar Object}",
      journal = {\aap},
         year = 2005,
        month = aug,
       volume = {438},
       number = {2},
        pages = {557-570},
          doi = {10.1051/0004-6361:20042219},
archivePrefix = {arXiv},
       eprint = {astro-ph/0503547},
 primaryClass = {astro-ph},
       adsurl = {https://ui.adsabs.harvard.edu/abs/2005A&A...438..557T}
}

@ARTICLE{Stapper2025A&A...693A.286S,
       author = {{Stapper}, L.~M. and {Hogerheijde}, M.~R. and {van Dishoeck}, E.~F. and {Vioque}, M. and {Williams}, J.~P. and {Ginski}, C.},
        title = "{Intermediate mass T Tauri disk masses and a comparison to their Herbig disk descendants}",
      journal = {\aap},
         year = 2025,
        month = jan,
       volume = {693},
          eid = {A286},
        pages = {A286},
          doi = {10.1051/0004-6361/202450260},
archivePrefix = {arXiv},
       eprint = {2411.08953},
 primaryClass = {astro-ph.EP},
       adsurl = {https://ui.adsabs.harvard.edu/abs/2025A&A...693A.286S}
}

@ARTICLE{2004AJ....128.1294C,
       author = {{Calvet}, Nuria and {Muzerolle}, James and {Brice{\~n}o}, C{\'e}sar and {Hern{\'a}ndez}, Jesus and {Hartmann}, Lee and {Saucedo}, Jos{\'e} Luis and {Gordon}, Karl D.},
        title = "{The Mass Accretion Rates of Intermediate-Mass T Tauri Stars}",
      journal = {\aj},
         year = 2004,
        month = sep,
       volume = {128},
       number = {3},
        pages = {1294-1318},
          doi = {10.1086/422733},
       adsurl = {https://ui.adsabs.harvard.edu/abs/2004AJ....128.1294C}
}

@ARTICLE{2023AJ....166..183V,
       author = {{Vioque}, Miguel and {Cavieres}, Manuel and {Pantaleoni Gonz{\'a}lez}, Michelangelo and {Ribas}, {\'A}lvaro and {Oudmaijer}, Ren{\'e} D. and {Mendigut{\'\i}a}, Ignacio and {Kilian}, Lena and {C{\'a}novas}, H{\'e}ctor and {Kuhn}, Michael A.},
        title = "{Clustering Properties of Intermediate and High-mass Young Stellar Objects}",
      journal = {\aj},
         year = 2023,
        month = nov,
       volume = {166},
       number = {5},
          eid = {183},
        pages = {183},
          doi = {10.3847/1538-3881/acf75f},
archivePrefix = {arXiv},
       eprint = {2309.00678},
 primaryClass = {astro-ph.SR},
       adsurl = {https://ui.adsabs.harvard.edu/abs/2023AJ....166..183V}
}

@article{herbig_spectra_1960,
    title = {The {Spectra} of {Be}- and {Ae}-{TYPE} {Stars} {Associated} with {Nebulosity}},
    volume = {4},
    doi = {10.1086/190050},
    journal = {\apjs},
    author = {Herbig, G. H.},
    month = mar,
    year = {1960},
    pages = {337},
}

@article{antonellini_understanding_2015,
       author = {{Antonellini}, S. and {Kamp}, I. and {Riviere-Marichalar}, P. and {Meijerink}, R. and {Woitke}, P. and {Thi}, W. -F. and {Spaans}, M. and {Aresu}, G. and {Lee}, E.},
        title = "{Understanding the water emission in the mid- and far-IR from protoplanetary disks around T Tauri stars}",
      journal = {\aap},
         year = 2015,
        month = oct,
       volume = {582},
          eid = {A105},
        pages = {A105},
          doi = {10.1051/0004-6361/201525724},
archivePrefix = {arXiv},
       eprint = {1510.01482},
 primaryClass = {astro-ph.SR},
       adsurl = {https://ui.adsabs.harvard.edu/abs/2015A&A...582A.105A}
}

@article{banzatti_observing_2018,
       author = {{Banzatti}, A. and {Garufi}, A. and {Kama}, M. and {Benisty}, M. and {Brittain}, S. and {Pontoppidan}, K.~M. and {Rayner}, J.},
        title = "{Observing the linked depletion of dust and CO gas at 0.1-10 au in disks of intermediate-mass stars}",
      journal = {\aap},
         year = 2018,
        month = feb,
       volume = {609},
          eid = {L2},
        pages = {L2},
          doi = {10.1051/0004-6361/201732034},
archivePrefix = {arXiv},
       eprint = {1711.09095},
 primaryClass = {astro-ph.EP},
       adsurl = {https://ui.adsabs.harvard.edu/abs/2018A&A...609L...2B}
}

@article{brittain_study_2016,
       author = {{Brittain}, Sean D. and {Najita}, Joan R. and {Carr}, John S. and {{\'A}d{\'a}mkovics}, M{\'a}t{\'e} and {Reynolds}, Nickalas},
        title = "{A Study of Ro-vibrational OH Emission from Herbig Ae/Be Stars}",
      journal = {\apj},
         year = 2016,
        month = oct,
       volume = {830},
       number = {2},
          eid = {112},
        pages = {112},
          doi = {10.3847/0004-637X/830/2/112},
archivePrefix = {arXiv},
       eprint = {1608.00986},
 primaryClass = {astro-ph.SR},
       adsurl = {https://ui.adsabs.harvard.edu/abs/2016ApJ...830..112B}
}

@article{fedele_water_2011,
    title = {Water {Depletion} in the {Disk} {Atmosphere} of {Herbig} {AeBe} {Stars}},
    volume = {732},
    doi = {10.1088/0004-637X/732/2/106},
    number = {2},
    journal = {\apj},
    author = {Fedele, D. and Pascucci, I. and Brittain, S. and Kamp, I. and Woitke, P. and Williams, J. P. and Dent, W. R. F. and Thi, W. -F.},
    month = may,
    year = {2011},
    note = {\_eprint: 1103.6039},
    pages = {106},
}

@article{pontoppidan_spitzer_2010,
    title = {A {Spitzer} {Survey} of {Mid}-infrared {Molecular} {Emission} from {Protoplanetary} {Disks}. {I}. {Detection} {Rates}},
    volume = {720},
    doi = {10.1088/0004-637X/720/1/887},
    number = {1},
    journal = {\apj},
    author = {Pontoppidan, Klaus M. and Salyk, Colette and Blake, Geoffrey A. and Meijerink, Rowin and Carr, John S. and Najita, Joan},
    month = sep,
    year = {2010},
    note = {\_eprint: 1006.4189},
    pages = {887--903},
}

@phdthesis{antonellini_water_2016,
       author = {{Antonellini}, Stefano},
        title = "{Water in protoplanetary disks: Line flux modeling and disk structure}",
       school = {University of Groningen, Netherlands},
         year = 2016,
        month = jan,
       adsurl = {https://ui.adsabs.harvard.edu/abs/2016PhDT.......274A}
}

@article{meeus_iso_2001,
       author = {{Meeus}, G. and {Waters}, L.~B.~F.~M. and {Bouwman}, J. and {van den Ancker}, M.~E. and {Waelkens}, C. and {Malfait}, K.},
        title = "{ISO spectroscopy of circumstellar dust in 14 Herbig Ae/Be systems: Towards an understanding of dust processing}",
      journal = {\aap},
         year = 2001,
        month = jan,
       volume = {365},
        pages = {476-490},
          doi = {10.1051/0004-6361:20000144},
archivePrefix = {arXiv},
       eprint = {astro-ph/0012295},
 primaryClass = {astro-ph},
       adsurl = {https://ui.adsabs.harvard.edu/abs/2001A&A...365..476M}
}

@article{brittain_warm_2007,
       author = {{Brittain}, Sean D. and {Simon}, Theodore and {Najita}, Joan R. and {Rettig}, Terrence W.},
        title = "{Warm Gas in the Inner Disks around Young Intermediate-Mass Stars}",
      journal = {\apj},
         year = 2007,
        month = apr,
       volume = {659},
       number = {1},
        pages = {685-704},
          doi = {10.1086/511255},
archivePrefix = {arXiv},
       eprint = {astro-ph/0612201},
 primaryClass = {astro-ph},
       adsurl = {https://ui.adsabs.harvard.edu/abs/2007ApJ...659..685B}
}

@article{gravity_collaboration_gravity_2019,
       author = {{GRAVITY Collaboration} and {Perraut}, K. and {Labadie}, L. and {Lazareff}, B. and {Klarmann}, L. and {Segura-Cox}, D. and {Benisty}, M. and {Bouvier}, J. and {Brandner}, W. and {Caratti O Garatti}, A. and {Caselli}, P. and {Dougados}, C. and {Garcia}, P. and {Garcia-Lopez}, R. and {Kendrew}, S. and {Koutoulaki}, M. and {Kervella}, P. and {Lin}, C. -C. and {Pineda}, J. and {Sanchez-Bermudez}, J. and {van Dishoeck}, E. and {Abuter}, R. and {Amorim}, A. and {Berger}, J. -P. and {Bonnet}, H. and {Buron}, A. and {Cantalloube}, F. and {Cl{\'e}net}, Y. and {Coud{\'e} Du Foresto}, V. and {Dexter}, J. and {de Zeeuw}, P.~T. and {Duvert}, G. and {Eckart}, A. and {Eisenhauer}, F. and {Eupen}, F. and {Gao}, F. and {Gendron}, E. and {Genzel}, R. and {Gillessen}, S. and {Gordo}, P. and {Grellmann}, R. and {Haubois}, X. and {Haussmann}, F. and {Henning}, T. and {Hippler}, S. and {Horrobin}, M. and {Hubert}, Z. and {Jocou}, L. and {Lacour}, S. and {Le Bouquin}, J. -B. and {L{\'e}na}, P. and {M{\'e}rand}, A. and {Ott}, T. and {Paumard}, T. and {Perrin}, G. and {Pfuhl}, O. and {Rabien}, S. and {Ray}, T. and {Rau}, C. and {Rousset}, G. and {Scheithauer}, S. and {Straub}, O. and {Straubmeier}, C. and {Sturm}, E. and {Vincent}, F. and {Waisberg}, I. and {Wank}, I. and {Widmann}, F. and {Wieprecht}, E. and {Wiest}, M. and {Wiezorrek}, E. and {Woillez}, J. and {Yazici}, S.},
        title = "{The GRAVITY Young Stellar Object survey. I. Probing the disks of Herbig Ae/Be stars in terrestrial orbits}",
      journal = {\aap},
         year = 2019,
        month = dec,
       volume = {632},
          eid = {A53},
        pages = {A53},
          doi = {10.1051/0004-6361/201936403},
archivePrefix = {arXiv},
       eprint = {1911.00611},
 primaryClass = {astro-ph.SR},
       adsurl = {https://ui.adsabs.harvard.edu/abs/2019A&A...632A..53G}
}

@software{pipeline,
       author = {{Bushouse}, Howard and {Eisenhamer}, Jonathan and {Dencheva}, Nadia and {Davies}, James and {Greenfield}, Perry and {Morrison}, Jane and {Hodge}, Phil and {Simon}, Bernie and {Grumm}, David and {Droettboom}, Michael and {Slavich}, Edward and {Sosey}, Megan and {Pauly}, Tyler and {Miller}, Todd and {Jedrzejewski}, Robert and {Hack}, Warren and {Davis}, David and {Crawford}, Steven and {Law}, David and {Gordon}, Karl and {Regan}, Michael and {Cara}, Mihai and {MacDonald}, Ken and {Bradley}, Larry and {Shanahan}, Clare and {Jamieson}, William and {Teodoro}, Mairan and {Williams}, Thomas and {Pena-Guerrero}, Maria and {Graham}, Brett and {Molter}, Edward and {Brandt}, Timothy and {Hayes}, Christian and {Cooper}, Rachel and {Clarke}, Melanie},
        title = "{JWST Calibration Pipeline}",
         year = 2024,
        month = nov,
          eid = {10.5281/zenodo.14153298},
          doi = {10.5281/zenodo.14153298},
      version = {1.16.1},
    publisher = {Zenodo},
       adsurl = {https://ui.adsabs.harvard.edu/abs/2024zndo..14153298B}
}

@article{Feroz2008,
   title={Multimodal nested sampling: an efficient and robust alternative to Markov Chain Monte Carlo methods for astronomical data analyses},
   volume={384},
   ISSN={1365-2966},
   url={http://dx.doi.org/10.1111/j.1365-2966.2007.12353.x},
   DOI={10.1111/j.1365-2966.2007.12353.x},
   number={2},
   journal={\mnras},
   publisher={Oxford University Press (OUP)},
   author={Feroz, F. and Hobson, M. P.},
   year={2008},
   month={Jan},
   pages={449–463} }

@article{Feroz2009,
   title={MultiNest: an efficient and robust Bayesian inference tool for cosmology and particle physics},
   volume={398},
   ISSN={1365-2966},
   url={http://dx.doi.org/10.1111/j.1365-2966.2009.14548.x},
   DOI={10.1111/j.1365-2966.2009.14548.x},
   number={4},
   journal={\mnras},
   publisher={Oxford University Press (OUP)},
   author={Feroz, F. and Hobson, M. P. and Bridges, M.},
   year={2009},
   month={Oct},
   pages={1601–1614} }

@article{Feroz2019,
   title={Importance Nested Sampling and the MultiNest Algorithm},
   volume={2},
   ISSN={2565-6120},
   url={http://dx.doi.org/10.21105/astro.1306.2144},
   DOI={10.21105/astro.1306.2144},
   number={1},
   journal={The Open Journal of Astrophysics},
   publisher={The Open Journal},
   author={Feroz, Farhan and Hobson, Michael P. and Cameron, Ewan and Pettitt, Anthony N.},
   year={2019},
   month={Nov} }

@article{Buchner2014,
	author = {{Buchner} and {Georgakakis} and {Nandra} and {Hsu} and {Rangel} and {Brightman} and {Merloni} and {Salvato} and {Donley} and {Kocevski}},
	title = {X-ray spectral modelling of the AGN obscuring region in the
          CDFS: Bayesian model selection and catalogue},
	DOI= "10.1051/0004-6361/201322971",
	url= "https://doi.org/10.1051/0004-6361/201322971",
	journal = {A\&A},
	year = 2014,
	volume = 564,
	pages = "A125",
	month = "",
}

@ARTICLE{Carnall2017,
       author = {{Carnall}, A.~C.},
        title = "{SpectRes: A Fast Spectral Resampling Tool in Python}",
      journal = {arXiv e-prints},
         year = 2017,
        month = may,
          eid = {arXiv:1705.05165},
        pages = {arXiv:1705.05165},
          doi = {10.48550/arXiv.1705.05165},
archivePrefix = {arXiv},
       eprint = {1705.05165},
 primaryClass = {astro-ph.IM},
       adsurl = {https://ui.adsabs.harvard.edu/abs/2017arXiv170505165C}
}

@ARTICLE{Fedele2016,
       author = {{Fedele}, D. and {van Dishoeck}, E.~F. and {Kama}, M. and {Bruderer}, S. and {Hogerheijde}, M.~R.},
        title = "{Probing the 2D temperature structure of protoplanetary disks with Herschel observations of high-J CO lines}",
      journal = {\aap},
         year = 2016,
        month = jun,
       volume = {591},
          eid = {A95},
        pages = {A95},
          doi = {10.1051/0004-6361/201526948},
archivePrefix = {arXiv},
       eprint = {1604.02055},
 primaryClass = {astro-ph.SR},
       adsurl = {https://ui.adsabs.harvard.edu/abs/2016A&A...591A..95F}
}

@ARTICLE{Trotta2008,
       author = {{Trotta}, Roberto},
        title = "{Bayes in the sky: Bayesian inference and model selection in cosmology}",
      journal = {Contemporary Physics},
         year = 2008,
        month = mar,
       volume = {49},
       number = {2},
        pages = {71-104},
          doi = {10.1080/00107510802066753},
archivePrefix = {arXiv},
       eprint = {0803.4089},
 primaryClass = {astro-ph},
       adsurl = {https://ui.adsabs.harvard.edu/abs/2008ConPh..49...71T}
}

@ARTICLE{Temmink2025,
       author = {{Temmink}, Milou and {Sellek}, Andrew D. and {Gasman}, Danny and {van Dishoeck}, Ewine F. and {Vlasblom}, Marissa and {Pranger}, Ang{\`e}l and {G{\"u}del}, Manuel and {Henning}, Thomas and {Lagage}, Pierre-Olivier and {Caratti o Garatti}, Alessio and {Kamp}, Inga and {Olofsson}, G{\"o}ran and {Arabhavi}, Aditya M. and {Grant}, Sierra L. and {Kaeufer}, Till and {Kurtovic}, Nicolas T. and {Perotti}, Giulia and {Samland}, Matthias and {Schwarz}, Kamber and {Tabone}, Beno{\^\i}t},
        title = "{MINDS: Water reservoirs of compact planet-forming dust discs: A diversity of H$_{2}$O distributions}",
      journal = {\aap},
         year = 2025,
        month = jul,
       volume = {699},
          eid = {A134},
        pages = {A134},
          doi = {10.1051/0004-6361/202554213},
archivePrefix = {arXiv},
       eprint = {2505.15237},
 primaryClass = {astro-ph.EP},
       adsurl = {https://ui.adsabs.harvard.edu/abs/2025A&A...699A.134T}
}

@ARTICLE{Jellison2024,
       author = {{Jellison}, Evan G. and {Banzatti}, Andrea and {Johnson}, Matthew B. and {Bruderer}, Simon},
        title = "{iSLAT: the Interactive Spectral-line Analysis Tool for JWST and Beyond}",
      journal = {\aj},
         year = 2024,
        month = sep,
       volume = {168},
       number = {3},
          eid = {99},
        pages = {99},
          doi = {10.3847/1538-3881/ad6142},
archivePrefix = {arXiv},
       eprint = {2402.04060},
 primaryClass = {astro-ph.IM},
       adsurl = {https://ui.adsabs.harvard.edu/abs/2024AJ....168...99J}
}

@ARTICLE{Tennyson2024,
       author = {{Tennyson}, Jonathan and {Yurchenko}, Sergei N. and {Zhang}, Jingxin and {Bowesman}, Charles A. and {Brady}, Ryan P. and {Buldyreva}, Jeanna and {Chubb}, Katy L. and {Gamache}, Robert R. and {Gorman}, Maire N. and {Guest}, Elizabeth R. and {Hill}, Christian and {Kefala}, Kyriaki and {Lynas-Gray}, A.~E. and {Mellor}, Thomas M. and {McKemmish}, Laura K. and {Mitev}, Georgi B. and {Mizus}, Irina I. and {Owens}, Alec and {Peng}, Zhijian and {Perri}, Armando N. and {Pezzella}, Marco and {Polyansky}, Oleg L. and {Qu}, Qianwei and {Semenov}, Mikhail and {Smola}, Oleksiy and {Solokov}, Andrei and {Somogyi}, Wilfrid and {Upadhyay}, Apoorva and {Wright}, Samuel O.~M. and {Zobov}, Nikolai F.},
        title = "{The 2024 release of the ExoMol database: Molecular line lists for exoplanet and other hot atmospheres}",
      journal = {\jqsrt},
         year = 2024,
        month = nov,
       volume = {326},
          eid = {109083},
        pages = {109083},
          doi = {10.1016/j.jqsrt.2024.109083},
archivePrefix = {arXiv},
       eprint = {2406.06347},
 primaryClass = {astro-ph.GA},
       adsurl = {https://ui.adsabs.harvard.edu/abs/2024JQSRT.32609083T}
}

@ARTICLE{Gaia2020,
       author = {{Gaia Collaboration}},
        title = "{VizieR Online Data Catalog: Gaia EDR3 (Gaia Collaboration, 2020)}",
      journal = {VizieR Online Data Catalog},
         year = 2020,
        month = nov,
          eid = {I/350},
        pages = {I/350},
          doi = {10.26093/cds/vizier.1350},
       adsurl = {https://ui.adsabs.harvard.edu/abs/2020yCat.1350....0G}
}

@ARTICLE{Kurucz1979,
       author = {{Kurucz}, R.~L.},
        title = "{Model atmospheres for G, F, A, B, and O stars.}",
      journal = {\apjs},
         year = 1979,
        month = may,
       volume = {40},
        pages = {1-340},
          doi = {10.1086/190589},
       adsurl = {https://ui.adsabs.harvard.edu/abs/1979ApJS...40....1K}
}

@ARTICLE{Samland2025,
       author = {{Samland}, Matthias and {Henning}, Thomas and {Caratti o Garatti}, Alessio and {Giannini}, Teresa and {Bouwman}, Jeroen and {Tabone}, Beno{\^\i}t and {Arabhavi}, Aditya M. and {Olofsson}, G{\"o}ran and {G{\"u}del}, Manuel and {Pawellek}, Nicole and {Kamp}, Inga and {Waters}, L.~B.~F.~M. and {Semenov}, Dmitry and {van Dishoeck}, Ewine F. and {Absil}, Olivier and {Barrado}, David and {Boccaletti}, Anthony and {Christiaens}, Valentin and {Gasman}, Danny and {Grant}, Sierra L. and {Jang}, Hyerin and {Kaeufer}, Till and {Kanwar}, Jayatee and {Perotti}, Giulia and {Schwarz}, Kamber and {Temmink}, Milou},
        title = "{MINDS: Detection of an Inner Gas Disk Caused by Evaporating Bodies around HD 172555}",
      journal = {\apj},
         year = 2025,
        month = aug,
       volume = {989},
       number = {2},
          eid = {132},
        pages = {132},
          doi = {10.3847/1538-4357/ade2db},
archivePrefix = {arXiv},
       eprint = {2506.09976},
 primaryClass = {astro-ph.EP},
       adsurl = {https://ui.adsabs.harvard.edu/abs/2025ApJ...989..132S}
}

@ARTICLE{Temmink_continuum_2024,
       author = {{Temmink}, Milou and {van Dishoeck}, Ewine F. and {Grant}, Sierra L. and {Tabone}, Beno{\^\i}t and {Gasman}, Danny and {Christiaens}, Valentin and {Samland}, Matthias and {Argyriou}, Ioannis and {Perotti}, Giulia and {G{\"u}del}, Manuel and {Henning}, Thomas and {Lagage}, Pierre-Olivier and {Abergel}, Alain and {Absil}, Olivier and {Barrado}, David and {Caratti o Garatti}, Alessio and {Glauser}, Adrian M. and {Kamp}, Inga and {Lahuis}, Fred and {Olofsson}, G{\"o}ran and {Ray}, Tom P. and {Scheithauer}, Silvia and {Vandenbussche}, Bart and {Waters}, L.~B.~F.~M. and {Arabhavi}, Aditya M. and {Jang}, Hyerin and {Kanwar}, Jayatee and {Morales-Calder{\'o}n}, Maria and {Rodgers-Lee}, Donna and {Schreiber}, J{\"u}rgen and {Schwarz}, Kamber and {Colina}, Luis},
        title = "{MINDS: The DR Tau disk. I. Combining JWST-MIRI data with high-resolution CO spectra to characterise the hot gas}",
      journal = {\aap},
         year = 2024,
        month = jun,
       volume = {686},
          eid = {A117},
        pages = {A117},
          doi = {10.1051/0004-6361/202348911},
archivePrefix = {arXiv},
       eprint = {2403.13591},
 primaryClass = {astro-ph.EP},
       adsurl = {https://ui.adsabs.harvard.edu/abs/2024A&A...686A.117T}
}

@ARTICLE{Ramirez-taunnus2025,
       author = {{Ram{\'\i}rez-Tannus}, Mar{\'\i}a Claudia and {Bik}, Arjan and {Getman}, Konstantin V. and {Waters}, Rens and {Portilla-Revelo}, Bayron and {G{\"o}ppl}, Christiane and {Winter}, Andrew J. and {Frediani}, Jenny and {Chaparro}, Germ{\'a}n and {Feigelson}, Eric D. and {Haworth}, Thomas J. and {Henning}, Thomas and {Hern{\'a}ndez}, Sebasti{\'a}n and {Lemus-Nemoc{\'o}n}, Maria Alejandra and {Kuhn}, Michael and {Preibisch}, Thomas and {Roccatagliata}, Veronica and {Sabbi}, Elena and {van Boekel}, Roy and {Zeidler}, Peter},
        title = "{XUE: JWST spectroscopy of externally irradiated disks around young intermediate-mass stars}",
      journal = {\aap},
         year = 2025,
        month = sep,
       volume = {701},
          eid = {A139},
        pages = {A139},
          doi = {10.1051/0004-6361/202555456},
archivePrefix = {arXiv},
       eprint = {2505.06093},
 primaryClass = {astro-ph.SR},
       adsurl = {https://ui.adsabs.harvard.edu/abs/2025A&A...701A.139R}
}

@ARTICLE{Banzatti2023,
       author = {{Banzatti}, Andrea and {Pontoppidan}, Klaus M. and {P{\'e}re Ch{\'a}vez}, Jos{\'e} and {Salyk}, Colette and {Diehl}, Lindsey and {Bruderer}, Simon and {Herczeg}, Gregory J. and {Carmona}, Andres and {Pascucci}, Ilaria and {Brittain}, Sean and {Jensen}, Stanley and {Grant}, Sierra and {van Dishoeck}, Ewine F. and {Kamp}, Inga and {Bosman}, Arthur D. and {{\"O}berg}, Karin I. and {Blake}, Geoff A. and {Meyer}, Michael R. and {Gaidos}, Eric and {Boogert}, Adwin and {Rayner}, John T. and {Wheeler}, Caleb},
        title = "{The Kinematics and Excitation of Infrared Water Vapor Emission from Planet-forming Disks: Results from Spectrally Resolved Surveys and Guidelines for JWST Spectra}",
      journal = {\aj},
         year = 2023,
        month = feb,
       volume = {165},
       number = {2},
          eid = {72},
        pages = {72},
          doi = {10.3847/1538-3881/aca80b},
archivePrefix = {arXiv},
       eprint = {2209.08216},
 primaryClass = {astro-ph.EP},
       adsurl = {https://ui.adsabs.harvard.edu/abs/2023AJ....165...72B}
}

@ARTICLE{Gordon2022,
       author = {{Gordon}, I.~E. and {Rothman}, L.~S. and {Hargreaves}, R.~J. and {Hashemi}, R. and {Karlovets}, E.~V. and {Skinner}, F.~M. and {Conway}, E.~K. and {Hill}, C. and {Kochanov}, R.~V. and {Tan}, Y. and {Wcis{\l}o}, P. and {Finenko}, A.~A. and {Nelson}, K. and {Bernath}, P.~F. and {Birk}, M. and {Boudon}, V. and {Campargue}, A. and {Chance}, K.~V. and {Coustenis}, A. and {Drouin}, B.~J. and {Flaud}, J. -M. and {Gamache}, R.~R. and {Hodges}, J.~T. and {Jacquemart}, D. and {Mlawer}, E.~J. and {Nikitin}, A.~V. and {Perevalov}, V.~I. and {Rotger}, M. and {Tennyson}, J. and {Toon}, G.~C. and {Tran}, H. and {Tyuterev}, V.~G. and {Adkins}, E.~M. and {Baker}, A. and {Barbe}, A. and {Can{\`e}}, E. and {Cs{\'a}sz{\'a}r}, A.~G. and {Dudaryonok}, A. and {Egorov}, O. and {Fleisher}, A.~J. and {Fleurbaey}, H. and {Foltynowicz}, A. and {Furtenbacher}, T. and {Harrison}, J.~J. and {Hartmann}, J. -M. and {Horneman}, V. -M. and {Huang}, X. and {Karman}, T. and {Karns}, J. and {Kassi}, S. and {Kleiner}, I. and {Kofman}, V. and {Kwabia-Tchana}, F. and {Lavrentieva}, N.~N. and {Lee}, T.~J. and {Long}, D.~A. and {Lukashevskaya}, A.~A. and {Lyulin}, O.~M. and {Makhnev}, V. Yu. and {Matt}, W. and {Massie}, S.~T. and {Melosso}, M. and {Mikhailenko}, S.~N. and {Mondelain}, D. and {M{\"u}ller}, H.~S.~P. and {Naumenko}, O.~V. and {Perrin}, A. and {Polyansky}, O.~L. and {Raddaoui}, E. and {Raston}, P.~L. and {Reed}, Z.~D. and {Rey}, M. and {Richard}, C. and {T{\'o}bi{\'a}s}, R. and {Sadiek}, I. and {Schwenke}, D.~W. and {Starikova}, E. and {Sung}, K. and {Tamassia}, F. and {Tashkun}, S.~A. and {Vander Auwera}, J. and {Vasilenko}, I.~A. and {Vigasin}, A.~A. and {Villanueva}, G.~L. and {Vispoel}, B. and {Wagner}, G. and {Yachmenev}, A. and {Yurchenko}, S.~N.},
        title = "{The HITRAN2020 molecular spectroscopic database}",
      journal = {\jqsrt},
         year = 2022,
        month = jan,
       volume = {277},
          eid = {107949},
        pages = {107949},
          doi = {10.1016/j.jqsrt.2021.107949},
       adsurl = {https://ui.adsabs.harvard.edu/abs/2022JQSRT.27707949G}
}

@article{Buldyreva2022,
    author = {Buldyreva, Jeanna and Yurchenko, Sergei N and Tennyson, Jonathan},
    title = {Simple semiclassical model of pressure-broadened infrared/microwave linewidths in the temperature range 200–3000K},
    journal = {RAS Techniques and Instruments},
    volume = {1},
    number = {1},
    pages = {43-47},
    year = {2022},
    month = {07},
    issn = {2752-8200},
    doi = {10.1093/rasti/rzac004},
    url = {https://doi.org/10.1093/rasti/rzac004},
    eprint = {https://academic.oup.com/rasti/article-pdf/1/1/43/50653360/rzac004.pdf},
}

@article{Guest2024,
title = {Predicting the rotational dependence of line broadening using machine learning},
journal = {Journal of Molecular Spectroscopy},
volume = {401},
pages = {111901},
year = {2024},
issn = {0022-2852},
doi = {https://doi.org/10.1016/j.jms.2024.111901},
url = {https://www.sciencedirect.com/science/article/pii/S0022285224000286},
author = {Elizabeth R. Guest and Jonathan Tennyson and Sergei N. Yurchenko}
}

@article{malfait_spectrum_1998,
       author = {{Malfait}, K. and {Waelkens}, C. and {Waters}, L.~B.~F.~M. and {Vandenbussche}, B. and {Huygen}, E. and {de Graauw}, M.~S.},
        title = "{The spectrum of the young star HD 100546 observed with the Infrared Space Observatory}",
      journal = {\aap},
         year = 1998,
        month = apr,
       volume = {332},
        pages = {L25-L28},
       adsurl = {https://ui.adsabs.harvard.edu/abs/1998A&A...332L..25M}
}

@article{mcclure_refractory_2025,
	title = {Refractory solid condensation detected in an embedded protoplanetary disk},
	volume = {643},
	issn = {0028-0836, 1476-4687},
	url = {https://www.nature.com/articles/s41586-025-09163-z},
	doi = {10.1038/s41586-025-09163-z},
	language = {en},
	number = {8072},
	urldate = {2025-07-31},
	journal = {Nature},
	author = {McClure, M. K. and Van’T Hoff, Merel and Francis, Logan and Bergin, Edwin and Rocha, Will R. M. and Sturm, J. A. and Harsono, Daniel and Van Dishoeck, Ewine F. and Black, John H. and Noble, J. A. and Qasim, D. and Dartois, E.},
	month = jul,
	year = {2025},
	pages = {649--653},
}

@article{guilloteau_first_1992,
       author = {{Guilloteau}, S. and {Bachiller}, R. and {Fuente}, A. and {Lucas}, R.},
        title = "{First observations of young bipolar outflows with the IRAM interferometer : 2'' resolution SiO images of the molecular jet in L 1448.}",
      journal = {\aap},
         year = 1992,
        month = nov,
       volume = {265},
        pages = {L49-L52},
       adsurl = {https://ui.adsabs.harvard.edu/abs/1992A&A...265L..49G}
}

@article{tabone_oh_2024,
	title = {{OH} mid-infrared emission as a diagnostic of {H2O} {UV} photodissociation - {III}. {Application} to planet-forming disks},
	volume = {691},
	copyright = {© The Authors 2024},
	issn = {0004-6361, 1432-0746},
	url = {https://www.aanda.org/articles/aa/abs/2024/11/aa48487-23/aa48487-23.html},
	doi = {10.1051/0004-6361/202348487},
	language = {en},
	urldate = {2025-07-31},
	journal = {\aap},
	author = {Tabone, Benoît and Dishoeck, Ewine F. van and Black, John H.},
	month = nov,
	year = {2024},
	note = {Publisher: EDP Sciences},
	pages = {A11},
}

@article{carr_oh_2014,
	title = {{THE} {OH} {ROTATIONAL} {POPULATION} {AND} {PHOTODISSOCIATION} {OF} {H2O} {IN} {DG} {Tauri}},
	volume = {788},
	issn = {0004-637X},
	url = {https://dx.doi.org/10.1088/0004-637X/788/1/66},
	doi = {10.1088/0004-637X/788/1/66},
	language = {en},
	number = {1},
	urldate = {2025-07-31},
	journal = {\apj},
	author = {Carr, John S. and Najita, Joan R.},
	month = may,
	year = {2014},
	note = {Publisher: The American Astronomical Society},
	pages = {66},
}

@article{banzatti_water_2025,
	title = {Water in {Protoplanetary} {Disks} with {JWST}-{MIRI}: {Spectral} {Excitation} {Atlas} and {Radial} {Distribution} from {Temperature} {Diagnostic} {Diagrams} and {Doppler} {Mapping}},
	volume = {169},
	issn = {1538-3881},
	shorttitle = {Water in {Protoplanetary} {Disks} with {JWST}-{MIRI}},
	url = {https://dx.doi.org/10.3847/1538-3881/ada962},
	doi = {10.3847/1538-3881/ada962},
	language = {en},
	number = {3},
	urldate = {2025-07-30},
	journal = {\aj},
	author = {Banzatti, Andrea and Salyk, Colette and Pontoppidan, Klaus M. and Carr, John S. and Zhang, Ke and Arulanantham, Nicole and Krijt, Sebastiaan and Öberg, Karin I. and Cleeves, L. Ilsedore and Najita, Joan R. and Pascucci, Ilaria and Blake, Geoffrey A. and Romero-Mirza, Carlos E. and Bergin, Edwin A. and Cieza, Lucas A. and Pinilla, Paola and Long, Feng and Mallaney, Patrick and Xie, Chengyan and Waggoner, Abygail R. and Kaeufer, Till and collaboration, the JDISCS},
	month = feb,
	year = {2025},
	note = {Publisher: The American Astronomical Society},
	pages = {165},
}

@article{argyriou_jwst_2023,
	title = {{JWST} {MIRI} flight performance: {The} {Medium}-{Resolution} {Spectrometer}},
	volume = {675},
	copyright = {© The Authors 2023},
	issn = {0004-6361, 1432-0746},
	shorttitle = {{JWST} {MIRI} flight performance},
	url = {https://www.aanda.org/articles/aa/abs/2023/07/aa46489-23/aa46489-23.html},
	doi = {10.1051/0004-6361/202346489},
	language = {en},
	urldate = {2025-07-30},
	journal = {\aap},
	author = {Argyriou, Ioannis and Glasse, Alistair and Law, David R. and Labiano, Alvaro and Álvarez-Márquez, Javier and Patapis, Polychronis and Kavanagh, Patrick J. and Gasman, Danny and Mueller, Michael and Larson, Kirsten and Vandenbussche, Bart and Glauser, Adrian M. and Royer, Pierre and Dicken, Daniel and Harkett, Jake and Sargent, Beth A. and Engesser, Michael and Jones, Olivia C. and Kendrew, Sarah and Noriega-Crespo, Alberto and Brandl, Bernhard and Rieke, George H. and Wright, Gillian S. and Lee, David and Wells, Martyn},
	month = jul,
	year = {2023},
	note = {Publisher: EDP Sciences},
	pages = {A111},
}

@article{wright_mid-infrared_2023,
	title = {The {Mid}-infrared {Instrument} for {JWST} and {Its} {In}-flight {Performance}},
	volume = {135},
	issn = {1538-3873},
	url = {https://dx.doi.org/10.1088/1538-3873/acbe66},
	doi = {10.1088/1538-3873/acbe66},
	language = {en},
	number = {1046},
	urldate = {2025-07-30},
	journal = {\pasp},
	author = {Wright, Gillian S. and Rieke, George H. and Glasse, Alistair and Ressler, Michael and García Marín, Macarena and Aguilar, Jonathan and Alberts, Stacey and Álvarez-Márquez, Javier and Argyriou, Ioannis and Banks, Kimberly and Baudoz, Pierre and Boccaletti, Anthony and Bouchet, Patrice and Bouwman, Jeroen and Brandl, Bernard R. and Breda, David and Bright, Stacey and Cale, Steven and Colina, Luis and Cossou, Christophe and Coulais, Alain and Cracraft, Misty and De Meester, Wim and Dicken, Daniel and Engesser, Michael and Etxaluze, Mireya and Fox, Ori D. and Friedman, Scott and Fu, Henry and Gasman, Danny and Gáspár, András and Gastaud, René and Geers, Vincent and Glauser, Adrian Michael and Gordon, Karl D. and Greene, Thomas and Greve, Thomas R. and Grundy, Timothy and Güdel, Manuel and Guillard, Pierre and Haderlein, Peter and Hashimoto, Ryan and Henning, Thomas and Hines, Dean and Holler, Bryan and Detre, Örs Hunor and Jahromi, Amir and James, Bryan and Jones, Olivia C. and Justtanont, Kay and Kavanagh, Patrick and Kendrew, Sarah and Klaassen, Pamela and Krause, Oliver and Labiano, Alvaro and Lagage, Pierre-Olivier and Lambros, Scott and Larson, Kirsten and Law, David and Lee, David and Libralato, Mattia and Alverez, Jose Lorenzo and Meixner, Margaret and Morrison, Jane and Mueller, Migo and Murray, Katherine and Mycroft, Matthew and Myers, Richard and Nayak, Omnarayani and Naylor, Bret and Nickson, Bryony and Noriega-Crespo, Alberto and Östlin, Göran and O’Sullivan, Brian and Ottens, Richard and Patapis, Polychronis and Penanen, Konstantin and Pietraszkiewicz, Martin and Ray, Tom and Regan, Michael and Roteliuk, Anthony and Royer, Pierre and Samara-Ratna, Piyal and Samuelson, Bridget and Sargent, Beth A. and Scheithauer, Silvia and Schneider, Analyn and Schreiber, Jürgen and Shaughnessy, Bryan and Sheehan, Evan and Shivaei, Irene and Sloan, G. C. and Tamas, Laszlo and Teague, Kelly and Temim, Tea and Tikkanen, Tuomo and Tustain, Samuel and van Dishoeck, Ewine F. and Vandenbussche, Bart and Weilert, Mark and Whitehouse, Paul and Wolff, Schuyler},
	month = may,
	year = {2023},
	note = {Publisher: The Astronomical Society of the Pacific},
	pages = {048003},
}

@article{colmenares_jwstmiri_2024,
	title = {{JWST}/{MIRI} {Detection} of a {Carbon}-rich {Chemistry} in the {Disk} of a {Solar} {Nebula} {Analog}},
	volume = {977},
	issn = {0004-637X},
	url = {https://dx.doi.org/10.3847/1538-4357/ad8b4f},
	doi = {10.3847/1538-4357/ad8b4f},
	language = {en},
	number = {2},
	urldate = {2025-07-30},
	journal = {\apj},
	author = {Colmenares, María José and Bergin, Edwin A. and Salyk, Colette and Pontoppidan, Klaus M. and Arulanantham, Nicole and Calahan, Jenny and Banzatti, Andrea and Andrews, Sean and Blake, Geoffrey A. and Ciesla, Fred and Green, Joel and Long, Feng and Lambrechts, Michiel and Najita, Joan and Pascucci, Ilaria and Pinilla, Paola and Krijt, Sebastiaan and Trapman, Leon and Collaboration, the JDISCS},
	month = dec,
	year = {2024},
	note = {Publisher: The American Astronomical Society},
	pages = {173},
}

@article{pontoppidan_high-contrast_2024,
	title = {High-contrast {JWST}-{MIRI} {Spectroscopy} of {Planet}-forming {Disks} for the {JDISC} {Survey}},
	volume = {963},
	issn = {0004-637X},
	url = {https://dx.doi.org/10.3847/1538-4357/ad20f0},
	doi = {10.3847/1538-4357/ad20f0},
	language = {en},
	number = {2},
	urldate = {2025-07-30},
	journal = {\apj},
	author = {Pontoppidan, Klaus M. and Salyk, Colette and Banzatti, Andrea and Zhang, Ke and Pascucci, Ilaria and Öberg, Karin I. and Long, Feng and Romero-Mirza, Carlos E. and Carr, John and Najita, Joan and Blake, Geoffrey A. and Arulanantham, Nicole and Andrews, Sean and Ballering, Nicholas P. and Bergin, Edwin and Calahan, Jenny and Cobb, Douglas and Colmenares, Maria Jose and Dickson-Vandervelde, Annie and Dignan, Anna and Green, Joel and Heretz, Phoebe and Herczeg, Gregory and Kalyaan, Anusha and Krijt, Sebastiaan and Pauly, Tyler and Pinilla, Paola and Trapman, Leon and Xie, Chengyan},
	month = mar,
	year = {2024},
	note = {Publisher: The American Astronomical Society},
	pages = {158},
}

@article{tabone_rich_2023,
	title = {A rich hydrocarbon chemistry and high {C} to {O} ratio in the inner disk around a very low-mass star},
	volume = {7},
	copyright = {2023 The Author(s), under exclusive licence to Springer Nature Limited},
	issn = {2397-3366},
	url = {https://www.nature.com/articles/s41550-023-01965-3},
	doi = {10.1038/s41550-023-01965-3},
	language = {en},
	number = {7},
	urldate = {2025-07-30},
	journal = {Nature Astronomy},
	author = {Tabone, B. and Bettoni, G. and van Dishoeck, E. F. and Arabhavi, A. M. and Grant, S. and Gasman, D. and Henning, Th and Kamp, I. and Güdel, M. and Lagage, P. O. and Ray, T. and Vandenbussche, B. and Abergel, A. and Absil, O. and Argyriou, I. and Barrado, D. and Boccaletti, A. and Bouwman, J. and Caratti o Garatti, A. and Geers, V. and Glauser, A. M. and Justannont, K. and Lahuis, F. and Mueller, M. and Nehmé, C. and Olofsson, G. and Pantin, E. and Scheithauer, S. and Waelkens, C. and Waters, L. B. F. M. and Black, J. H. and Christiaens, V. and Guadarrama, R. and Morales-Calderón, M. and Jang, H. and Kanwar, J. and Pawellek, N. and Perotti, G. and Perrin, A. and Rodgers-Lee, D. and Samland, M. and Schreiber, J. and Schwarz, K. and Colina, L. and Östlin, G. and Wright, G.},
	month = jul,
	year = {2023},
	note = {Publisher: Nature Publishing Group},
	pages = {805--814},
}

@article{grant_minds_2023,
	title = {{MINDS}. {The} {Detection} of {13CO2} with {JWST}-{MIRI} {Indicates} {Abundant} {CO2} in a {Protoplanetary} {Disk}},
	volume = {947},
	issn = {2041-8205},
	url = {https://dx.doi.org/10.3847/2041-8213/acc44b},
	doi = {10.3847/2041-8213/acc44b},
	language = {en},
	number = {1},
	urldate = {2025-07-30},
	journal = {\apjl},
	author = {Grant, Sierra L. and van Dishoeck, Ewine F. and Tabone, Benoît and Gasman, Danny and Henning, Thomas and Kamp, Inga and Güdel, Manuel and Lagage, Pierre-Olivier and Bettoni, Giulio and Perotti, Giulia and Christiaens, Valentin and Samland, Matthias and Arabhavi, Aditya M. and Argyriou, Ioannis and Abergel, Alain and Absil, Olivier and Barrado, David and Boccaletti, Anthony and Bouwman, Jeroen and o Garatti, Alessio Caratti and Geers, Vincent and Glauser, Adrian M. and Guadarrama, Rodrigo and Jang, Hyerin and Kanwar, Jayatee and Lahuis, Fred and Morales-Calderón, Maria and Mueller, Michael and Nehmé, Cyrine and Olofsson, Göran and Pantin, Eric and Pawellek, Nicole and Ray, Tom P. and Rodgers-Lee, Donna and Scheithauer, Silvia and Schreiber, Jürgen and Schwarz, Kamber and Temmink, Milou and Vandenbussche, Bart and Vlasblom, Marissa and Waters, L. B. F. M. and Wright, Gillian and Colina, Luis and Greve, Thomas R. and Justannont, Kay and Östlin, Göran},
	month = apr,
	year = {2023},
	note = {Publisher: The American Astronomical Society},
	pages = {L6},
}

@article{arabhavi_abundant_2024,
	title = {Abundant hydrocarbons in the disk around a very-low-mass star},
	volume = {384},
	url = {https://www.science.org/doi/10.1126/science.adi8147},
	doi = {10.1126/science.adi8147},
	number = {6700},
	urldate = {2025-07-30},
	journal = {Science},
	author = {Arabhavi, A. M. and Kamp, I. and Henning, Th. and van Dishoeck, E. F. and Christiaens, V. and Gasman, D. and Perrin, A. and Güdel, M. and Tabone, B. and Kanwar, J. and Waters, L. B. F. M. and Pascucci, I. and Samland, M. and Perotti, G. and Bettoni, G. and Grant, S. L. and Lagage, P. O. and Ray, T. P. and Vandenbussche, B. and Absil, O. and Argyriou, I. and Barrado, D. and Boccaletti, A. and Bouwman, J. and Caratti o Garatti, A. and Glauser, A. M. and Lahuis, F. and Mueller, M. and Olofsson, G. and Pantin, E. and Scheithauer, S. and Morales-Calderón, M. and Franceschi, R. and Jang, H. and Pawellek, N. and Rodgers-Lee, D. and Schreiber, J. and Schwarz, K. and Temmink, M. and Vlasblom, M. and Wright, G. and Colina, L. and Östlin, G.},
	month = jun,
	year = {2024},
	note = {Publisher: American Association for the Advancement of Science},
	pages = {1086--1090},
}

@article{temmink_minds_2024,
	title = {{MINDS}: {The} {DR} {Tau} disk - {II}. {Probing} the hot and cold {H2O} reservoirs in the {JWST}-{MIRI} spectrum},
	volume = {689},
	copyright = {© The Authors 2024},
	issn = {0004-6361, 1432-0746},
	shorttitle = {{MINDS}},
	url = {https://www.aanda.org/articles/aa/abs/2024/09/aa50355-24/aa50355-24.html},
	doi = {10.1051/0004-6361/202450355},
	language = {en},
	urldate = {2025-07-30},
	journal = {\aap},
	author = {Temmink, Milou and Dishoeck, Ewine F. van and Gasman, Danny and Grant, Sierra L. and Tabone, Benoît and Güdel, Manuel and Henning, Thomas and Barrado, David and Garatti, Alessio Caratti o and Glauser, Adrian M. and Kamp, Inga and Arabhavi, Aditya M. and Jang, Hyerin and Kurtovic, Nicolas and Perotti, Giulia and Schwarz, Kamber and Vlasblom, Marissa},
	month = sep,
	year = {2024},
	note = {Publisher: EDP Sciences},
	pages = {A330},
}

@article{arabhavi_minds_2025,
	title = {{MINDS}: {The} very low-mass star and brown dwarf sample - {Detections} and trends in the inner disk gas},
	volume = {699},
	copyright = {© The Authors 2025},
	issn = {0004-6361, 1432-0746},
	shorttitle = {{MINDS}},
	url = {https://www.aanda.org/articles/aa/abs/2025/07/aa54109-25/aa54109-25.html},
	doi = {10.1051/0004-6361/202554109},
	language = {en},
	urldate = {2025-07-30},
	journal = {\aap},
	author = {Arabhavi, A. M. and Kamp, I. and Henning, Th and Dishoeck, E. F. van and Jang, H. and Waters, L. B. F. M. and Christiaens, V. and Gasman, D. and Pascucci, I. and Perotti, G. and Grant, S. L. and Güdel, M. and Lagage, P.-O. and Barrado, D. and Garatti, A. Caratti o and Lahuis, F. and Kaeufer, T. and Kanwar, J. and Morales-Calderón, M. and Schwarz, K. and Sellek, A. D. and Tabone, B. and Temmink, M. and Vlasblom, M. and Patapis, P.},
	month = jul,
	year = {2025},
	note = {Publisher: EDP Sciences},
	pages = {A194},
}

@article{blake_high-resolution_2004,
	title = {High-{Resolution} 4.7 {Micron} {Keck}/{NIRSPEC} {Spectroscopy} of the {CO} {Emission} from the {Disks} {Surrounding} {Herbig} {Ae} {Stars}},
	volume = {606},
	issn = {0004-637X},
	url = {https://iopscience.iop.org/article/10.1086/421082/meta},
	doi = {10.1086/421082},
	language = {en},
	number = {1},
	urldate = {2025-07-30},
	journal = {\apj},
	author = {Blake, Geoffrey A. and Boogert, A. C. A.},
	month = apr,
	year = {2004},
	note = {Publisher: IOP Publishing},
	pages = {L73},
}

@article{najita_high-resolution_2009,
	title = {{HIGH}-{RESOLUTION} {K}-{BAND} {SPECTROSCOPY} {OF} {MWC} 480 {AND} {V1331} {Cyg}*},
	volume = {691},
	issn = {0004-637X},
	url = {https://dx.doi.org/10.1088/0004-637X/691/1/738},
	doi = {10.1088/0004-637X/691/1/738},
	language = {en},
	number = {1},
	urldate = {2025-07-15},
	journal = {\apj},
	author = {Najita, Joan R. and Doppmann, Greg W. and Carr, John S. and Graham, James R. and Eisner, J. A.},
	month = jan,
	year = {2009},
	note = {Publisher: The American Astronomical Society},
	pages = {738},
}

@article{eisner_water_2007,
	title = {Water vapour and hydrogen in the terrestrial-planet-forming region of a protoplanetary disk},
	volume = {447},
	copyright = {http://www.springer.com/tdm},
	issn = {0028-0836, 1476-4687},
	url = {https://www.nature.com/articles/nature05867},
	doi = {10.1038/nature05867},
	language = {en},
	number = {7144},
	urldate = {2025-07-15},
	journal = {Nature},
	author = {Eisner, J. A.},
	month = may,
	year = {2007},
	note = {Publisher: Springer Science and Business Media LLC},
	pages = {562--564},
}

@article{benisty_strong_2010,
	title = {Strong near-infrared emission in the sub-{AU} disk of the {Herbig} {Ae} star {HD} 163296: evidence of refractory dust?},
	volume = {511},
	copyright = {© ESO, 2010},
	issn = {0004-6361, 1432-0746},
	shorttitle = {Strong near-infrared emission in the sub-{AU} disk of the {Herbig} {Ae} star {HD} 163296},
	url = {https://www.aanda.org/articles/aa/abs/2010/03/aa12898-09/aa12898-09.html},
	doi = {10.1051/0004-6361/200912898},
	language = {en},
	urldate = {2025-07-15},
	journal = {\aap},
	author = {Benisty, M. and Natta, A. and Isella, A. and Berger, J.-P. and Massi, F. and Bouquin, J.-B. Le and Mérand, A. and Duvert, G. and Kraus, S. and Malbet, F. and Olofsson, J. and Robbe-Dubois, S. and Testi, L. and Vannier, M. and Weigelt, G.},
	month = feb,
	year = {2010},
	note = {Publisher: EDP Sciences},
	pages = {A74},
}

@article{grady_ensuremathbeta_1996,
       author = {{Grady}, C.~A. and {Perez}, M.~R. and {Talavera}, A. and {Bjorkman}, K.~S. and {de Winter}, D. and {The}, P. -S. and {Molster}, F.~J. and {van den Ancker}, M.~E. and {Sitko}, M.~L. and {Morrison}, N.~D. and {Beaver}, M.~L. and {McCollum}, B. and {Castelaz}, M.~W.},
        title = "{The {\ensuremath{\beta}} Pictoris phenomenon among Herbig Ae/Be stars. UV and optical high dispersion spectra.}",
      journal = {\aaps},
         year = 1996,
        month = nov,
       volume = {120},
        pages = {157-177},
       adsurl = {https://ui.adsabs.harvard.edu/abs/1996A&AS..120..157G}
}

@article{fedele_structure_2008,
	title = {The structure of the protoplanetary disk surrounding three young intermediate mass stars - {II}. {Spatially} resolved dust and gas distribution},
	volume = {491},
	copyright = {© ESO, 2008},
	issn = {0004-6361, 1432-0746},
	url = {https://www.aanda.org/articles/aa/abs/2008/45/aa10126-08/aa10126-08.html},
	doi = {10.1051/0004-6361:200810126},
	language = {en},
	number = {3},
	urldate = {2025-06-23},
	journal = {\aap},
	author = {Fedele, D. and Ancker, M. E. van den and Acke, B. and Plas, G. van der and Boekel, R. van and Wittkowski, M. and Henning, Th and Bouwman, J. and Meeus, G. and Rafanelli, P.},
	month = dec,
	year = {2008},
	note = {Number: 3
Publisher: EDP Sciences},
	pages = {809--820},
}

@INPROCEEDINGS{rivinius_classical_2024,
       author = {{Rivinius}, Thomas and {Klement}, Robert},
        title = "{Classical Be stars}",
    booktitle = {Encyclopedia of Astrophysics},
         year = 2026,
       volume = {2},
        month = jan,
        pages = {430-448},
          doi = {10.1016/B978-0-443-21439-4.00042-0},
archivePrefix = {arXiv},
       eprint = {2411.06882},
 primaryClass = {astro-ph.SR},
       adsurl = {https://ui.adsabs.harvard.edu/abs/2026enap....2..430R}
}

@article{ancker_composition_2001,
	title = {The composition of circumstellar gas and dust in 51 {Oph}},
	volume = {369},
	copyright = {© ESO, 2001},
	issn = {0004-6361, 1432-0746},
	url = {https://www.aanda.org/articles/aa/abs/2001/14/aada022/aada022.html},
	doi = {10.1051/0004-6361:20010245},
	language = {en},
	number = {2},
	urldate = {2025-06-13},
	journal = {\aap},
	author = {Ancker, M. E. van den and Meeus, G. and Cami, J. and Waters, L. B. F. M. and Waelkens, C.},
	month = apr,
	year = {2001},
	note = {Number: 2
Publisher: EDP Sciences},
	pages = {L17--L21},
}

@article{grant_dotmmdisk_2023,
	title = {The {\textbackslash}{dotM}–{Mdisk} {Relationship} for {Herbig} {Ae}/{Be} {Stars}: {A} {Lifetime} {Problem} for {Disks} with {Low} {Masses}?},
	volume = {166},
	issn = {1538-3881},
	shorttitle = {The {\textbackslash}{dotM}–{Mdisk} {Relationship} for {Herbig} {Ae}/{Be} {Stars}},
	url = {https://dx.doi.org/10.3847/1538-3881/acf128},
	doi = {10.3847/1538-3881/acf128},
	language = {en},
	number = {4},
	urldate = {2025-06-13},
	journal = {\aj},
	author = {Grant, Sierra L. and Stapper, Lucas M. and Hogerheijde, Michiel R. and van Dishoeck, Ewine F. and Brittain, Sean and Vioque, Miguel},
	month = sep,
	year = {2023},
	note = {Publisher: The American Astronomical Society},
	pages = {147},
}

@article{lisse_abundant_2009,
	title = {{ABUNDANT} {CIRCUMS}℡{LAR} {SILICA} {DUST} {AND} {SiO} {GAS} {CREATED} {BY} {A} {GIANT} {HYPERVELOCITY} {COLLISION} {IN} {THE} ∼12 {MYR} {HD172555} {SYSTEM}},
	volume = {701},
	issn = {0004-637X},
	url = {https://dx.doi.org/10.1088/0004-637X/701/2/2019},
	doi = {10.1088/0004-637X/701/2/2019},
	language = {en},
	number = {2},
	urldate = {2025-06-06},
	journal = {\apj},
	author = {Lisse, C. M. and Chen, C. H. and Wyatt, M. C. and Morlok, A. and Song, I. and Bryden, G. and Sheehan, P.},
	month = aug,
	year = {2009},
	note = {Publisher: The American Astronomical Society},
	pages = {2019},
}

@article{woitke_equilibrium_2018,
	title = {Equilibrium chemistry down to 100 {K}: {Impact} of silicates and phyllosilicates on the carbon to oxygen ratio},
	volume = {614},
	copyright = {https://www.edpsciences.org/en/authors/copyright-and-licensing},
	issn = {0004-6361, 1432-0746},
	shorttitle = {Equilibrium chemistry down to 100 {K}},
	url = {https://www.aanda.org/10.1051/0004-6361/201732193},
	doi = {10.1051/0004-6361/201732193},
	language = {en},
	urldate = {2025-06-06},
	journal = {\aap},
	author = {Woitke, P. and Helling, Ch. and Hunter, G. H. and Millard, J. D. and Turner, G. E. and Worters, M. and Blecic, J. and Stock, J. W.},
	month = jun,
	year = {2018},
	pages = {A1},
}

@article{bally_protostellar_2016,
	title = {Protostellar {Outflows}},
	volume = {54},
	issn = {0066-4146, 1545-4282},
	url = {https://www.annualreviews.org/doi/10.1146/annurev-astro-081915-023341},
	doi = {10.1146/annurev-astro-081915-023341},
	language = {en},
	number = {1},
	urldate = {2025-06-06},
	journal = {\araa},
	author = {Bally, John},
	month = sep,
	year = {2016},
	pages = {491--528},
}

@article{kaeufer_disentangling_2024,
       author = {{Kaeufer}, T. and {Woitke}, P. and {Kamp}, I. and {Kanwar}, J. and {Min}, M.},
        title = "{Disentangling the dust and gas contributions of the JWST/MIRI spectrum of Sz 28}",
      journal = {\aap},
         year = 2024,
        month = oct,
       volume = {690},
          eid = {A100},
        pages = {A100},
          doi = {10.1051/0004-6361/202450891},
archivePrefix = {arXiv},
       eprint = {2408.06077},
 primaryClass = {astro-ph.EP},
       adsurl = {https://ui.adsabs.harvard.edu/abs/2024A&A...690A.100K}
}

@article{kaeufer_bayesian_2024,
       author = {{Kaeufer}, T. and {Min}, M. and {Woitke}, P. and {Kamp}, I. and {Arabhavi}, A.~M.},
        title = "{Bayesian analysis of the molecular emission and dust continuum of protoplanetary disks}",
      journal = {\aap},
         year = 2024,
        month = jul,
       volume = {687},
          eid = {A209},
        pages = {A209},
          doi = {10.1051/0004-6361/202449936},
archivePrefix = {arXiv},
       eprint = {2405.06486},
 primaryClass = {astro-ph.EP},
       adsurl = {https://ui.adsabs.harvard.edu/abs/2024A&A...687A.209K}
}

@article{krasnokutski_formation_2014,
	title = {{FORMATION} {OF} {SILICON} {OXIDE} {GRAINS} {AT} {LOW} {TEMPERATURE}},
	volume = {782},
	issn = {0004-637X},
	url = {https://dx.doi.org/10.1088/0004-637X/782/1/15},
	doi = {10.1088/0004-637X/782/1/15},
	language = {en},
	number = {1},
	urldate = {2025-05-22},
	journal = {\apj},
	author = {Krasnokutski, S. A. and Rouillé, G. and Jäger, C. and Huisken, F. and Zhukovska, S. and Henning, Th.},
	month = jan,
	year = {2014},
	note = {Publisher: The American Astronomical Society},
	pages = {15},
}

@article{reber_sio_2008,
	title = {From {SiO} {Molecules} to {Silicates} in {Circumstellar} {Space}: {Atomic} {Structures}, {Growth} {Patterns}, and {Optical} {Signatures} of {SinOm} {Clusters}},
	volume = {2},
	issn = {1936-0851},
	shorttitle = {From {SiO} {Molecules} to {Silicates} in {Circumstellar} {Space}},
	url = {https://doi.org/10.1021/nn7003958},
	doi = {10.1021/nn7003958},
	number = {8},
	urldate = {2025-05-22},
	journal = {ACS Nano},
	author = {Reber, Arthur C. and Paranthaman, Selvarengan and Clayborne, Andre Z. and Khanna, Shiv N. and Castleman, A. Welford Jr.},
	month = aug,
	year = {2008},
	note = {Publisher: American Chemical Society},
	pages = {1729--1737},
}

@article{yang_directed_2018,
	title = {Directed gas phase formation of silicon dioxide and implications for the formation of interstellar silicates},
	volume = {9},
	copyright = {2018 The Author(s)},
	issn = {2041-1723},
	url = {https://www.nature.com/articles/s41467-018-03172-5},
	doi = {10.1038/s41467-018-03172-5},
	language = {en},
	number = {1},
	urldate = {2025-05-22},
	journal = {Nature Communications},
	author = {Yang, Tao and Thomas, Aaron M. and Dangi, Beni B. and Kaiser, Ralf I. and Mebel, Alexander M. and Millar, Tom J.},
	month = feb,
	year = {2018},
	note = {Publisher: Nature Publishing Group},
	pages = {774},
}

@article{gravity_collaboration_gravity_2024,
	title = {The {GRAVITY} young stellar object survey: {XII}. {The} hot gas disk component in {Herbig} {Ae}/{Be} stars},
	volume = {684},
	copyright = {https://creativecommons.org/licenses/by/4.0},
	issn = {0004-6361, 1432-0746},
	shorttitle = {The {GRAVITY} young stellar object survey},
	url = {https://www.aanda.org/10.1051/0004-6361/202245804},
	doi = {10.1051/0004-6361/202245804},
	language = {en},
	urldate = {2025-03-28},
	journal = {\aap},
	author = {{GRAVITY Collaboration} and Garcia Lopez, R. and Natta, A. and Fedriani, R. and Caratti O Garatti, A. and Sanchez-Bermudez, J. and Perraut, K. and Dougados, C. and Bouarour, Y.-I. and Bouvier, J. and Brandner, W. and Garcia, P. and Koutoulaki, M. and Labadie, L. and Linz, H. and Alécian, E. and Benisty, M. and Berger, J.-P. and Bourdarot, G. and Caselli, P. and Clénet, Y. and De Zeeuw, P. T. and Davies, R. and Eckart, A. and Eisenhauer, F. and Förster-Schreiber, N. M. and Gendron, E. and Gillessen, S. and Grant, S. and Henning, Th. and Kervella, P. and Lacour, S. and Lapeyrère, V. and Le Bouquin, J.-B. and Lutz, D. and Mang, F. and Nowacki, H. and Ott, T. and Paumard, T. and Perrin, G. and Shangguan, J. and Shimizu, T. and Soulain, A. and Straubmeier, C. and Sturm, E. and Tacconi, L. and Van Dishoeck, E. F. and Vincent, F. and Widmann, F.},
	month = apr,
	year = {2024},
	pages = {A43},
}

@article{vines_span_2022,
	title = {{\textless}span style="font-variant:small-caps;"{\textgreater}ariadne{\textless}/span{\textgreater} : measuring accurate and precise stellar parameters through {SED} fitting},
	volume = {513},
	copyright = {https://academic.oup.com/journals/pages/open\_access/funder\_policies/chorus/standard\_publication\_model},
	issn = {0035-8711, 1365-2966},
	shorttitle = {{\textless}span style="font-variant},
	url = {https://academic.oup.com/mnras/article/513/2/2719/6564732},
	doi = {10.1093/mnras/stac956},
	language = {en},
	number = {2},
	urldate = {2025-03-19},
	journal = {\mnras},
	author = {Vines, Jose I and Jenkins, James S},
	month = may,
	year = {2022},
	pages = {2719--2731},
}

@article{kamp_chemical_2023,
	title = {The chemical inventory of the inner regions of planet-forming disks – the {JWST}/{MINDS} program},
	volume = {245},
	url = {https://ui.adsabs.harvard.edu/abs/2023FaDi..245..112K},
	doi = {10.1039/D3FD00013C},
	urldate = {2025-03-17},
	journal = {Faraday Discussions},
	author = {Kamp, Inga and Henning, Thomas and Arabhavi, Aditya M. and Bettoni, Giulio and Christiaens, Valentin and Gasman, Danny and Grant, Sierra L. and Morales-Calderón, Maria and Tabone, Benoît and Abergel, Alain and Absil, Olivier and Argyriou, Ioannis and Barrado, David and Boccaletti, Anthony and Bouwman, Jeroen and Caratti o Garatti, Alessio and van Dishoeck, Ewine F. and Geers, Vincent and Glauser, Adrian M. and Güdel, Manuel and Guadarrama, Rodrigo and Jang, Hyerin and Kanwar, Jayatee and Lagage, Pierre-Olivier and Lahuis, Fred and Mueller, Michael and Nehmé, Cyrine and Olofsson, Göran and Pantin, Eric and Pawellek, Nicole and Perotti, Giulia and Ray, Tom P. and Rodgers-Lee, Donna and Samland, Matthias and Scheithauer, Silvia and Schreiber, Jürgen and Schwarz, Kamber and Temmink, Milou and Vandenbussche, Bart and Vlasblom, Marissa and Waelkens, Christoffel and Waters, L. B. F. M. and Wright, Gillian},
	month = sep,
	year = {2023},
	note = {ADS Bibcode: 2023FaDi..245..112K},
	pages = {112--137},
}

@article{henning_minds_2024,
	title = {{MINDS}: {The} {JWST} {MIRI} {Mid}-{INfrared} {Disk} {Survey}},
	volume = {136},
	doi = {10.1088/1538-3873/ad3455},
	number = {5},
	journal = {\pasp},
	author = {Henning, Thomas and Kamp, Inga and Samland, Matthias and Arabhavi, Aditya M. and Kanwar, Jayatee and van Dishoeck, Ewine F. and Güdel, Manuel and Lagage, Pierre-Olivier and Waelkens, Christoffel and Abergel, Alain and Absil, Olivier and Barrado, David and Boccaletti, Anthony and Bouwman, Jeroen and Caratti o Garatti, Alessio and Geers, Vincent and Glauser, Adrian M. and Lahuis, Fred and Mueller, Michael and Nehmé, Cyrine and Olofsson, Göran and Pantin, Eric and Ray, Tom P. and Scheithauer, Silvia and Vandenbussche, Bart and Waters, L. B. F. M. and Wright, Gillian and Argyriou, Ioannis and Christiaens, Valentin and Franceschi, Riccardo and Gasman, Danny and Grant, Sierra L. and Guadarrama, Rodrigo and Jang, Hyerin and Morales-Calderón, Maria and Pawellek, Nicole and Perotti, Giulia and Rodgers-Lee, Donna and Schreiber, Jürgen and Schwarz, Kamber and Tabone, Benoît and Temmink, Milou and Vlasblom, Marissa and Colina, Luis and Greve, Thomas R. and Östlin, Göran},
	month = may,
	year = {2024},
	note = {\_eprint: 2403.09210},
	pages = {054302},
}

@article{brittain_herbig_2023,
	title = {Herbig {Stars}},
	volume = {219},
	issn = {0038-6308},
	url = {https://ui.adsabs.harvard.edu/abs/2023SSRv..219....7B},
	doi = {10.1007/s11214-023-00949-z},
	urldate = {2025-03-17},
	journal = {\ssr},
	author = {Brittain, Sean D. and Kamp, Inga and Meeus, Gwendolyn and Oudmaijer, René D. and Waters, L. B. F. M.},
	month = feb,
	year = {2023},
	note = {ADS Bibcode: 2023SSRv..219....7B},
	pages = {7},
}

@article{van_der_plas_structure_2008,
	title = {The structure of protoplanetary disks surrounding three young intermediate mass stars. {I}. {Resolving} the disk rotation in the [{OI}] 6300 Å line},
	volume = {485},
	issn = {0004-6361},
	url = {https://ui.adsabs.harvard.edu/abs/2008A&A...485..487V},
	doi = {10.1051/0004-6361:20078867},
	urldate = {2025-03-17},
	journal = {\aap},
	author = {van der Plas, G. and van den Ancker, M. E. and Fedele, D. and Acke, B. and Dominik, C. and Waters, L. B. F. M. and Bouwman, J.},
	month = jul,
	year = {2008},
	note = {ADS Bibcode: 2008A\&A...485..487V},
	pages = {487--495},
}

@article{miroshnichenko_fundamental_2004,
	title = {Fundamental parameters and evolutionary state of the {Herbig} {Ae} starcandidate {HD} 35929},
	volume = {427},
	issn = {0004-6361, 1432-0746},
	url = {http://www.aanda.org/10.1051/0004-6361:20041618},
	doi = {10.1051/0004-6361:20041618},
	language = {en},
	number = {3},
	urldate = {2025-03-14},
	journal = {\aap},
	author = {Miroshnichenko, A. S. and Gray, R. O. and Klochkova, V. G. and Bjorkman, K. S. and Kuratov, K. S.},
	month = dec,
	year = {2004},
	pages = {937--944},
}

@article{heays_photodissociation_2017,
	title = {Photodissociation and photoionisation of atoms and molecules of astrophysical interest},
	volume = {602},
	copyright = {© ESO, 2017},
	issn = {0004-6361, 1432-0746},
	url = {https://www.aanda.org/articles/aa/abs/2017/06/aa28742-16/aa28742-16.html},
	doi = {10.1051/0004-6361/201628742},
	language = {en},
	urldate = {2025-03-13},
	journal = {\aap},
	author = {Heays, A. N. and Bosman, A. D. and Dishoeck, E. F. van},
	month = jun,
	year = {2017},
	note = {Publisher: EDP Sciences},
	pages = {A105},
}

@article{gelder_jwst_2024,
	title = {{JWST} {Observations} of {Young} {protoStars} ({JOYS}) - {Overview} of gaseous molecular emission and absorption in low-mass protostars},
	volume = {692},
	copyright = {© The Authors 2024},
	issn = {0004-6361, 1432-0746},
	url = {https://www.aanda.org/articles/aa/abs/2024/12/aa51967-24/aa51967-24.html},
	doi = {10.1051/0004-6361/202451967},
	language = {en},
	urldate = {2025-03-13},
	journal = {\aap},
	author = {Gelder, M. L. van and Francis, L. and Dishoeck, E. F. van and Tychoniec, Ł and Ray, T. P. and Beuther, H. and Garatti, A. Caratti o and Chen, Y. and Devaraj, R. and Gieser, C. and Justtanont, K. and Kavanagh, P. J. and Nazari, P. and Reyes, S. and Rocha, W. R. M. and Slavicinska, K. and Güdel, M. and Henning, Th and Lagage, P.-O. and Wright, G.},
	month = dec,
	year = {2024},
	note = {Publisher: EDP Sciences},
	pages = {A197},
}

@article{gusdorf_sio_2008,
	title = {{SiO} line emission from {C}-type shock waves: interstellar jets and outflows},
	volume = {482},
	issn = {0004-6361, 1432-0746},
	shorttitle = {{SiO} line emission from {C}-type shock waves},
	url = {http://www.aanda.org/10.1051/0004-6361:20078900},
	doi = {10.1051/0004-6361:20078900},
	language = {en},
	number = {3},
	urldate = {2025-03-13},
	journal = {\aap},
	author = {Gusdorf, A. and Cabrit, S. and Flower, D. R. and Pineau Des Forêts, G.},
	month = may,
	year = {2008},
	pages = {809--829},
}

@ARTICLE{marconi_pulsation_2000,
       author = {{Marconi}, M. and {Ripepi}, V. and {Alcal{\'a}}, J.~M. and {Covino}, E. and {Palla}, F. and {Terranegra}, L.},
        title = "{Pulsation in two Herbig Ae stars: HD 35929 and V351 Ori}",
      journal = {\aap},
         year = 2000,
        month = mar,
       volume = {355},
        pages = {L35-L38},
          doi = {10.48550/arXiv.astro-ph/0002466},
archivePrefix = {arXiv},
       eprint = {astro-ph/0002466},
 primaryClass = {astro-ph},
       adsurl = {https://ui.adsabs.harvard.edu/abs/2000A&A...355L..35M}
}

@article{johnson_self-consistent_2012,
	title = {A {SELF}-{CONSISTENT} {MODEL} {OF} {THE} {CIRCUMS}℡{LAR} {DEBRIS} {CREATED} {BY} {A} {GIANT} {HYPERVELOCITY} {IMPACT} {IN} {THE} {HD} 172555 {SYSTEM}},
	volume = {761},
	issn = {0004-637X},
	url = {https://dx.doi.org/10.1088/0004-637X/761/1/45},
	doi = {10.1088/0004-637X/761/1/45},
	language = {en},
	number = {1},
	urldate = {2025-01-13},
	journal = {\apj},
	author = {Johnson, B. C. and Lisse, C. M. and Chen, C. H. and Melosh, H. J. and Wyatt, M. C. and Thebault, P. and Henning, W. G. and Gaidos, E. and Elkins-Tanton, L. T. and Bridges, J. C. and Morlok, A.},
	month = nov,
	year = {2012},
	note = {Publisher: The American Astronomical Society},
	pages = {45},
}

@article{lazareff_structure_2017,
	title = {Structure of {Herbig} {AeBe} disks at the milliarcsecond scale: {A} statistical survey in the \textit{{H}} band using {PIONIER}-{VLTI}},
	volume = {599},
	copyright = {https://www.edpsciences.org/en/authors/copyright-and-licensing},
	issn = {0004-6361, 1432-0746},
	shorttitle = {Structure of {Herbig} {AeBe} disks at the milliarcsecond scale},
	url = {http://www.aanda.org/10.1051/0004-6361/201629305},
	doi = {10.1051/0004-6361/201629305},
	language = {en},
	urldate = {2024-12-19},
	journal = {\aap},
	author = {Lazareff, B. and Berger, J.-P. and Kluska, J. and Le Bouquin, J.-B. and Benisty, M. and Malbet, F. and Koen, C. and Pinte, C. and Thi, W.-F. and Absil, O. and Baron, F. and Delboulbé, A. and Duvert, G. and Isella, A. and Jocou, L. and Juhasz, A. and Kraus, S. and Lachaume, R. and Ménard, F. and Millan-Gabet, R. and Monnier, J. D. and Moulin, T. and Perraut, K. and Rochat, S. and Soulez, F. and Tallon, M. and Thiébaut, E. and Traub, W. and Zins, G.},
	month = mar,
	year = {2017},
	pages = {A85},
}

@article{adams_water_2019,
	title = {Water and {OH} {Emission} from the {Inner} {Disk} of a {Herbig} {Ae}/{Be} {Star}},
	volume = {871},
	issn = {0004-637X},
	url = {https://dx.doi.org/10.3847/1538-4357/aaf9a4},
	doi = {10.3847/1538-4357/aaf9a4},
	language = {en},
	number = {2},
	urldate = {2024-10-31},
	journal = {\apj},
	author = {Adams, Steven C. and Ádámkovics, Máté and Carr, John S. and Najita, Joan R. and Brittain, Sean D.},
	month = jan,
	year = {2019},
	note = {Publisher: The American Astronomical Society},
	pages = {173},
}

@article{banzatti_scanning_2022,
	title = {Scanning {Disk} {Rings} and {Winds} in {CO} at 0.01-10 au: {A} {High}-resolution {M}-band {Spectroscopy} {Survey} with {IRTF}-{iSHELL}},
	volume = {163},
	doi = {10.3847/1538-3881/ac52f0},
	number = {4},
	journal = {\aj},
	author = {{Banzatti}, A. and Abernathy, Kirsten M. and Brittain, Sean and Bosman, Arthur D. and Pontoppidan, Klaus M. and Boogert, Adwin and Jensen, Stanley and Carr, John and Najita, Joan and Grant, Sierra and Sigler, Rocio M. and Sanchez, Michael A. and Kern, Joshua and Rayner, John T.},
	month = apr,
	year = {2022},
	note = {\_eprint: 2202.03438},
	pages = {174},
}

@article{valegard_what_2021,
       author = {{Valeg{\r{a}}rd}, P. -G. and {Waters}, L.~B.~F.~M. and {Dominik}, C.},
        title = "{What happened before?. Disks around the precursors of young Herbig Ae/Be stars}",
      journal = {\aap},
         year = 2021,
        month = aug,
       volume = {652},
          eid = {A133},
        pages = {A133},
          doi = {10.1051/0004-6361/202039802},
archivePrefix = {arXiv},
       eprint = {2104.14212},
 primaryClass = {astro-ph.SR},
       adsurl = {https://ui.adsabs.harvard.edu/abs/2021A&A...652A.133V}
}

@article{sargent_silica_2008,
	title = {{SILICA} {IN} {PROTOPLANETARY} {DISKS}},
	volume = {690},
	issn = {0004-637X},
	url = {https://dx.doi.org/10.1088/0004-637X/690/2/1193},
	doi = {10.1088/0004-637X/690/2/1193},
	language = {en},
	number = {2},
	urldate = {2024-08-26},
	journal = {\apj},
	author = {Sargent, B. A. and Forrest, W. J. and Tayrien, C. and McClure, M. K. and Li, A. and Basu, A. R. and Manoj, P. and Watson, D. M. and Bohac, C. J. and Furlan, E. and Kim, K. H. and Green, J. D. and Sloan, G. C.},
	month = dec,
	year = {2008},
	note = {Publisher: The American Astronomical Society},
	pages = {1193},
}

@article{gravity_collaboration_gravity_2021,
       author = {{GRAVITY Collaboration} and {Koutoulaki}, M. and {Garcia Lopez}, R. and {Natta}, A. and {Fedriani}, R. and {Caratti O Garatti}, A. and {Ray}, T.~P. and {Coffey}, D. and {Brandner}, W. and {Dougados}, C. and {Garcia}, P.~J.~V. and {Klarmann}, L. and {Labadie}, L. and {Perraut}, K. and {Sanchez-Bermudez}, J. and {Lin}, C. -C. and {Amorim}, A. and {Baub{\"o}ck}, M. and {Benisty}, M. and {Berger}, J.~P. and {Buron}, A. and {Caselli}, P. and {Cl{\'e}net}, Y. and {Coud{\'e} Du Foresto}, V. and {de Zeeuw}, P.~T. and {Duvert}, G. and {de Wit}, W. and {Eckart}, A. and {Eisenhauer}, F. and {Filho}, M. and {Gao}, F. and {Gendron}, E. and {Genzel}, R. and {Gillessen}, S. and {Grellmann}, R. and {Habibi}, M. and {Haubois}, X. and {Haussmann}, F. and {Henning}, T. and {Hippler}, S. and {Hubert}, Z. and {Horrobin}, M. and {Jimenez Rosales}, A. and {Jocou}, L. and {Kervella}, P. and {Kolb}, J. and {Lacour}, S. and {Le Bouquin}, J. -B. and {L{\'e}na}, P. and {Linz}, H. and {Ott}, T. and {Paumard}, T. and {Perrin}, G. and {Pfuhl}, O. and {Ram{\'\i}rez-Tannus}, M.~C. and {Rau}, C. and {Rousset}, G. and {Scheithauer}, S. and {Shangguan}, J. and {Stadler}, J. and {Straub}, O. and {Straubmeier}, C. and {Sturm}, E. and {van Dishoeck}, E. and {Vincent}, F. and {von Fellenberg}, S. and {Widmann}, F. and {Wieprecht}, E. and {Wiest}, M. and {Wiezorrek}, E. and {Yazici}, S. and {Zins}, G.},
        title = "{The GRAVITY young stellar object survey. IV. The CO overtone emission in 51 Oph at sub-au scales}",
      journal = {\aap},
         year = 2021,
        month = jan,
       volume = {645},
          eid = {A50},
        pages = {A50},
          doi = {10.1051/0004-6361/202038000},
archivePrefix = {arXiv},
       eprint = {2011.05955},
 primaryClass = {astro-ph.SR},
       adsurl = {https://ui.adsabs.harvard.edu/abs/2021A&A...645A..50G}
}

@article{valenti_iue_2000,
	title = {An {IUE} {Atlas} of {Pre}-{Main}-{Sequence} {Stars}. {I}. {Co}-added {Final} {Archive} {Spectra} from the {SWP} {Camera}},
	volume = {129},
	doi = {10.1086/313408},
	number = {1},
	journal = {\apjs},
	author = {Valenti, Jeff A. and Johns-Krull, Christopher M. and Linsky, Jeffrey L.},
	month = jul,
	year = {2000},
	pages = {399--420},
}

@ARTICLE{fairlamb_spectroscopic_2015,
       author = {{Fairlamb}, J.~R. and {Oudmaijer}, R.~D. and {Mendigut{\'\i}a}, I. and {Ilee}, J.~D. and {van den Ancker}, M.~E.},
        title = "{A spectroscopic survey of Herbig Ae/Be stars with X-shooter - I. Stellar parameters and accretion rates}",
      journal = {\mnras},
         year = 2015,
        month = oct,
       volume = {453},
       number = {1},
        pages = {976-1001},
          doi = {10.1093/mnras/stv1576},
archivePrefix = {arXiv},
       eprint = {1507.05967},
 primaryClass = {astro-ph.SR},
       adsurl = {https://ui.adsabs.harvard.edu/abs/2015MNRAS.453..976F}
}

@article{acke_spitzers_2010,
       author = {{Acke}, B. and {Bouwman}, J. and {Juh{\'a}sz}, A. and {Henning}, Th. and {van den Ancker}, M.~E. and {Meeus}, G. and {Tielens}, A.~G.~G.~M. and {Waters}, L.~B.~F.~M.},
        title = "{Spitzer's View on Aromatic and Aliphatic Hydrocarbon Emission in Herbig Ae Stars}",
      journal = {\apj},
         year = 2010,
        month = jul,
       volume = {718},
       number = {1},
        pages = {558-574},
          doi = {10.1088/0004-637X/718/1/558},
archivePrefix = {arXiv},
       eprint = {1006.1130},
 primaryClass = {astro-ph.SR},
       adsurl = {https://ui.adsabs.harvard.edu/abs/2010ApJ...718..558A}
}

@article{juhasz_dust_2010,
       author = {{Juh{\'a}sz}, A. and {Bouwman}, J. and {Henning}, Th. and {Acke}, B. and {van den Ancker}, M.~E. and {Meeus}, G. and {Dominik}, C. and {Min}, M. and {Tielens}, A.~G.~G.~M. and {Waters}, L.~B.~F.~M.},
        title = "{Dust Evolution in Protoplanetary Disks Around Herbig Ae/Be Stars{\textemdash}the Spitzer View}",
      journal = {\apj},
         year = 2010,
        month = sep,
       volume = {721},
       number = {1},
        pages = {431-455},
          doi = {10.1088/0004-637X/721/1/431},
archivePrefix = {arXiv},
       eprint = {1008.0083},
 primaryClass = {astro-ph.SR},
       adsurl = {https://ui.adsabs.harvard.edu/abs/2010ApJ...721..431J}
}

@article{ilee_investigating_2014,
	title = {Investigating the inner discs of {Herbig} {Ae}/{Be} stars with {CO} bandhead and {Br}{\textbackslash}ensuremath{\textbackslash}gamma emission},
	volume = {445},
	doi = {10.1093/mnras/stu1942},
	number = {4},
	journal = {\mnras},
	author = {Ilee, J. D. and Fairlamb, J. and Oudmaijer, R. D. and Mendigutía, I. and van den Ancker, M. E. and Kraus, S. and Wheelwright, H. E.},
	month = dec,
	year = {2014},
	note = {\_eprint: 1409.4897},
	pages = {3723--3736},
}

@article{vioque_gaia_2018,
       author = {{Vioque}, M. and {Oudmaijer}, R.~D. and {Baines}, D. and {Mendigut{\'\i}a}, I. and {P{\'e}rez-Mart{\'\i}nez}, R.},
        title = "{Gaia DR2 study of Herbig Ae/Be stars}",
      journal = {\aap},
         year = 2018,
        month = dec,
       volume = {620},
          eid = {A128},
        pages = {A128},
          doi = {10.1051/0004-6361/201832870},
archivePrefix = {arXiv},
       eprint = {1808.00476},
 primaryClass = {astro-ph.SR},
       adsurl = {https://ui.adsabs.harvard.edu/abs/2018A&A...620A.128V}
}

@article{wichittanakom_accretion_2020,
	title = {The accretion rates and mechanisms of {Herbig} {Ae}/{Be} stars},
	volume = {493},
	doi = {10.1093/mnras/staa169},
	number = {1},
	journal = {\mnras},
	author = {Wichittanakom, C. and Oudmaijer, R. D. and Fairlamb, J. R. and Mendigutía, I. and Vioque, M. and Ababakr, K. M.},
	month = mar,
	year = {2020},
	note = {\_eprint: 2001.05971},
	pages = {234--249},
}

@article{barcelo_forteza_unveiling_2020,
       author = {{Barcel{\'o} Forteza}, S. and {Moya}, A. and {Barrado}, D. and {Solano}, E. and {Mart{\'\i}n-Ruiz}, S. and {Su{\'a}rez}, J.~C. and {Garc{\'\i}a Hern{\'a}ndez}, A.},
        title = "{Unveiling the power spectra of {\ensuremath{\delta}} Scuti stars with TESS. The temperature, gravity, and frequency scaling relation}",
      journal = {\aap},
         year = 2020,
        month = jun,
       volume = {638},
          eid = {A59},
        pages = {A59},
          doi = {10.1051/0004-6361/201937262},
archivePrefix = {arXiv},
       eprint = {2004.07647},
 primaryClass = {astro-ph.SR},
       adsurl = {https://ui.adsabs.harvard.edu/abs/2020A&A...638A..59B}
}

@article{grant_tracing_2022,
       author = {{Grant}, Sierra L. and {Espaillat}, Catherine C. and {Brittain}, Sean and {Scott-Joseph}, Caleb and {Calvet}, Nuria},
        title = "{Tracing Accretion onto Herbig Ae/Be Stars Using the Br{\ensuremath{\gamma}} Line}",
      journal = {\apj},
         year = 2022,
        month = feb,
       volume = {926},
       number = {2},
          eid = {229},
        pages = {229},
          doi = {10.3847/1538-4357/ac450a},
archivePrefix = {arXiv},
       eprint = {2112.10428},
 primaryClass = {astro-ph.SR},
       adsurl = {https://ui.adsabs.harvard.edu/abs/2022ApJ...926..229G}
}

@article{stapper_complete_2025,
       author = {{Stapper}, L.~M. and {Hogerheijde}, M.~R. and {van Dishoeck}, E.~F. and {Booth}, A.~S. and {Grant}, S.~L. and {van Terwisga}, S.~E.},
        title = "{A complete Herbig disk mass survey in Orion}",
      journal = {\aap},
         year = 2025,
        month = jan,
       volume = {693},
          eid = {A49},
        pages = {A49},
          doi = {10.1051/0004-6361/202450678},
archivePrefix = {arXiv},
       eprint = {2411.08959},
 primaryClass = {astro-ph.EP},
       adsurl = {https://ui.adsabs.harvard.edu/abs/2025A&A...693A..49S}
}

@ARTICLE{woitke_2024,
       author = {{Woitke}, P. and {Thi}, W. -F. and {Arabhavi}, A.~M. and {Kamp}, I. and {K{\'o}sp{\'a}l}, {\'A}. and {{\'A}brah{\'a}m}, P.},
        title = "{2D disc modelling of the JWST line spectrum of EX Lupi}",
      journal = {\aap},
         year = 2024,
        month = mar,
       volume = {683},
          eid = {A219},
        pages = {A219},
          doi = {10.1051/0004-6361/202347730},
archivePrefix = {arXiv},
       eprint = {2311.18321},
 primaryClass = {astro-ph.EP},
       adsurl = {https://ui.adsabs.harvard.edu/abs/2024A&A...683A.219W}
}

@ARTICLE{Romero_2024,
       author = {{Romero-Mirza}, Carlos E. and {Banzatti}, Andrea and {{\"O}berg}, Karin I. and {Pontoppidan}, Klaus M. and {Salyk}, Colette and {Najita}, Joan and {Blake}, Geoffrey A. and {Krijt}, Sebastiaan and {Arulanantham}, Nicole and {Pinilla}, Paola and {Long}, Feng and {Rosotti}, Giovanni and {Andrews}, Sean M. and {Wilner}, David J. and {Calahan}, Jenny and {The Jdiscs Collaboration}},
        title = "{Retrieval of Thermally Resolved Water Vapor Distributions in Disks Observed with JWST-MIRI}",
      journal = {\apj},
         year = 2024,
        month = nov,
       volume = {975},
       number = {1},
          eid = {78},
        pages = {78},
          doi = {10.3847/1538-4357/ad769e},
archivePrefix = {arXiv},
       eprint = {2409.03831},
 primaryClass = {astro-ph.EP},
       adsurl = {https://ui.adsabs.harvard.edu/abs/2024ApJ...975...78R}
}

@ARTICLE{Thi_2005,
       author = {{Thi}, W. -F. and {van Dalen}, B. and {Bik}, A. and {Waters}, L.~B.~F.~M.},
        title = "{Evidence for a hot dust-free inner disk around 51 Oph}",
      journal = {\aap},
         year = 2005,
        month = jan,
       volume = {430},
        pages = {L61-L64},
          doi = {10.1051/0004-6361:200400132},
archivePrefix = {arXiv},
       eprint = {astro-ph/0412514},
 primaryClass = {astro-ph},
       adsurl = {https://ui.adsabs.harvard.edu/abs/2005A&A...430L..61T}
}

@article{dullemond_passive_2001,
       author = {{Dullemond}, C.~P. and {Dominik}, C. and {Natta}, A.},
        title = "{Passive Irradiated Circumstellar Disks with an Inner Hole}",
      journal = {\apj},
         year = 2001,
        month = oct,
       volume = {560},
       number = {2},
        pages = {957-969},
          doi = {10.1086/323057},
archivePrefix = {arXiv},
       eprint = {astro-ph/0106470},
 primaryClass = {astro-ph},
       adsurl = {https://ui.adsabs.harvard.edu/abs/2001ApJ...560..957D}
}

@ARTICLE{hein_bertelsen_proposed_2016,
       author = {{Hein Bertelsen}, R.~P. and {Kamp}, I. and {van der Plas}, G. and {van den Ancker}, M.~E. and {Waters}, L.~B.~F.~M. and {Thi}, W. -F. and {Woitke}, P.},
        title = "{A proposed new diagnostic for Herbig disc geometry. FWHM versus J of CO ro-vibrational lines}",
      journal = {\aap},
         year = 2016,
        month = may,
       volume = {590},
          eid = {A98},
        pages = {A98},
          doi = {10.1051/0004-6361/201527652},
archivePrefix = {arXiv},
       eprint = {1603.03546},
 primaryClass = {astro-ph.SR},
       adsurl = {https://ui.adsabs.harvard.edu/abs/2016A&A...590A..98H}
}

@ARTICLE{Perotti2023,
       author = {{Perotti}, G. and {Christiaens}, V. and {Henning}, Th. and {Tabone}, B. and {Waters}, L.~B.~F.~M. and {Kamp}, I. and {Olofsson}, G. and {Grant}, S.~L. and {Gasman}, D. and {Bouwman}, J. and {Samland}, M. and {Franceschi}, R. and {van Dishoeck}, E.~F. and {Schwarz}, K. and {G{\"u}del}, M. and {Lagage}, P. -O. and {Ray}, T.~P. and {Vandenbussche}, B. and {Abergel}, A. and {Absil}, O. and {Arabhavi}, A.~M. and {Argyriou}, I. and {Barrado}, D. and {Boccaletti}, A. and {Caratti o Garatti}, A. and {Geers}, V. and {Glauser}, A.~M. and {Justannont}, K. and {Lahuis}, F. and {Mueller}, M. and {Nehm{\'e}}, C. and {Pantin}, E. and {Scheithauer}, S. and {Waelkens}, C. and {Guadarrama}, R. and {Jang}, H. and {Kanwar}, J. and {Morales-Calder{\'o}n}, M. and {Pawellek}, N. and {Rodgers-Lee}, D. and {Schreiber}, J. and {Colina}, L. and {Greve}, T.~R. and {{\"O}stlin}, G. and {Wright}, G.},
        title = "{Water in the terrestrial planet-forming zone of the PDS 70 disk}",
      journal = {\nat},
         year = 2023,
        month = aug,
       volume = {620},
       number = {7974},
        pages = {516-520},
          doi = {10.1038/s41586-023-06317-9},
archivePrefix = {arXiv},
       eprint = {2307.12040},
 primaryClass = {astro-ph.EP},
       adsurl = {https://ui.adsabs.harvard.edu/abs/2023Natur.620..516P}
}

@ARTICLE{Arulanantham2025,
       author = {{Arulanantham}, Nicole and {Salyk}, Colette and {Pontoppidan}, Klaus and {Banzatti}, Andrea and {Zhang}, Ke and {{\"O}berg}, Karin and {Long}, Feng and {Carr}, John and {Najita}, Joan and {Pascucci}, Ilaria and {Colmenares}, Mar{\'\i}a Jos{\'e} and {Xie}, Chengyan and {Huang}, Jane and {Green}, Joel and {Andrews}, Sean M. and {Blake}, Geoffrey A. and {Bergin}, Edwin A. and {Pinilla}, Paola and {Vioque}, Miguel and {Dahl}, Emma and {Raul}, Eshan and {Krijt}, Sebastiaan and {The Jdiscs Collaboration}},
        title = "{The JDISC Survey: Linking the Physics and Chemistry of Inner and Outer Protoplanetary Disk Zones}",
      journal = {\aj},
         year = 2025,
        month = aug,
       volume = {170},
       number = {2},
          eid = {67},
        pages = {67},
          doi = {10.3847/1538-3881/addd01},
archivePrefix = {arXiv},
       eprint = {2505.07562},
 primaryClass = {astro-ph.SR},
       adsurl = {https://ui.adsabs.harvard.edu/abs/2025AJ....170...67A}
}

@ARTICLE{Kanwar2025,
       author = {{Kanwar}, Jayatee and {Woitke}, Peter and {Kamp}, Inga and {Rimmer}, Paul and {Helling}, Christiane},
        title = "{Can thermodynamic equilibrium be established in planet-forming disks?}",
      journal = {\aap},
         year = 2025,
        month = jun,
       volume = {698},
          eid = {A294},
        pages = {A294},
          doi = {10.1051/0004-6361/202452249},
archivePrefix = {arXiv},
       eprint = {2505.13705},
 primaryClass = {astro-ph.EP},
       adsurl = {https://ui.adsabs.harvard.edu/abs/2025A&A...698A.294K}
}

@ARTICLE{Salyk2025,
       author = {{Salyk}, Colette and {Pontoppidan}, Klaus M. and {Banzatti}, Andrea and {Bergin}, Edwin and {Arulanantham}, Nicole and {Najita}, Joan and {Blake}, Geoffrey A. and {Carr}, John and {Zhang}, Ke and {Xie}, Chengyan},
        title = "{Emission from Multiple Molecular Isotopologues in a High-inclination Protoplanetary Disk}",
      journal = {\aj},
         year = 2025,
        month = mar,
       volume = {169},
       number = {3},
          eid = {184},
        pages = {184},
          doi = {10.3847/1538-3881/adb397},
archivePrefix = {arXiv},
       eprint = {2502.05061},
 primaryClass = {astro-ph.SR},
       adsurl = {https://ui.adsabs.harvard.edu/abs/2025AJ....169..184S}
}

@ARTICLE{Woods2009,
       author = {{Woods}, Paul M. and {Willacy}, Karen},
        title = "{Carbon Isotope Fractionation in Protoplanetary Disks}",
      journal = {\apj},
         year = 2009,
        month = mar,
       volume = {693},
       number = {2},
        pages = {1360-1378},
          doi = {10.1088/0004-637X/693/2/1360},
archivePrefix = {arXiv},
       eprint = {0812.0269},
 primaryClass = {astro-ph},
       adsurl = {https://ui.adsabs.harvard.edu/abs/2009ApJ...693.1360W}
}

@ARTICLE{Wilson1999,
       author = {{Wilson}, T.~L.},
        title = "{Isotopes in the interstellar medium and circumstellar envelopes}",
      journal = {Reports on Progress in Physics},
         year = 1999,
        month = feb,
       volume = {62},
       number = {2},
        pages = {143-185},
          doi = {10.1088/0034-4885/62/2/002},
       adsurl = {https://ui.adsabs.harvard.edu/abs/1999RPPh...62..143W}
}

@ARTICLE{Gasman2023,
       author = {{Gasman}, Danny and {van Dishoeck}, Ewine F. and {Grant}, Sierra L. and {Temmink}, Milou and {Tabone}, Beno{\^\i}t and {Henning}, Thomas and {Kamp}, Inga and {G{\"u}del}, Manuel and {Lagage}, Pierre-Olivier and {Perotti}, Giulia and {Christiaens}, Valentin and {Samland}, Matthias and {Arabhavi}, Aditya M. and {Argyriou}, Ioannis and {Abergel}, Alain and {Absil}, Olivier and {Barrado}, David and {Boccaletti}, Anthony and {Bouwman}, Jeroen and {Caratti o Garatti}, Alessio and {Geers}, Vincent and {Glauser}, Adrian M. and {Guadarrama}, Rodrigo and {Jang}, Hyerin and {Kanwar}, Jayatee and {Lahuis}, Fred and {Morales-Calder{\'o}n}, Maria and {Mueller}, Michael and {Nehm{\'e}}, Cyrine and {Olofsson}, G{\"o}ran and {Pantin}, {\'E}ric and {Pawellek}, Nicole and {Ray}, Tom P. and {Rodgers-Lee}, Donna and {Scheithauer}, Silvia and {Schreiber}, J{\"u}rgen and {Schwarz}, Kamber and {Vandenbussche}, Bart and {Vlasblom}, Marissa and {Waters}, Rens L.~B.~F.~M. and {Wright}, Gillian and {Colina}, Luis and {Greve}, Thomas R. and {{\"O}stlin}, G{\"o}ran},
        title = "{MINDS. Abundant water and varying C/O across the disk of Sz 98 as seen by JWST/MIRI}",
      journal = {\aap},
         year = 2023,
        month = nov,
       volume = {679},
          eid = {A117},
        pages = {A117},
          doi = {10.1051/0004-6361/202347005},
archivePrefix = {arXiv},
       eprint = {2307.09301},
 primaryClass = {astro-ph.EP},
       adsurl = {https://ui.adsabs.harvard.edu/abs/2023A&A...679A.117G}
}




\newpage
\appendix

\section{CO ro-vibrational line emission}
\label{sec:CO}

\begin{figure}
    \centering
    \includegraphics[width=1.0\linewidth]{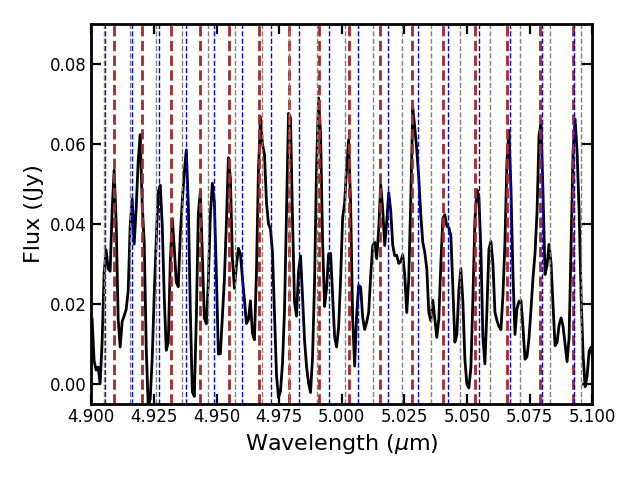}
    \caption{CO ro-vibrational emission lines in HD~35929. The dashed lines indicate the positions of the v=1-0 (red), v=2-1 (blue) and v=3-2 (grey) rotational lines. The observed spectrum has been corrected for the radial velocity of the star of $22\,\rm km/s$. }
    \label{fig:rot-co}
\end{figure}

In Fig.~\ref{fig:rot-co} we show the continuum subtracted MIRI spectrum of HD\,35929 between $4.9\,\rm \mu m$ and $5.1\,\rm \mu m$. This regions shows CO emission that is significantly broadened and blended as discussed in Sect.~\ref{sec:kinematics}. 

\section{Comparing HD\,35929 to water-rich T\,Tauri discs}
\label{sec:compare-ttauri}

\begin{figure*}
    \centering
    \includegraphics[width=1.0\linewidth]{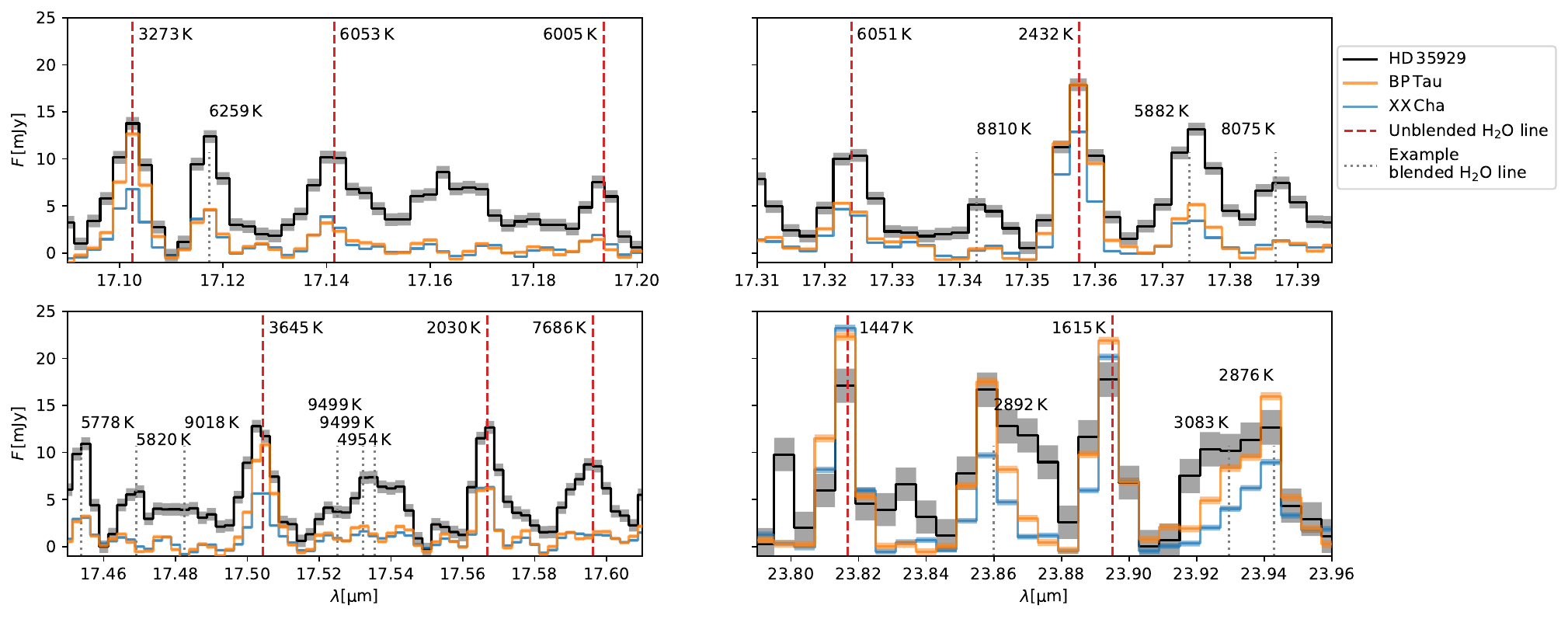}
    \caption{Selected wavelength windows showing the JWST/MIRI spectra of HD\,35929 (black) and the two T\,Tauri discs BP\,Tau and XX\,Cha \citep[orange and blue,][]{Temmink2025}, scaled to the same distance. A few unblended water lines \citep[red, as identified by][]{banzatti_water_2025} and examples of blended water lines (grey) are labelled with their upper level energy.}
    \label{fig:compare_ttauri}
\end{figure*}

This section compares the rotational water spectrum of HD\,35929 to the spectra of the water-rich T\,Tauri discs of BP\,Tau and XX\,Cha \citep{Temmink2025} which we show to be cold-water analogues. The spectra are both scaled to the distance of HD\,35929 ($380\,\rm pc$) and selected wavelength windows are compared in Fig.~\ref{fig:compare_ttauri}. The cold water lines look remarkably similar for all objects. This includes the coldest water lines (upper level energies of less than $2000\,\rm K$) around $23.82\,\rm \mu m$ and $23.90\,\rm \mu m$, but also the unblended water line at $17.36\,\rm \mu m$. The only exception seems to be the unblended line at $17.57\,\rm \mu m$, which is roughly a factor of $2$ stronger in HD\,35929. Unblended warm water lines (e.g. at $17.10\,\rm \mu m$ and $17.50\,\rm \mu m$) also exhibit strong similarities, especially for BP\,Tau. The relative enhancement of cold water lines compared to warm water lines of XX\,Cha have been noted by \cite{Temmink2025} already. We conclude that the emitting conditions of the cold (and warm) water reservoirs must be very similar between all objects.

\cite{Temmink2025} fitted BP\,Tau and XX\,Cha with a temperature and column density power law (among other profiles) allowing for a direct comparison between the retrieved conditions and HD\,35929. The profiles of both discs extend outwards to $\sim 2.0 \,\rm au$ with near constant column densities of $10^{18}\,\rm cm^{-2}$, which coincides extremely well with the cold water conditions retrieved for HD\,35929 (see Fig.~\ref{fig:water-con}). However, as shown in Fig.~\ref{fig:compare_ttauri}, the unblended water lines with upper level energies above $6000\,\rm K$, tracing the hot water component, are significantly stronger for HD\,35929 (seen at $17.14\,\rm \mu m$, $17.19\,\rm \mu m$, $17.32\,\rm \mu m$, and $17.60\,\rm \mu m$). Additionally, most of the wavelength ranges between the unblended water lines which show significantly more flux for HD\,35929 are dominated by blended lines with high upper level energies. This conclusively shows that the column densities of hot water must be higher than the values retrieved by \cite{Temmink2025} for BP\,Tau and XX\,Cha ($\sim 10^{18}\,\rm cm^{-2}$).

We note that even though many lines will be optically thick at their line centres, an increase in column density still results in an increase in line flux due to the increase in flux in the optically thin line wings. For example, a slab model with a temperature of $800\,\rm K$ shows an increase of flux at the hot water line at $17.32\,\rm \mu m$ (Einstein-A coefficient of $41.53$) by a factor of $1.87$ when increasing the column density from $10^{18}\,\rm cm^{-2}$ to $10^{19}\,\rm cm^{-2}$.

\section{Fitting the rotational water lines}
\label{sec:rot-water}
\begin{figure}
    \centering
    \includegraphics[width=1.0\linewidth]{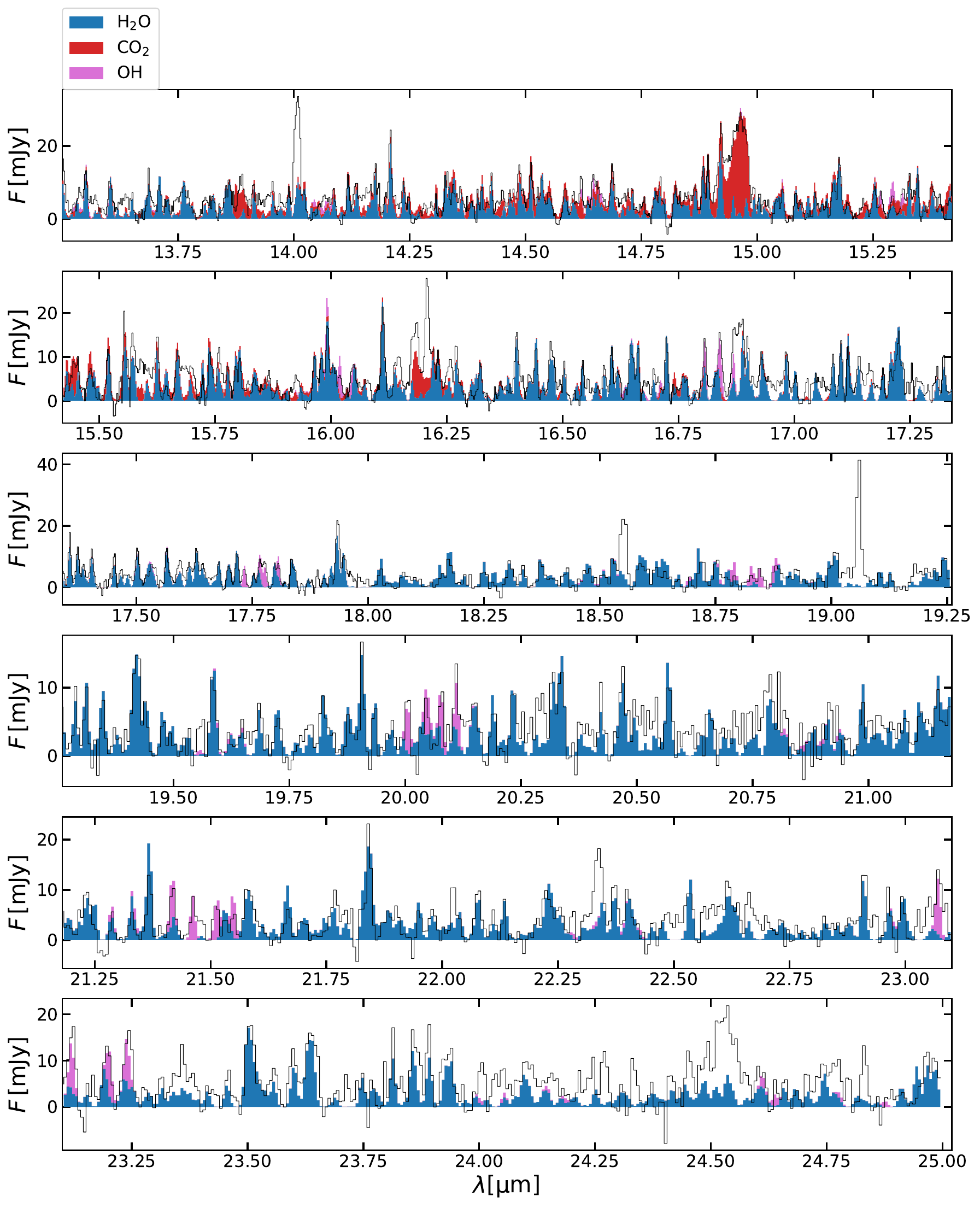}
    \caption{Fitting the rotational water lines in the continuum subtracted MIRI spectrum from $13.5\,\rm \mu m$ to $25.0\,\rm \mu m$. Overplotted are the contributions of H$_2$O, CO$_2$, and OH from the median probability model.}
    \label{fig:rot-water-fit}
\end{figure}

\begin{figure}
    \centering
    \includegraphics[width=1.0\linewidth]{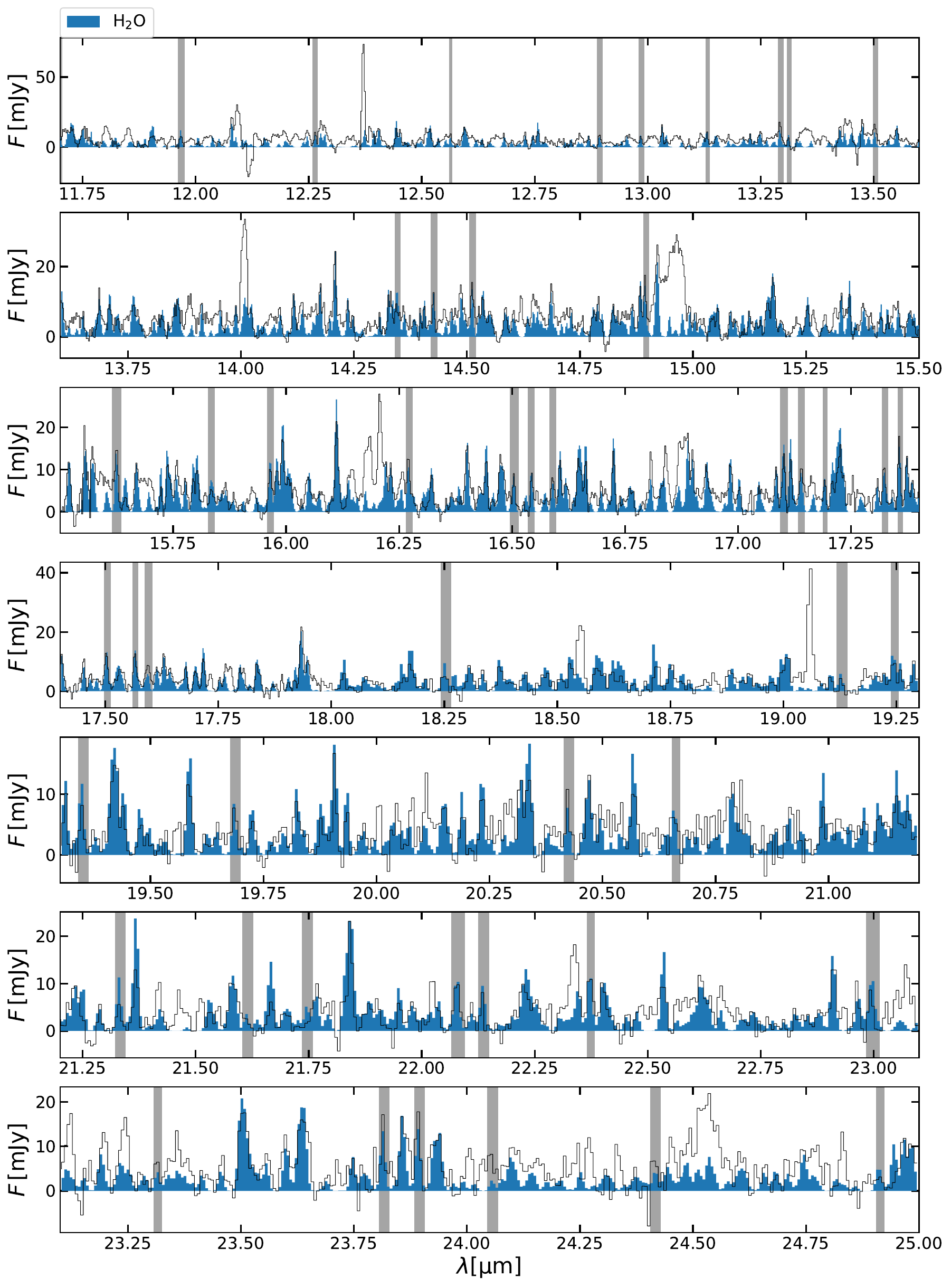}
    \caption{Fitting the unblended rotational water lines in the continuum subtracted MIRI spectrum. Overplotted is the contribution of H$_2$O from the median probability model. The shaded regions are the wavelengths windows used during fitting.}
    \label{fig:unblended-water-fit}
\end{figure}

This section shows the fits or the rotational and unblended water lines fits. The fits are explained in Sect.~\ref{sec:mol_conditions}. Fig.~\ref{fig:rot-water-fit} shows the fit of the rotational water lines in the range of $13.5-25.0\,\rm \mu m$ and Fig.~\ref{fig:unblended-water-fit} displays a fit to the unblended water lines between $11.7\,\rm \mu m$ and $25.0\,\rm \mu m$.

\section{Non-detections of isotopologues}
\label{sec:isos}
Given the high retrieved column densities, naturally the question arises if isotopologues of the fitted molecules are visible in the spectrum as well. Therefore, we investigate the possibility of features by \ce{^{13}CO2} and $\rm{H_2}^{18}\rm{O}$.

Multiple JWST/MIRI spectra of protoplanetary discs show a clear feature of \ce{^{13}CO2} around $15.4\,\rm \mu m$ \citep[e.g.][]{grant_minds_2023,Salyk2025}. The spectrum of HD\,35929 does not exhibit any sign of it (e.g. Fig.~\ref{fig:mol-fit}). If an abundance fraction of $70$ is assumed between \ce{^{12}CO2} and \ce{^{13}CO2} \citep{Woods2009} and line overlap between the isotopologues is taken into account, the model fitted to the spectrum (Fig.~\ref{fig:mol-fit}) would show a peak \ce{^{13}CO2} line flux of $13.44\,\rm mJy$, which would be detectable. We can only speculate regarding the reasons for the observed non-detection. \ce{^{13}CO2} is optically thinner than \ce{^{12}CO2} and therefore potentially traces deeper disc layers with lower temperatures. This effect has been observed by \cite{grant_minds_2023}, with a temperature difference of $75\,\rm K$. For the retrieved model, decreasing the temperature by $200\,\rm K$ results in peak fluxes of $6.88\,\rm mJy$ which even though significantly larger than the estimated noise by the JWST Exposure Time Calculator (ETC) at this wavelength ($0.7\,\rm mJy$) might result in a non-detection due to the uncertainties of the continuum subtraction and interference with other molecular features. Similarly, it is possible that the column density of \ce{^{13}CO2} is lower than expected. This can be due to a lower \ce{^{13}CO2} to \ce{^{12}CO2} ratio or due to an overestimation of the \ce{^{12}CO2} column density. As discussed in Section~\ref{sec:mol_conditions}, fitting only the Q-branch results in lower column densities by approximately a factor of $10$. Progressing this to \ce{^{13}CO2} would result in a hard-to-detect peak flux of $4.74\,\rm mJy$.

Water isotopologues are difficult to detect in JWST/MIRI spectra \citep{Gasman2023,temmink_minds_2024}, due to the strong overlap with water lines and the low isotopologue ratio \citep[$^{16}\rm O$/$^{18}\rm O$ of 550,][]{Wilson1999}. Salyk et al. (in prep.) identified several isolated $\rm{H_2}^{18}\rm{O}$ lines that will assist with future detections. All lines are longwards of $22.7\,\rm \mu m$ and have upper level energies below $3300\,\rm K$. A visual inspection of the spectrum did not result in conclusive evidence for any $\rm{H_2}^{18}\rm{O}$ line. However, at these long wavelength the spectrum of HD\,35929 is dominated by noise. At $24\,\rm \mu m$ and $26\,\rm \mu m$  the ETC estimated noise is $1.84\,\rm mJy$ and $3.89\,\rm mJy$, respectively. Additional uncertainty sources like the continuum subtraction complicate the isotopologue detection further.

\section{Measured emission lines.}
\label{sec:meas-emis-line}

\begin{table*}
\caption{Measured data of selected molecular emission lines and strong HI recombination lines, using iSLAT. Given are the quantum levels of the upper and lower levels, the rest wavelength of the transition, the line flux and error from the Gaussian fit, the measured Full Width at Half Maximum (FWHM) and its error, corrected for the MIRI instrument spectral resolution, and the Doppler shift of the line, assuming a stellar radial velocity of 22 km/s.}  
\label{tab:lines}
\begin{tabular}{llllllll}

\hline
\hline
         Upper level &                Lower level &   Wavelength & Line flux&    Line flux error &    FWHM & FWHM error  &  Doppler shift \\
           &    &  $\mu$m &  \multicolumn{2}{c}{erg/cm$^{2}$ s} &  \multicolumn{2}{c}{km/s} & km/s \\ \hline

\\
\multicolumn{8}{c}{CO} \\
\\
\hline
             1 &                 0|P-28 &    4.94336 &   1.64e-14 &   2.75e-15 &      115.1 &        27.3 &     -9.8 \\
             1 &                 0|P-31 &    4.97878 &   2.59e-14 &   8.15e-16 &      135.7 &         5.3 &      3.1 \\
             2 &                 1|P-26 &    4.98310 &   1.41e-14 &   2.42e-15 &      196.1 &        34.9 &    -21.4 \\
             3 &                 2|P-26 &    5.04724 &   9.69e-15 &   1.36e-15 &      125.4 &        22.7 &    -14.7 \\
             3 &                 2|P-27 &    5.05907 &   1.62e-14 &   2.60e-15 &      194.8 &        31.5 &     11.9 \\
             1 &                 0|P-49 &    5.21766 &   1.37e-14 &   5.38e-15 &      303.5 &       115.2 &    -15.1 \\
             1 &                 0|P-52 &    5.26221 &   6.02e-15 &   7.23e-16 &       85.4 &        18.5 &    -10.4 \\
\hline
\\
\multicolumn{8}{c}{SiO v=1-0} \\
\\
\hline
             56 &                     55 &    7.72822 &   6.47e-15 &   1.38e-15 &      215.4 &        42.1 &   -197.9 \\
             44 &                     43 &    7.79399 &   3.47e-15 &   3.63e-16 &       76.7 &        13.9 &     -1.7 \\
             33 &                     32 &    7.86396 &   1.14e-14 &   3.74e-15 &      263.3 &        78.2 &      7.0 \\
             36 &                     35 &    7.84394 &   3.81e-15 &   4.42e-16 &      115.6 &        15.9 &     26.7 \\
             17 &                     16 &    7.98289 &   5.70e-15 &   1.50e-15 &      133.0 &        38.0 &      3.5 \\
              6 &                      5 &    8.07693 &   5.28e-15 &   5.37e-16 &      104.2 &        13.4 &     -8.4 \\
              8 &                      9 &    8.22220 &   2.82e-15 &   5.93e-16 &       64.9 &        26.9 &     10.6 \\
             30 &                     31 &    8.47350 &   4.59e-15 &   3.83e-16 &      143.8 &        11.9 &    -17.1 \\
             34 &                     35 &    8.52443 &   4.55e-15 &   2.21e-15 &      219.1 &       102.2 &    -63.3 \\
             35 &                     36 &    8.53742 &   5.15e-15 &   1.08e-15 &      192.2 &        36.9 &     -7.0 \\
             39 &                     40 &    8.59049 &   5.14e-15 &   9.33e-16 &      129.2 &        24.4 &     13.2 \\
\hline
\\
\multicolumn{8}{c}{OH} \\
\\
\hline
    X3/2\_0|15.0 &  X3/2\_0|RR\_13.5ff\_14.0 &   20.00854 &   8.55e-16 &   2.54e-16 &      100.5 &        61.0 &     -6.0 \\
    X3/2\_0|15.0 &  X3/2\_0|RR\_13.5ee\_14.0 &   20.04975 &   9.79e-16 &   3.91e-16 &      201.0 &        85.2 &    -25.5 \\
    X1/2\_0|14.0 &  X1/2\_0|RR\_12.5ee\_13.0 &   20.08194 &   8.69e-16 &   2.12e-16 &       88.6 &        51.5 &    -11.5 \\
    X3/2\_0|14.0 &  X3/2\_0|RR\_12.5ff\_13.0 &   21.41989 &   1.10e-15 &   1.09e-16 &      149.5 &        22.0 &     20.2 \\
    X1/2\_0|13.0 &  X1/2\_0|RR\_11.5ee\_12.0 &   21.51689 &   7.71e-16 &   3.22e-16 &      162.6 &        92.9 &    -26.1 \\
\hline
\\
\multicolumn{8}{c}{HI} \\
\\
\hline
n$_{\rm up}$ & n$_{\rm low}$ & & & & & & \\
\hline
              6 &                      5 &    7.45988 &   3.47e-14 &   2.12e-15 &      125.1 &         8.0 &     -4.6 \\
             10 &                      7 &    8.76006 &   1.11e-14 &   1.03e-15 &      131.0 &        12.5 &     13.3 \\
              6 &                      5 &    7.45988 &   3.47e-14 &   2.04e-15 &      125.0 &         7.6 &     -4.6 \\
             10 &                      7 &    8.76006 &   1.12e-14 &   1.20e-15 &      134.7 &        14.6 &     13.8 \\
              9 &                      7 &   11.30869 &   6.58e-15 &   1.61e-16 &      141.0 &         4.2 &     -7.0 \\
              7 &                      6 &   12.37192 &   1.07e-14 &   6.97e-16 &      113.7 &        10.9 &      2.7 \\
             10 &                      8 &   16.20910 &   3.64e-15 &   2.44e-16 &      167.7 &        13.2 &     -5.7 \\
              8 &                      7 &   19.06189 &   4.49e-15 &   3.94e-16 &      122.7 &        19.1 &      0.2 \\
             11 &                      9 &   22.34046 &   2.75e-15 &   9.97e-17 &      287.1 &        10.0 &     -3.3 \\

\hline
\end{tabular}
\end{table*}

\addtocounter{table}{-1}

\begin{table*}
\caption{continued.}  
\begin{tabular}{llllllll}

\hline
\hline
         Upper level &                Lower level &   Wavelength & Line flux&    Line flux error &    FWHM & FWHM error  &  Doppler shift \\
           \multicolumn{2}{c}{$v_1 v_2 v_3~|~J_{\:K_a \:K_c}$}    &  $\mu$m &  \multicolumn{2}{c}{erg/cm$^{2}$ s} &  \multicolumn{2}{c}{km/s} & km/s \\ \hline

\hline
\\
\multicolumn{8}{c}{H$_2$O} \\
\\
\hline
    010|$2_{\:2\:1 }$&            000|$2_{\:1\:2 }$&    6.01392 &   1.31e-14 &   2.54e-15 &      359.7 &        65.2 &    210.7 \\
    010|$3_{\:2\:1 }$&            000|$3_{\:1\:2 }$&    6.07545 &   6.18e-15 &   8.84e-16 &      193.6 &        26.3 &     14.6 \\
    010|$1_{\:0\:1 }$&            000|$1_{\:1\:0 }$&    6.34443 &   5.29e-15 &   3.75e-16 &      177.5 &        11.8 &     -1.6 \\
    010|$5_{\:3\:2 }$&            000|$6_{\:4\:3 }$&    7.27924 &   4.41e-15 &   4.24e-16 &      121.7 &        12.8 &    -24.2 \\
    010|$5_{\:3\:3 }$&            000|$6_{\:4\:2 }$&    7.30659 &   5.55e-15 &   3.28e-16 &      192.1 &        10.8 &     12.1 \\
   000|$14_{\:8\:7 }$&           000|$13_{\:5\:8 }$&   11.96812 &   1.28e-15 &   2.95e-16 &      122.4 &        42.5 &     13.9 \\
  000|$18_{\:7\:12 }$&          000|$17_{\:4\:13 }$&   12.26544 &   1.44e-15 &   2.06e-16 &      111.9 &        24.8 &      8.6 \\
   000|$10_{\:6\:5 }$&            000|$9_{\:1\:8 }$&   12.56450 &   1.68e-15 &   5.80e-16 &      186.4 &        75.0 &    -21.6 \\
   000|$12_{\:5\:7 }$&          000|$11_{\:2\:10 }$&   12.89409 &   1.21e-15 &   2.47e-16 &      127.6 &        31.6 &     21.6 \\
   000|$12_{\:7\:5 }$&           000|$11_{\:4\:8 }$&   12.98575 &   1.08e-15 &   2.50e-16 &      167.0 &        39.9 &      5.4 \\
  000|$16_{\:7\:10 }$&          000|$15_{\:4\:11 }$&   13.13243 &   1.27e-15 &   2.61e-16 &      128.3 &        31.1 &     -0.1 \\
  000|$15_{\:3\:12 }$&          000|$14_{\:2\:13 }$&   13.29319 &   2.81e-15 &   3.72e-16 &      190.4 &        24.6 &      0.7 \\
  000|$16_{\:4\:12 }$&          000|$15_{\:3\:13 }$&   13.31231 &   5.59e-16 &   2.04e-16 &        8.9 &       266.6 &    -10.5 \\
   000|$11_{\:7\:4 }$&           000|$10_{\:4\:7 }$&   13.50312 &   3.03e-15 &   2.25e-16 &      214.1 &        17.8 &     -5.3 \\
  000|$14_{\:3\:11 }$&          000|$13_{\:2\:12 }$&   14.34608 &   2.03e-15 &   7.45e-16 &      320.2 &       122.0 &     20.4 \\
  000|$15_{\:4\:11 }$&          000|$14_{\:3\:12 }$&   14.42757 &   1.28e-15 &   3.17e-16 &      102.6 &        41.7 &    -16.5 \\
  000|$13_{\:2\:11 }$&          000|$12_{\:1\:12 }$&   14.51301 &   2.39e-15 &   3.80e-16 &      213.8 &        34.5 &     -0.5 \\
  000|$14_{\:5\:10 }$&          000|$13_{\:2\:11 }$&   14.89513 &   1.89e-15 &   3.94e-16 &      133.7 &        35.4 &     -0.7 \\
  000|$13_{\:3\:10 }$&          000|$12_{\:2\:11 }$&   15.62568 &   2.25e-15 &   4.24e-16 &      210.9 &        41.4 &    -16.4 \\
  000|$18_{\:8\:10 }$&          000|$17_{\:7\:11 }$&   15.83495 &   8.64e-16 &   1.21e-16 &      221.6 &        34.0 &     32.2 \\
   000|$13_{\:5\:9 }$&          000|$12_{\:2\:10 }$&   15.96622 &   1.24e-15 &   1.75e-16 &      143.6 &        26.5 &      4.7 \\
  000|$15_{\:5\:10 }$&          000|$14_{\:4\:11 }$&   16.27136 &   1.51e-15 &   2.40e-16 &      153.7 &        29.7 &      3.1 \\
  000|$17_{\:7\:10 }$&          000|$16_{\:6\:11 }$&   16.50525 &   1.02e-15 &   3.57e-16 &      213.4 &        74.5 &    -24.0 \\
   000|$11_{\:6\:6 }$&           000|$10_{\:3\:7 }$&   16.54402 &   7.30e-16 &   1.12e-16 &       48.4 &        35.1 &     17.9 \\
   000|$16_{\:9\:7 }$&           000|$15_{\:8\:8 }$&   16.59123 &   9.77e-16 &   6.91e-17 &      161.2 &        13.3 &     -0.1 \\
   000|$12_{\:5\:8 }$&           000|$11_{\:2\:9 }$&   17.10254 &   1.17e-15 &   1.27e-16 &       97.7 &        17.1 &    -11.0 \\
   000|$16_{\:8\:8 }$&           000|$15_{\:7\:9 }$&   17.14148 &   1.38e-15 &   1.69e-16 &      199.6 &        25.9 &      1.6 \\
   010|$13_{\:4\:9 }$&          010|$12_{\:3\:10 }$&   17.19352 &   5.98e-16 &   7.29e-17 &       77.1 &        20.3 &    -16.0 \\
   000|$16_{\:8\:9 }$&           000|$15_{\:7\:8 }$&   17.32395 &   9.06e-16 &   9.77e-17 &       93.9 &        16.6 &      1.1 \\
   000|$11_{\:2\:9 }$&          000|$10_{\:1\:10 }$&   17.35766 &   1.18e-15 &   9.13e-17 &       47.4 &        15.3 &     -4.2 \\
   000|$13_{\:4\:9 }$&          000|$12_{\:3\:10 }$&   17.50436 &   1.30e-15 &   1.01e-16 &      135.2 &        12.8 &    -20.0 \\
    000|$8_{\:6\:3 }$&            000|$7_{\:3\:4 }$&   17.56683 &   1.13e-15 &   9.30e-17 &      106.4 &        12.7 &      0.4 \\
   010|$15_{\:7\:8 }$&           010|$14_{\:6\:9 }$&   17.59626 &   1.31e-15 &   1.00e-16 &      234.4 &        17.6 &    -17.2 \\
   000|$11_{\:5\:7 }$&           000|$10_{\:2\:8 }$&   18.25429 &   7.79e-16 &   2.68e-16 &      175.7 &        81.1 &    -19.6 \\
   000|$11_{\:3\:8 }$&           000|$10_{\:2\:9 }$&   19.24597 &   1.23e-15 &   1.14e-16 &      162.6 &        20.1 &      8.7 \\
   000|$10_{\:5\:6 }$&            000|$9_{\:2\:7 }$&   19.34995 &   6.99e-16 &   2.72e-16 &       62.7 &       103.3 &     16.1 \\
   000|$14_{\:7\:8 }$&           000|$13_{\:6\:7 }$&   19.68805 &   1.02e-15 &   3.38e-16 &      227.5 &        78.0 &     -9.0 \\
   000|$13_{\:7\:7 }$&           000|$12_{\:6\:6 }$&   20.42595 &   8.34e-16 &   2.73e-16 &       90.5 &        62.1 &     -3.3 \\
    000|$7_{\:4\:3 }$&            000|$6_{\:1\:6 }$&   20.66181 &   1.71e-15 &   1.15e-16 &      518.0 &        36.3 &    -79.6 \\
   000|$12_{\:7\:6 }$&           000|$11_{\:6\:5 }$&   21.33317 &   4.62e-16 &   7.63e-17 &       83.0 &        43.0 &     -5.2 \\
   000|$11_{\:4\:7 }$&           000|$10_{\:3\:8 }$&   22.08091 &   1.11e-15 &   1.12e-16 &      177.4 &        22.2 &     -3.6 \\
   000|$11_{\:7\:4 }$&           000|$10_{\:6\:5 }$&   22.37473 &   1.22e-15 &   1.19e-16 &      168.8 &        21.9 &      7.8 \\
   000|$12_{\:6\:7 }$&           000|$11_{\:5\:6 }$&   22.99881 &   6.15e-16 &   1.64e-16 &       64.3 &        66.6 &    -34.7 \\
   000|$10_{\:6\:5 }$&           000|$10_{\:3\:8}$ &   23.31846 &   5.79e-16 &   8.78e-17 &      123.2 &        30.4 &    -31.9 \\
    000|$8_{\:4\:5 }$&            000|$7_{\:1\:6 }$&   23.89518 &   1.23e-15 &   6.41e-17 &       87.6 &        10.4 &    -14.8 \\
  010|$16_{\:5\:12 }$&          010|$15_{\:4\:11 }$&   24.05845 &   1.10e-15 &   4.39e-16 &      300.3 &       122.3 &    -40.4 \\
    010|$9_{\:6\:3 }$&            010|$8_{\:5\:4 }$&   24.91403 &   3.98e-16 &   1.57e-16 &      268.3 &       137.1 &     -4.6 \\

\end{tabular}
\end{table*}


\bsp	
\label{lastpage}
\end{document}